\documentclass[a4paper,fleqn,usenatbib]{mnras}

\usepackage[T1]{fontenc}
\usepackage{ae,aecompl}
\usepackage{array}
\usepackage{xspace}
\usepackage{graphicx}	
\usepackage{amsmath}	
\usepackage{amssymb}	

\title[WARPFIELD-POPulation Synthesis]{WARPFIELD Population Synthesis: The physics of (extra-)Galactic star formation and feedback driven cloud structure and emission from sub-to-kpc scales}

\author[Pellegrini and Reissl]{%
Eric W. Pellegrini$^{1}$\thanks{E-mail: eric.pellegrini@uni-heidelberg.de} \& Stefan Reissl$^{1}$\thanks{E-mail: reissl@uni-heidelberg.de}, 
\newauthor
and Daniel Rahner$^{1}$, Ralf S. Klessen$^{1,2}$, Simon C. O. Glover$^{1}$,  R{\"u}diger Pakmor$^{3}$,\newauthor Rodrigo Herrera-Camus$^{4}$ and Robert J. J. Grand$^3$\\
$^{1}$Universit{\"a}t Heidelberg, Zentrum f{\"u}r Astronomie, Institut f{\"u}r Theoretische Astrophysik, Albert-Ueberle-Str. 2, 69120 Heidelberg, Germany\\
$^{2}$Universit{\"a}t Heidelberg, Interdisziplin{\"a}res Zentrum f{\"u}r Wissenschaftliches Rechnen, Im Neuenheimer Feld 205,  69120 Heidelberg, Germany\\
$^{3}$Max-Planck-Institut f{\"u}r Astrophysik, Karl-Schwarzschild-Str. 1
85748 Garching, Germany\\
$^{4}$Departamento de Astronom\'ia, Universidad de Concepci\'on, Casilla 160-C, Concepci\'on, Chile
}

\date{Accepted XXX. Received YYY; in original form ZZZ}

\pubyear{2018}

\newcommand{\hii}{H$\,${\sc ii}\xspace}

\newcommand{\Ha}{H$\rm \alpha$\xspace}
\newcommand{\Hb}{H$\beta$\xspace}

\newcommand{\SIII}{[\ion{S}{III}]\xspace}

\newcommand{\OIII}{[\ion{O}{III}]\xspace}

\hypersetup{draft}

\begin{document}
\label{firstpage}
\pagerange{\pageref{firstpage}--\pageref{lastpage}}
\maketitle

\begin{abstract}

We present a novel method to model galactic scale star formation, including the resulting emission of star clusters and from the multi-phase interstellar medium. We combine global parameters, including star formation rate and metallicity, with  {\sc warpfield} which determines the feedback-driven evolution of individual star-forming regions. Our approach includes stellar evolution, stellar winds, radiation pressure, supernovae, all of which couple to the dynamical evolution of the parental cloud in a highly non-linear fashion. The heating of the diffuse galactic gas and dust component is calculated self-consistently with the age, mass and density dependent escape fractions of photons from local star-forming regions. From this we construct the interstellar radiation field, and we employ the multi-frequency Monte Carlo radiative transfer code  {\sc polaris} to produce synthetic emission maps for the one-to-one comparison with observations. 

We apply this to a Milky Way like galaxy built-up in a high-resolution MHD simulation of cosmic structure formation. We give three examples of applications of the method. First, we produce the multi-scale distribution of electron density and temperature and compute the resulting synthesized all-sky spatial distribution of H$\alpha$ emission. We use a multipole expansion to show that the resulting maps reproduce all observed statistical emission characteristics. Second, we predict the expected \SIII 9530~\AA\ emission, a key line that will be observed in several large forthcoming surveys. It suffers less extinction than other diagnostic lines and provides information about star formation in very dense environments that are otherwise observationally inaccessible in the optical. Third,  we explore the effects of differential extinction as seen by an extragalactic observer, and discuss the consequences for the correct interpretation of \Ha emission at different viewing angles.

\end{abstract}



\section{Introduction}

Recent years have seen dramatic improvements in our ability to model the formation and evolution of realistic spiral galaxies within large cosmological simulations 
\citep[e.g.][]{Grand2017,Hopkins2018}. At the same time, the advent of ALMA and of large integral field unit (IFU) spectrographs such as MUSE \citep{MUSE} or SITELLE \citep{Sitelle} has for the first time made it possible to map both gas and star formation on small ($\sim 100\ \mathrm{pc}$) scales within a large sample of local galaxies. \citep[e.g.][]{Kreckel2018, Rousseau2018}

An obvious next step is to compare the predictions of the simulations with observations of real galaxies, but this remains a highly challenging problem. Although high-resolution cosmological simulations can now resolve individual star-forming regions on scales of 10--100s of parsecs, this is not the same as being able to resolve individual young stellar clusters. In addition, the small scale physics of the interstellar medium (ISM) and stellar feedback is often treated in a highly idealized fashion in these simulations, and so the information necessary for making realistic synthetic emission maps based on the simulations is frequently not directly available (see e.g.\ the Illustris or Illustris-TNG simulations, introduced in \citealt{Vogelsberger2014} and \citealt{Springel2018}, respectively, which use the effective equation of state from \citealt{Springel2003} to model the pressure of the dense interstellar medium and hence do not directly follow the dense gas temperature).

On the observational side, even at the resolution achievable with ALMA and MUSE, individual star-forming regions are not fully resolved (pc scale), unless one focusses solely on very nearby galaxies such as the Magellanic clouds or M31.
Observations of more distant galaxies convolve together light from multiple stellar clusters of different masses, ages and potentially also metallicities within a single aperture. Observations of emission lines within these apertures therefore probe gas with a range of different physical conditions, exposed to a variety of radiation fields, greatly complicating efforts to compare the emission line strengths with the predictions of simple single-component photoionization and photodissociation region (PDR) models. 
Depending on the wavelength considered, these measurements are often also contaminated by diffuse emission powered by the field star population or by ionizing photons that manage to escape from the immediate vicinity of star-forming clouds \citep[e.g.][]{Medling2018,Tomicic2019}.

In addition, some quantities of great observational interest, such as the distribution of Faraday rotation measures \citep[see e.g.][]{Oppermann2012}, are sensitive to the small-scale distribution of young massive clusters via their impact on the galactic free electron distribution, but also depend directly on large-scale features such as the structure and strength of the magnetic field. Making realistic predictions for these quantities therefore requires us to take a holistic view of a model galaxy, rather than one focused on individual star-forming regions.

Since cosmological simulations do not currently have sufficient resolution to directly predict the locations or masses of young massive clusters, if we want to use these simulations to make observational predictions of e.g.\ star formation diagnostics or Faraday rotation measures, it is necessary to adopt a population synthesis approach in which we add a population of  clusters to the galactic gas distribution provided by the simulation and then use the resulting combined model of stars and gas to generate our synthetic observables. However, an important consideration when constructing such a population synthesis model is the relationship between the gas and the young stars. Given an overall star formation rate for a model galaxy, or a coarse-resolution map of the star formation rate surface density, it is straightforward to generate a population of young clusters by sampling from an appropriate cluster mass function and depositing the clusters randomly within the model galaxy. However, such an approach results in a cluster distribution which takes no account of the gas distribution. For old stellar clusters, this may be a reasonable approximation, but for the young clusters that contribute most of the stellar feedback, it is a poor approximation, since we know that these clusters must have formed within gravitationally unstable clouds of gas.

In this paper we present a new method for building up a population of clusters within a simulated galaxy that accounts for the link between the gas distribution and the cluster locations, and that allows us to model not just the direct emission from the clusters but also the diffuse emission from the ISM. Our method is based upon the {\sc warpfield-emp} code (Pellegrini et al., in prep.). This combines the {\sc warpfield} stellar feedback model, described in \citet{Rahner2017,Rahner2018b}, with the {\sc cloudy} PDR code\footnote{\url{http://www.nublado.org/}} \citep[][]{Ferland2017} and the {\sc polaris} radiative transfer (RT) code\footnote{\url{http://www1.astrophysik.uni-kiel.de/~polaris}} \citep{Reissl2016}. {\sc warpfield} models the impact of a stellar cluster on its surrounding cloud, accounting for a wide range of different feedback processes (radiation in ionizing and non-ionizing wavebands, stellar winds, supernovae) and solving for the dynamical evolution of the gas in spherical symmetry. The results of the model are then post-processed using {\sc cloudy} (to generate emissivities) and {\sc polaris} (to account for line RT, dust absorption, and synthetic observations), yielding predictions for the emission from the cloud/cluster system in the continuum and a large number of lines (Pellegrini et al., in prep.). The method is fast and computationally efficient, and thus allows us to put together a large database of cloud/cluster models that cover the entire parameter space relevant for normal spiral galaxies. Hence, for any combination of cloud masses, gas densities, and most importantly star formation efficiencies, we can describe the state and lifetime of individual clouds. 

Here, we connect our {\sc warpfield-emp} models to a Milky Way-like galaxy produced within a cosmological simulation taken from the Auriga project \citep{Grand2017}. However, we note that in principle our new population synthesis method is compatible with any type of mock or simulated galaxy. In the proof of concept presented here, we restrict our attention to a subset of the observational tracers that can be studied using the model. 

We generate synthetic maps of \Ha, \Hb, and \SIII~9530~\AA\xspace line emission as well as the Faraday rotation measure (published in a companion paper), considering both what would be seen by an observer within the galaxy and also what would be seen by an external observer.  We focus on \SIII rather than the more commonly used \OIII~5007\AA\xspace line because although both have similar ionization potentials and hence trace similar regions of massive star formation, the longer wavelength of the \SIII line means that it is less affected by dust extinction, allowing it to probe more distant or more embedded \hii regions than \OIII. Nevertheless, our method can also be used to make maps of the \OIII line, 
as well as many other different observational tracers, ranging from 
polarized dust emission to atomic and molecular line emission.

\begin{figure*}
	\centering
	\includegraphics[width=0.75\linewidth]{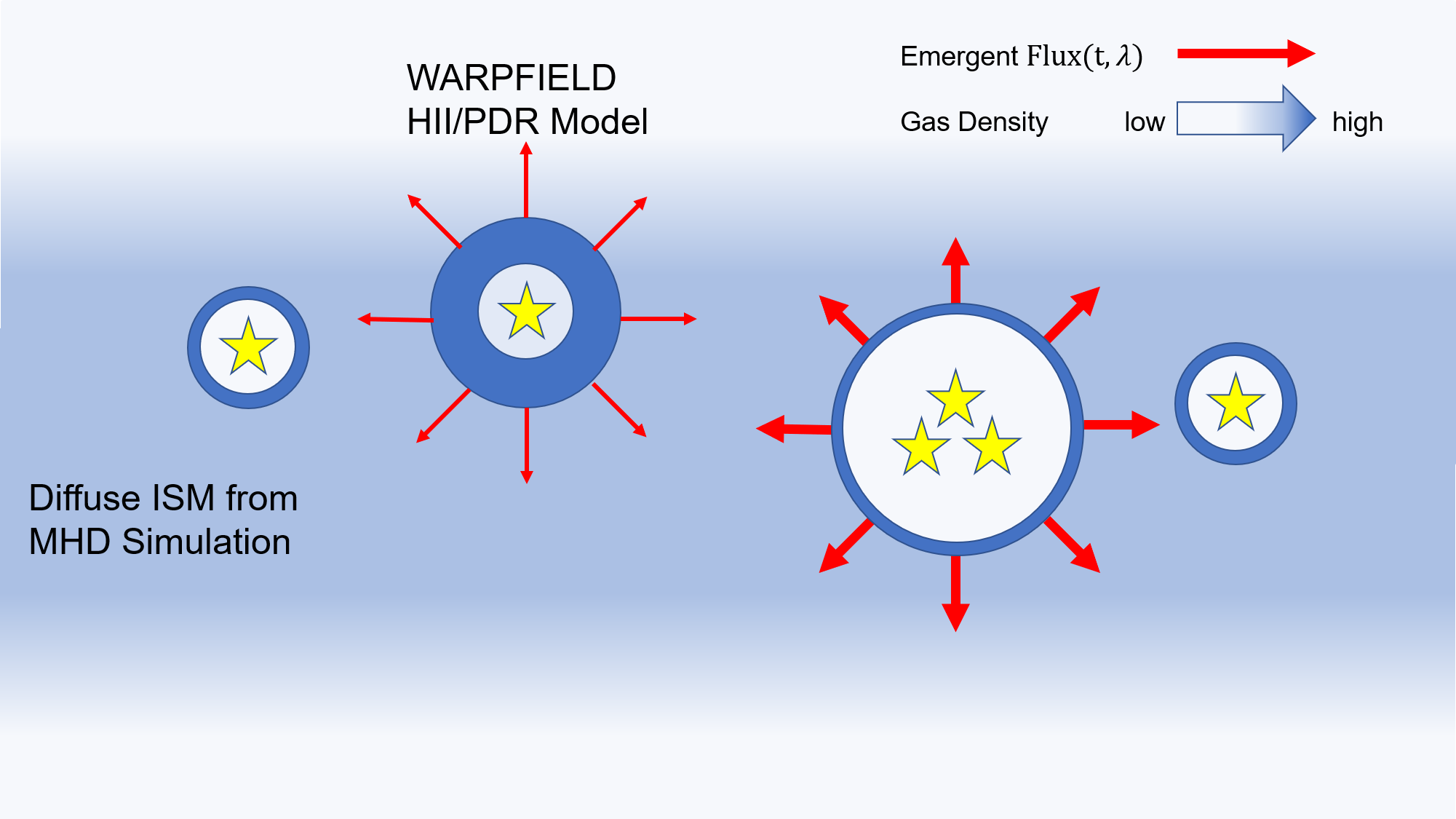}
	\caption{A schematic of the model presented. A distribution of cluster ages is derived from a star formation history and a cluster mass function. The associated \hii region and PDR evolution is determined at each evolutionary point with {\sc warpfield-emp}, including the line emission. Finally the emergent radiation from each region is then injected back into the galaxy as a source of heating and ionization, added to a diffuse galactic radiation field.}
	\label{fig:cartoon-aurandwarpfield}
\end{figure*}

Using Figure~\ref{fig:cartoon-aurandwarpfield} as a guide, our paper is structured in the following manner. In Section~\ref{sect:TestSetup} we describe the cosmological simulation of a Milky Way type galaxy (diffuse gas) that we populate with {\sc warpfield} star cluster models, shown as clusters surrounded by expanding shells. 
 In Section~\ref{sec:popsyn}, we present the main features of the {\sc warpfield-pop} model. In particular, in  Section~\ref{sect:ClusterMassDistribution} we describe our population synthesis model, which depends on an input star formation rate and cluster mass function, and in Section~\ref{sect:Photoionization} we describe how we use the emergent radiation  (red arrows emerging from the model) computed by the {\sc warpfield-emp} models to photo-ionize the disk. Line emission from the \hii region complexes and diffuse gas, as well as its transfer through the galaxy is treated in Section~\ref{sect:RT}, followed an in depth analysis of the synthetic observations in Section~\ref{sect:Observations}. We discuss some of our main results in Section~\ref{sect:results} and close with a brief summary in Section~\ref{sect:Summary}.

\section{Cosmological simulation}
\label{sect:TestSetup}
The model galaxy that we post-process using the {\sc warpfield} population synthesis method is taken from a simulation carried out as part of the Auriga project. It comes from a set of 30 cosmological magnetohydrodynamical (MHD) zoom-in simulations of isolated Milky Way-like galaxies. The simulations assume $\Lambda$CDM, with cosmological parameters taken from  the Planck Collaboration (\citeyear{Planck2014B}). They begin at a redshift of $z=127$ and are evolved all the way to $z=0$. The initial conditions were generated by selecting regions in the dark matter only version of the EAGLE simulation \citep[][]{McAlpine2016} that form dark matter halos with $M_{\rm vir} \sim 10^{12} \: {\rm M_{\odot}}$, with the restriction that the halos should not be too close to other halos of comparable mass. The selection criteria (described in more detail in \citealt{Grand2017}) yielded 174 candidate regions, of which 30 were selected for further study using a standard zoom-in approach. The zoom-in simulations include both gas and dark matter and account for a wide variety of physical processes, including primordial and metal line cooling (with a correction for self-shielding), the influence of the extragalactic UV background, star formation, stellar evolution and metal return \citep[][]{Vogelsberger2013}. Moreover, the simulations employ an effective model for Galactic winds \citep[][]{Marinacci2014, Grand2017}, and a prescription for the formation and growth of black holes and their feedback \citep[][]{Grand2017}. All simulations include magnetic fields that are seeded with small amplitudes at $z=127$ and are self-consistently evolved until $z=0$ \citep[][]{Pakmor2013, Pakmor2014, Pakmor2017}.

The simulations employ the moving mesh code {\sc Arepo} that solves the equations of MHD coupled with self-gravity on an unstructured Voronoi grid that evolves with time in a quasi-Lagrangian fashion \citep{Springel2010}. The Auriga simulation suite focuses on two sets of simulations at different resolution:  $30$ galaxies at standard resolution (level 4, $M_\mathrm{baryon} = 2\times 10^4 - 5\times 10^4 \,\mathrm{M_\odot}$) and 6 halos at high resolution (level 3 and level 4, $M_\mathrm{baryon} = 3\times 10^3 - 6\times 10^3 \,\mathrm{M_\odot}$). Explicit refinement and de-refinement is used to keep cells in the high resolution region within a factor of two of the target mass resolution. The high resolution region is made sufficiently large that there is no contamination within $1\ \mathrm{Mpc}$ of the main halo at $z=0$, i.e.\ there are no low resolution elements closer than $1\ \mathrm{Mpc}$.

The highest gas density at $z=0$ reached within the high resolution (level 3) simulations is $n \sim 10 \: {\rm cm^{-3}}$, corresponding to a cell size $\sim 25\ \mathrm{pc}$. 
Resolving structure in the gas distribution requires a few cells per dimension, and so the effective spatial resolution of the simulation is at best around $100\ \mathrm{pc}$ (comparable to the gravitational softening length for the gas) at the highest densities. Because of this limited resolution (which is, nevertheless, very good by the standards of cosmological simulations), the Auriga simulations cannot follow the formation of individual star clusters and cannot model the effects of stellar feedback with the same fidelity as our {\sc warpfield} models. Instead, the impact of stellar feedback is accounted for in a sub-grid fashion following the prescription introduced by \citet{Springel2003}. Gas above a density threshold of $n = 0.13 \: {\rm cm^{-3}}$ is artificially pressurized and is taken to represent some unresolved mix of cold/warm gas with medium to high densities and hot, supernova-heated gas with a low density. 
When we  post-process the gas distribution, however, we ignore this complication and treat the gas as if it were all at the specified cell density. This is a reasonable assumption, since most of the mass will be in the denser gas and hence the mass-weighted mean density in these regions should lie close to the value in the simulations.

For the purposes of this paper, we post-process only one of the 30 galaxies simulated in the Auriga project. Our selected galaxy, Au-6, is modelled in one of the high resolution (level 3) runs and has a halo mass of $10^{12}\,\mathrm{M_\odot}$ and a stellar mass of $6 \times 10^{10}\,\mathrm{M_\odot}$. This galaxy is similar to the Milky Way in many respects, including the properties of the stellar disk \citep[][]{Grand2017, Grand2018}, the gas disk \citep[][]{Marinacci2017}, the stellar halo \citep[][]{Monachesi2016}, the magnetic field structure \citep[][]{Pakmor2017}, and the population of satellite galaxies \citep[][]{Simpson2018}. This makes the Auriga galaxy Au-6 an excellent test-case for understanding the physical processes that govern the formation and evolution of the Milky Way. It is also a good starting point for our population synthesis model. We take the gas density and other galaxy properties of Au-6 as input for deriving the cluster mass distribution as well as for synthesizing electron densities, electron temperatures, and emissivities, and finally, the calculation of synthetic line emission, as we discuss in the sections below.

\section{Population synthesis modeling}
\label{sec:popsyn}
Our objective in this paper is to present a method for forward modelling the stellar population and emission of a region, be it an entire galaxy or a kpc-scale sub-region within a galaxy, that is described by a star formation rate, a metallicity, and a characteristic environmental density and instantaneous star formation efficiency. Within such a region, we expect to find many different clouds and star clusters\footnote{We use the words `star cluster' or `cluster'  for brevity, but in actual fact our model is agnostic on the issue of whether the stars are located in a gravitationally bound cluster or a gravitationally unbound association, so long as they remain reasonably localized in space for the $\sim 20 \: {\rm Myr}$ period during which they contribute to the emission of the galaxy in the tracers of interest in this study.} with a range of different masses and ages, and so a key part of this method is the generation of an appropriate sample of clouds and clusters. 

This task is made more difficult by the potential complexity of the evolution of the individual star-forming regions. As we have explored in earlier papers, the interplay of cooling, gravity, and the time-varying energy and momentum input from an evolving stellar population yields a range of different dynamical outcomes \citep[see e.g.][]{Rahner2017, Rahner2019} that are not self-similar between objects of different mass, density or metallicity. While all star-forming regions undergo an initial feedback-driven expansion, the combination of hot gas cooling and escape of radiation alone are enough to cause the expansion of some regions to stall. Clouds in this regime will often recollapse under their own self-gravity, forming a second generation of stars. As a result, the star formation efficiency of a given cloud can depend on whether it is destroyed by its initial burst of star formation, or whether feedback is initially unable to destroy it, resulting in star formation continuing over a more extended period. Because of this, it is difficult to predict {\it a priori} the contribution of a given cloud to the global star formation rate (SFR), as this depends on the cloud's dynamical history and on whether it undergoes one or multiple bursts of star formation. 

In this section, we outline how we avoid this problem and generate a sample of clouds and clusters that are consistent with a specified global SFR, while still accounting for the fact that some clouds may form multiple populations of stars. 
Armed with this sample, we then explore how we can use it to make predictions about the observable properties of the region as a whole.

\subsection{Sub-grid models of star-forming clouds}
\label{sect:subgrid}

To calculate the evolution of the natal cloud around each of our model clusters we use {\sc warpfield}. In this introductory study, we assume for simplicity that all clouds share the same average natal cloud density $n = 100 \, {\rm cm^{-3}}$ and star formation efficiency $\epsilon = 1\%$, but we note that the model can easily be extended to consider complex distributions of both of these properties. Since we know the cluster mass $M_{\rm cl}$, the cloud mass then follows simply as $M_{\rm cloud} = \epsilon^{-1} M_{\rm cl}$. The remaining {\sc warpfield} input parameter is the metallicity. This could in principle be adopted from the cosmological simulation, but in the present paper we assume, again for simplicity, that it has the solar value. The time evolution of the cloud, as modelled by {\sc warpfield}, provides us with the internal pressure and the radiation field which uniquely determine the properties of the hydrostatic HII region as it expands into the natal cloud structure and beyond. These properties are then fed into {\sc cloudy}, along with the spectral energy distribution of the star cluster, allowing us to solve  simultaneously for the emissivity and the full attenuated/reprocessed spectrum emerging from the star-forming region, as described in more detail in our companion paper on {\sc warpfield-emp} (Pellegrini et al., in prep.). 

Our model makes the assumption that the duration of any individual burst of star formation within a cloud is short compared to the evolutionary timescale of the cloud, so that we can treat it as instantaneous.\footnote{Strictly speaking, we only require that the massive stars that dominate the feedback form rapidly, and so our assumption is not inconsistent with scenarios in which low-mass stars form over a more extended period.} We justify this assumption as a consequence of the effectiveness of feedback in cluster-forming regions, as explored in \citet{Rahner2017, Rahner2019}. The young massive clusters that we are primarily concerned with here quickly clear out gas from their immediate vicinity, driving expanding shells into the surrounding cloud and the larger-scale ISM. 
During the expansion phase, the cloud is subjected to intense radiation, and we find that molecular gas is destroyed rapidly, temperatures rise and the cloud becomes partially to fully ionized. In these extreme conditions, star formation is unlikely to occur over extended periods of time. 

For more massive clusters, it can sometimes be the case that feedback from the initial burst of star formation is unable to completely disrupt the cloud. In that case, as the cluster ages and its feedback becomes less effective, the expansion of the feedback-driven shell stalls. Following this, the gas undergoes renewed collapse due to its own self-gravity, a phenomenon we refer to as `recollapse' (\citealt{Rahner2017}; see also \citealt{Rahner2018a} and \citealt{Rugel2018} for examples of some clusters where there is good evidence that recollapse is occurring or has occurred). In mass bins in which recollapse occurs, equilibrium between cluster formation and cluster death takes longer to establish, but for the current 
experiment, tailored to Milky Way conditions, we find the cluster population and the ionizing output reach equilibrium after around 20~Myr, 
which is the time we choose to present here.

\subsection{Generating the Cluster Mass and Age Distribution}
\label{sect:ClusterMassDistribution}

In order to generate a sample of clusters in a region of interest, we need to know two things: the initial mass function of the star clusters and the rate at which gas is converted into stars within that region. We assume that the initial cluster mass function is a power law with exponent $\beta$, where
\begin{equation}
\log_{10}\left( \frac{dN_\mathrm{cl}}{dM_\mathrm{cl}} \right) \propto -\beta \log_{10}(M_\mathrm{cl})\;,
\end{equation}
and which extends between a minimum cluster mass $M_{\rm cl, min} = 10^{2} \: {\rm M_{\odot}}$ and a maximum cluster mass $M_{\rm cl, max} = 10^{7} \: {\rm M_{\odot}}$, consistent with observations in nearby galaxies \citep{Zhang1999, Lada2003, Portegies2010, Krumholz2018, Gouliermis2018} Observationally determined values of $\beta$ typically lie in the range $\beta = 2.0 \pm 0.2$  and so in our fiducial model we set $\beta = 2.0$. However, our method is readily generalizable to the case where $\beta \neq 2.0$.

The other important parameter is the star formation rate (SFR). One option is to take this directly from the adopted cosmological simulation, but it can also be specified directly by the user. For example, in the models presented in this paper, we make the simplifying assumption that the SFR is constant over times that are short compared to the molecular gas depletion time ($\tau_{\rm depl} \equiv M_{\rm mol} / \mbox{SFR}$, where $M_{\rm mol}$ is the molecular gas mass). We also assume that the SFR is directly proportional to the local gas mass. We measure this gas mass by dividing up the galaxy in the Au-6 simulation into a set of linearly spaced radial bins originating at the galactic center-of-mass and summing up the gas mass within each bin. We limit the vertical extent of each bin to $\pm 5$~kpc from the disk midplane. 

If the gas mass in the $i$th bin is $M_{{\rm gas}, i}$, then the corresponding SFR is given by
\begin{equation}
{\rm SFR}_\mathrm{i} = M_{\mathrm{gas,i}} \times f_{\mathrm{dense}} / \tau_{\mathrm{depl}} \;,
\label{eq:SFR}
\end{equation}
where $f_\mathrm{dense}$ is fraction of dense, cold gas, and $\tau_\mathrm{depl}$ is the assumed depletion time due to star formation. We take values of $f_\mathrm{dense} = 25\ \%$ \citep[][]{Klessen2016} and $\tau_\mathrm{depl} = 2.0\ \mathrm{Gyr}$ \citep[][]{Bigiel2008}, reasonable values for a Milky Way like system. In the Au-6 simulation, roughly 40\% of the gas mass is above the \citet{Springel2003} density threshold and hence is potentially dense and cold. However, the total mass of the Au-6 galaxy also is approximately twice that of the MW, so by using a smaller value for $f_{\rm dense}$, we obtain a total SFR in better agreement with the Galactic value. This enables us to make a more meaningful comparison between the results we obtain for this galaxy and observations of the real Milky Way (see Section~\ref{sect:Observations} below).

\begin{figure}
	\centering
	\includegraphics[width=1.0\linewidth]{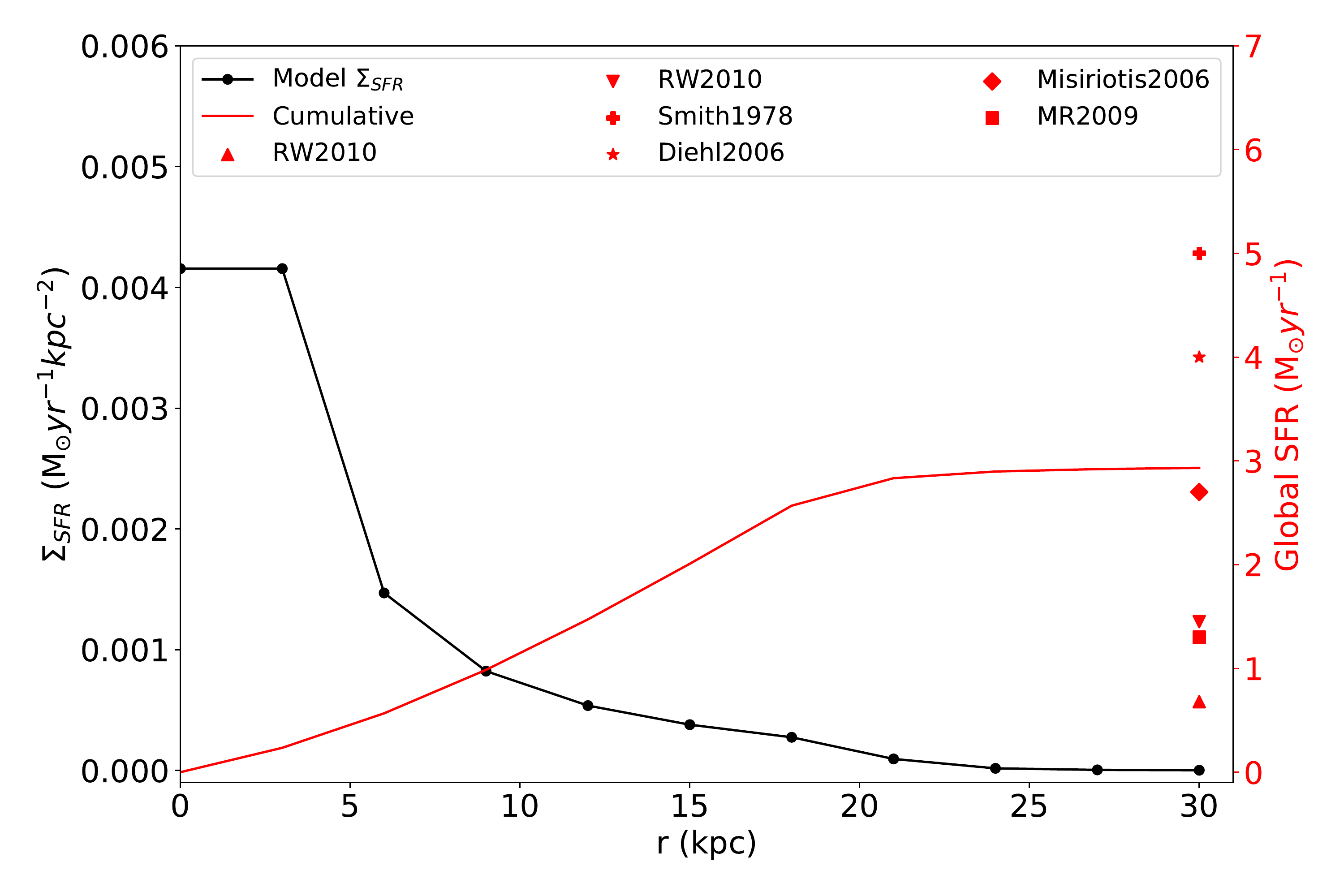}
	\caption{The radial star formation rate surface density (black) of our model galaxy, computed using Eq.~\ref{eq:SFR} for 10 evenly spaced radial bins. Also shown is the cumulative star formation rate (red) of the entire galaxy as a function of radius. Various estimates of the global Milky Way star formation rate are indicated as symbols at the outermost radii for comparison. These are taken from 
	 \citeauthor{RW2010}~(2010; RW2010), who give both upper and lower limits, \citet{Smith1978}, \citet{Diehl2006}, \citet{Misiriotis2006} and  \citeauthor{Murray2009}~(2009; MR2009).}
	\label{fig:modelmwsfr}
\end{figure}

In the model presented in this paper, we calculate the star formation rate in 10 annular bins, resulting in the radial profile shown in Figure~\ref{fig:modelmwsfr}. Also shown are various observational estimates for the total SFR in the Milky Way. These span values from $\sim 1 \: {\rm M_{\odot} \: yr^{-1}}$ to $5  \: {\rm M_{\odot} \: yr^{-1}}$, and the number of $2.9 \: {\rm M_{\odot} \: yr^{-1}}$ that we obtain with our simple SFR prescription lies well within this range. Indeed, it agrees well with the value that \citet{Diehl2006} infer based on their \Ha observations of the Milky Way. 

\begin{figure}
	\centering
	\includegraphics[width=1.0\linewidth]{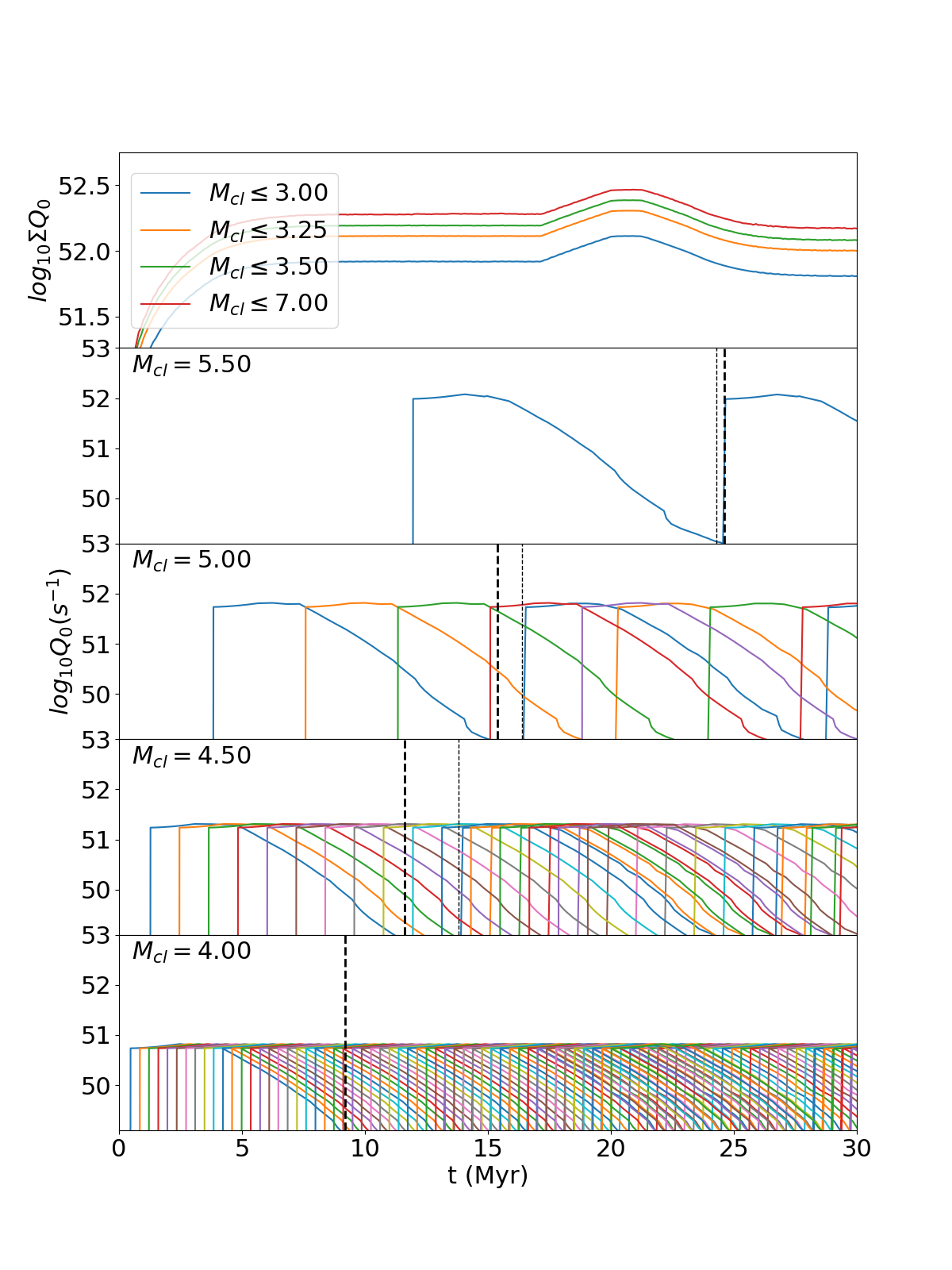}
	\caption{The time evolution of the cluster population in the central annulus of the galaxy. From bottom to top are clusters in the $\log_{10} M_{\rm cl} = 4$, 4.5, 5 and 5.0 bins. The heavy dashed line marks the time when the  ionizing luminosity of the first cluster formed in each bin decreases to equal $Q_0 = 10^{49.0} \, \rm s^{-1}$, equivalent to a single O star. This is also the time when the ionizing photon flux of a system with constant SFR reaches steady state. The light dashed line indicates the occurance of a recollapse event forming a new stellar population in an existing cluster. }
	\label{fig:sfe1}
\end{figure}

Given the initial cluster mass function and the star formation rate, we proceed as follows. We consider only clusters formed within the last 50 Myr\footnote{Older clusters make a negligible contribution to the emission of the galaxy in the tracers of interest in this study.} and split up this period into 50 uniform time-steps, each with length $\Delta t = 1.0 \: {\rm Myr}$. We calculate the gas mass converted into stars in each time-step. For time-step $n$, spanning the period $t_{n} \rightarrow t_{n} + \Delta t$, we have:
\begin{equation}
\Delta M = \int\limits_{t_{n}}^{t_{n}+\Delta t} {\rm SFR}(t) \, {\rm d}t, 
\end{equation}
where ${\rm SFR}(t)$ is the value of the SFR at time $t$. In the simple case of a constant star formation rate, this reduces to $\Delta M_{n} = {\rm SFR} \times \Delta t$.

We next assign this mass to a set of discrete mass bins of 0.25~dex, evenly spaced in logarithm between $M_{\rm cl, min}$ and $M_{\rm cl, max}$. We use discrete mass bins matched to our existing {\sc warpfield-emp} database instead of randomly sampling the mass function to limit the computational cost of the model. Each bin receives a fraction of $\Delta M$ corresponding to the slope of the initial cluster mass function that is located in that bin mass. In the simple case of $\beta = 2.0$ considered in this paper, each logarithmic bin has an equal fraction of the mass in the initial cluster mass function and hence receives an equal fraction of $\Delta M$ on each time-step.

At the end of each time-step, we check for each bin $M_{i}$ whether the accumulated mass is sufficient to form a cluster of mass $M_{i}$. If so, then we add a new cluster of that mass to our population at that time-step and decrease the mass in the bin by $M_{i}$. The newly-created cluster is assigned an age drawn randomly from the time interval $[t \rightarrow t + \Delta t]$. If the mass in the cluster is sufficient to form more than one new cluster, then we simply repeat this procedure as many times as necessary. On the other hand, if the mass in the bin is insufficient to form even one new cluster, then we retain all of it for the next time-step. 

Up to this point in calculating an  SFR we have assumed that each cloud undergoes only a single burst of star formation. As shown in \citet{Rahner2018a} and \citet{Rugel2018}, this is not necessarily the case. Once a stellar population becomes older than $\sim 3 \: {\rm Myr}$, its output of energy in the form of stellar winds and radiation steadily declines as its most massive stars begin to die. Consequently, the cluster becomes a less effective source of stellar feedback as it ages. This is of little importance if the cloud has already been destroyed, but for some combinations of parameters, the cloud may not yet have been completely dissolved at the time that the cluster feedback becomes ineffective. In this case, recollapse of the cloud may occur, leading to a new burst of star formation. To determine whether this occurs, we run a {\sc warpfield} model for each new cluster, using the appropriate cluster and cloud masses. This allows us to identify the cases in which recollapse occurs and also provides us with the time at which it occurs and the mass of new stars formed in each recollapsing cloud. 

We chose a fiducial reference point of 20~Myr to compute the SFR, having found that at this point the cluster population has reached a roughly steady state. 
The contribution to the total SFR coming from recollapsed clusters is 
\begin{equation}
{\rm SFR}_{\rm rec, 20} = \frac{M_{\rm rec, 20}}{20 \: {\rm Myr}}.
\end{equation}
This needs to be added to the SFR of newly formed clusters. 
Since this is initially chosen to be the same as our desired SFR, whenever there is recollapse the total SFR is initially larger than our desired value. To account for this effect, we adjust the input SFR downwards and repeat the whole procedure, resulting in a new value of SFR$_{\rm rec, 20}$. We continue like this using an iterative shooting method until we match the desired SFR calculated from the gas distribution. We note that this is a highly non-linear process: changing the SFR changes the integer number of clusters at a given mass which have formed. Since the recollapse timescale and the question of whether or not a given cloud recollapses both depend on cloud mass, instantaneous star formation efficiency, and cloud density, changing the number of new clusters formed during a given timestep inevitably has a knock-on effect on SFR$_{\rm rec, 20}$. 
With each iteration, we therefore use the first derivative of the change in SFR$_{\rm rec, 20}$ to anticipate the input SFR which will produce the desired total SFR. In most cases, we find that no more than 4 iterations are necessary to reproduce a desired SFR to 1\% accuracy. For clouds with low densities, as studied here, the contribution made by recollapsing clouds to the total SFR is below $40\%$. A full exploration of the effects of recollapse on the shape and normalization of the resulting cluster mass function will be the subject of a follow-up study. 

In Figure~\ref{fig:sfe1} we show the time evolution of the ionizing output from our cluster population. To make it visually understandable, in the bottom four panels we show the time evolution of a single mass bin. Each cluster is shown with a new line. Statistically, the combination of an SFR and a cluster mass function will inevitably lead to an average formation rate of a cluster at a given mass. For our purposes, the cluster ``dies'' when its ionizing luminosity drops to undetectable levels.\footnote{Note that this is a separate issue from whether or not the cluster survives for an extended period of time as a gravitationally bound structure following gas expulsion. Note also that the emission from old clusters is considered to be accounted for by our diffuse interstellar radiation field. In order to avoid overcounting their contribution, once they become faint in ionizing radiation we stop tracking them as individual sources.} In the present case, we assume that this occurs when the ionizing photon flux $Q_{0}$ drops below $Q_0 = 10^{49.0} \, {\rm s^{-1}}$, the equivalent of a single O6.5 star, and roughly corresponding to the ionizing photon flux of the Orion Nebula. This time is marked with a heavy vertical dashed line in each panel. It occurs earlier for lower-mass, less luminous clusters. 

\subsection{Photoionization calculation}
\label{sect:Photoionization}

We now have the age and mass distribution of the clusters as a function of galactic radius, but we still need to determine their spatial distribution. To do this for a given cluster, we begin by picking a random location in the disk, with a probability distribution set by the SFR/gas mass profile (see Figure~\ref{fig:modelmwsfr}). We then check whether the total gas mass within a distance of 50~pc fulfills the condition:
\begin{equation}
M_{r < 50 \: {\rm pc}} \geq M_{\rm cloud}.
\end{equation}
If this condition is fulfilled, we calculate the center-of-mass of the gas within 50~pc of our randomly selected point and place our cluster there. Otherwise, we repeat this procedure with a new randomly selected point. 
This algorithm ensures that our star clusters are distributed in positions close to, but not necessarily on top of,  density peaks of the gas distribution. 
This is consistent with observations where young clusters are seen in the vicinity of dense molecular gas, but are no longer deeply embedded in their parental clouds owing to efficient stellar feedback.

Once we have placed each of our clusters in the density distribution, we next compute the  density profile of the gas. For each cluster we randomly select a set of piecewise orthogonal basis vectors. This yields 6 cardinal directions around each cluster along which we trace $n_\mathrm{g}(r)$, stopping once we reach the boundary of the grid.\footnote{A more accurate approach would be to carry out the photoionization calculation along a much larger set of rays sampling the space around each cluster. Unfortunately, the computational cost of the calculation and the fact that we need to repeat it for a large number of clusters renders this approach computationally unfeasible at the present time.} Our {\sc warpfield} models provide us with the flux of ionizing photons escaping from each cluster at any given moment. We use this as input to a set of photoionization calculations in which we obtain the ionization state, and hence the thermal electron number density $n_\mathrm{th}$ and electron temperature $T_\mathrm{e}$, as a function of distance from the cluster. Finally, we assume that in other radial directions from the cluster, we can obtain $n_\mathrm{th}$ and $T_\mathrm{e}$ as a function of distance by interpolating between the results of our six calculations. We do this by projecting the unit vector in the radial direction of interest onto our set of basis vectors. This yields a set of three coefficients: the dot products of the radial unit vector with the three basis vectors. We use the magnitude of each coefficient as the weight for that component in the interpolation, while the sign tells us whether we should take the solution in the positive or the negative direction along that basis vector.

The photoionization calculations are carried out using the patched version of {\sc cloudy} $\rm v17$ \citep[][]{Ferland2017}, and so as a byproduct we also obtain the emissivities of any emission lines of interest within this region. As an example, in this paper we present results for \Ha and the \SIII 9530~\AA\xspace line. We stress that the method itself is not limited to these lines, but instead can be used to model any line emission process that {\sc cloudy} or {\sc polaris} can internally handle.
Each {\sc cloudy} calculation assumes Milky Way-like values for the atomic hydrogen cosmic ray ionization rate ($\zeta_{\rm H} = 2\times10^{-16} \, {\rm s}^{-1}$; \citealt{Indriolo2007}), and the strength and spectrum of the diffuse interstellar radiation field from old stars. Each calculation also adopts solar values for the metal abundances.
The actual sphere of influence of the individual clusters  typically does not extend  beyond $\approx 75\ \mathrm{pc}$ in the plane of the disk, outside of which cosmic ray ionization and interstellar radiation from older stars dominate. 

At this point, we have a prediction for what the temperature, ionization state, and emissivity of the gas would be as a function of distance from each cluster if that cluster were the only source of radiation (besides the diffuse interstellar background radiation field). We now want to translate this information back onto the Voronoi grid from the Au-6 simulation, which means that we need to decide how to combine these individual radial predictions. To do this, we loop over the Voronoi grid cells. For each cell, we first check whether it overlaps a region directly represented by one of our {\sc warpfield} models. If it does, then we populate it with the appropriate profile of emission  taken from that model. If it does not, then we identify which cluster contributes the largest flux of ionizing photons at this point in space and take the necessary values from the radial solution \footnote{We use the radius measured in the plane of the disk with an emissivity solution dependent also on z-height above the disk.}  that we have computed for that cluster. In other words, we make the approximation that at any given point in space, the local temperature and ionization are determined solely by the contribution from a single cluster. In practice, we find that this is generally a very good approximation in our models, since only a few cells see roughly equal ionizing photon fluxes from different clusters. We have experimented with using a more accurate iterative approach to combine the effects of the different clusters on their surroundings, but find that although it is vastly more expensive to calculate, it does not offer significant improvement for our results.

\begin{figure*}
 \begin{center}
         \begin{minipage}[c]{1.0\linewidth}
                         \begin{center}
                                 \includegraphics[width=0.49\textwidth]{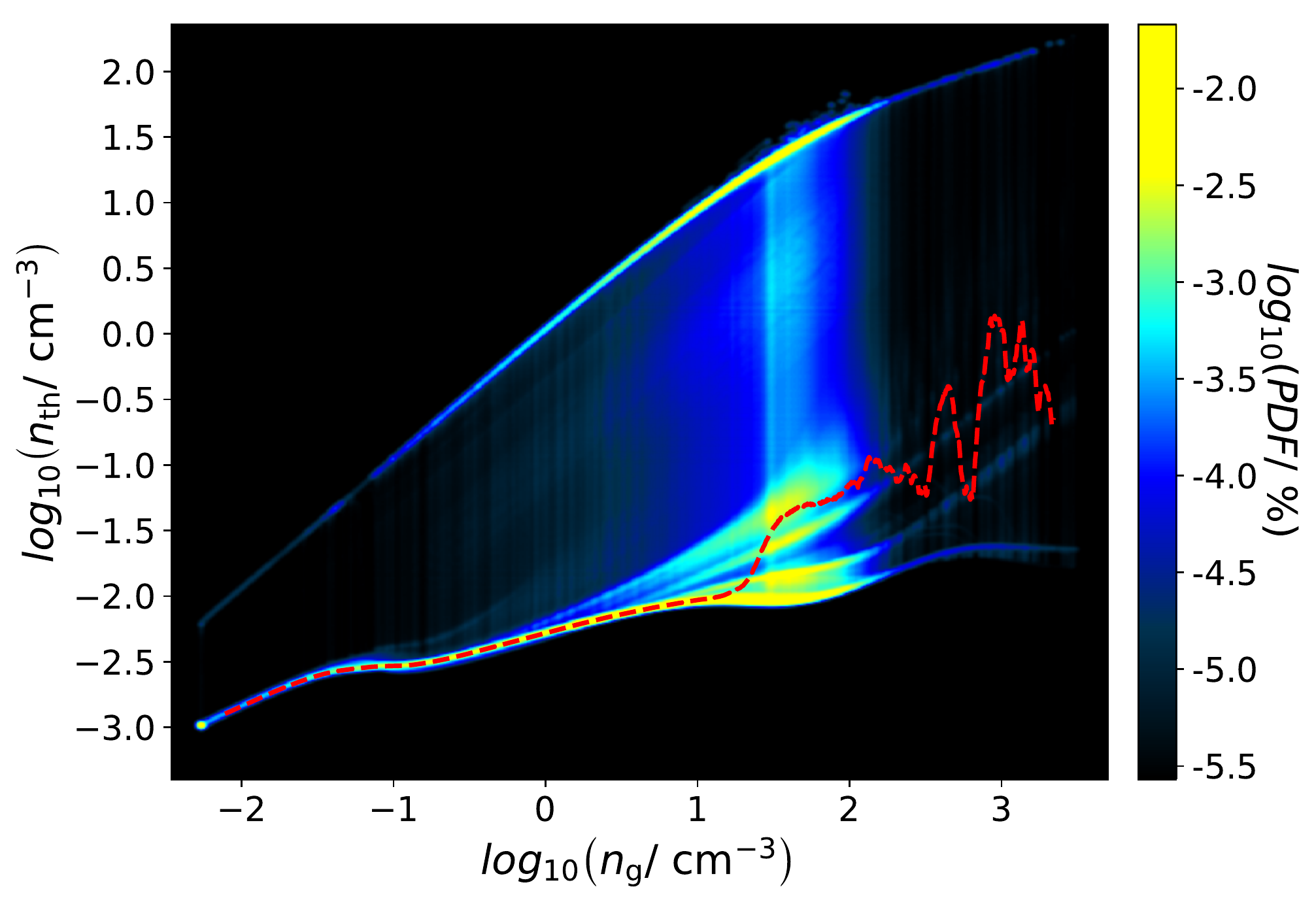}
                                 \includegraphics[width=0.49\textwidth]{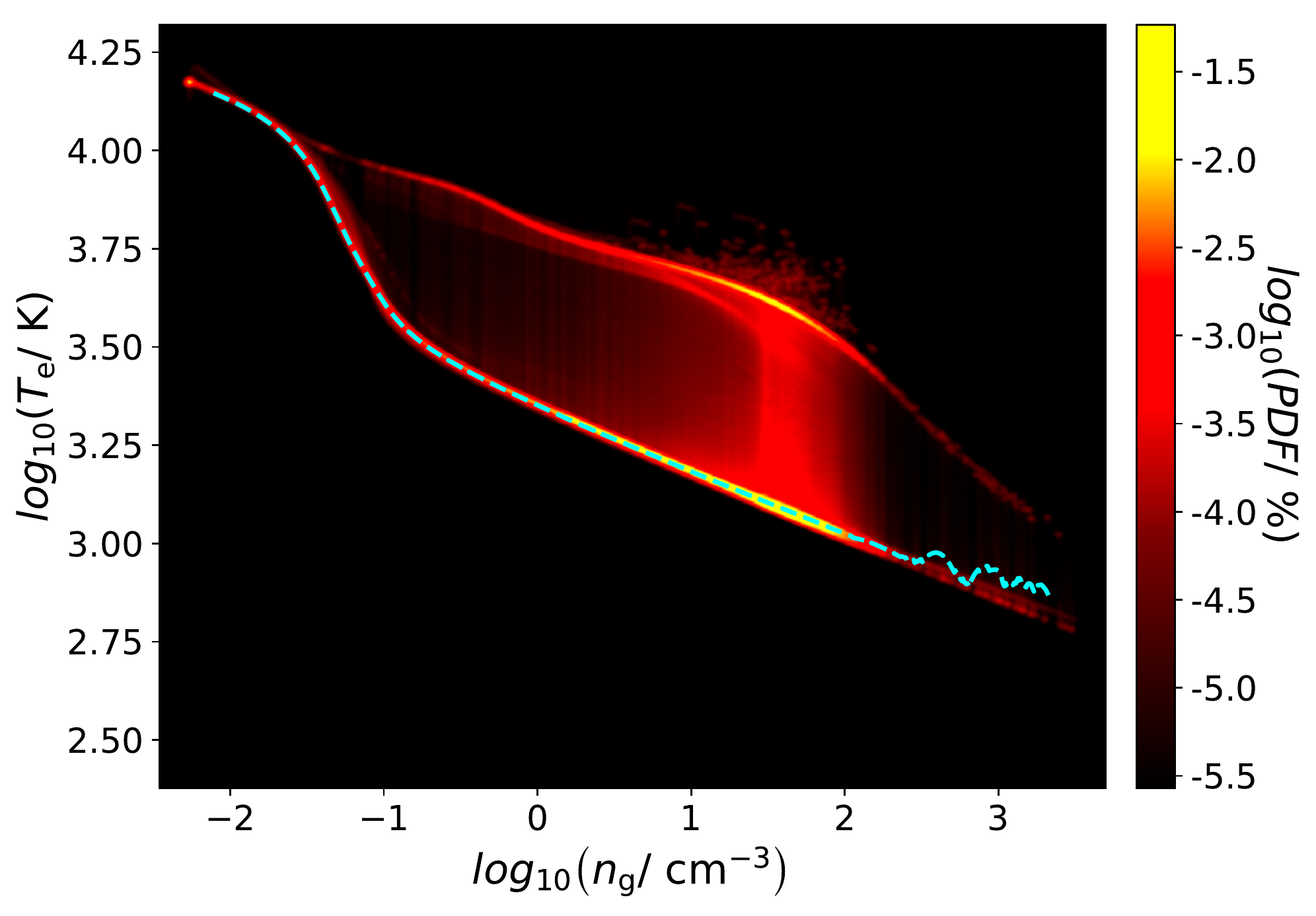}

                         \end{center}
                       \end{minipage} 
         \begin{minipage}[c]{1.0\linewidth}
                         \begin{center}
                                 \includegraphics[width=0.49\textwidth]{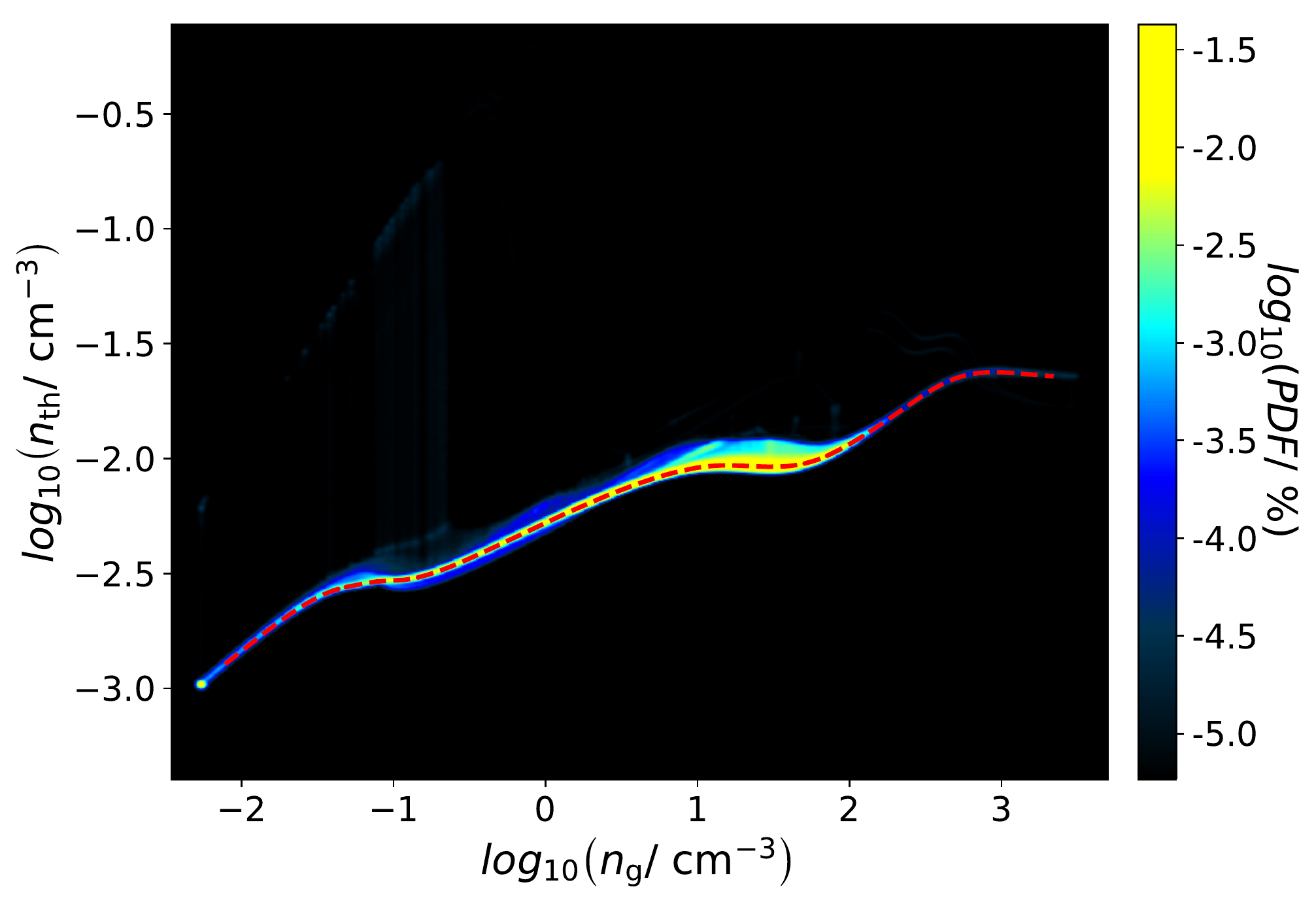}
                                 \includegraphics[width=0.49\textwidth]{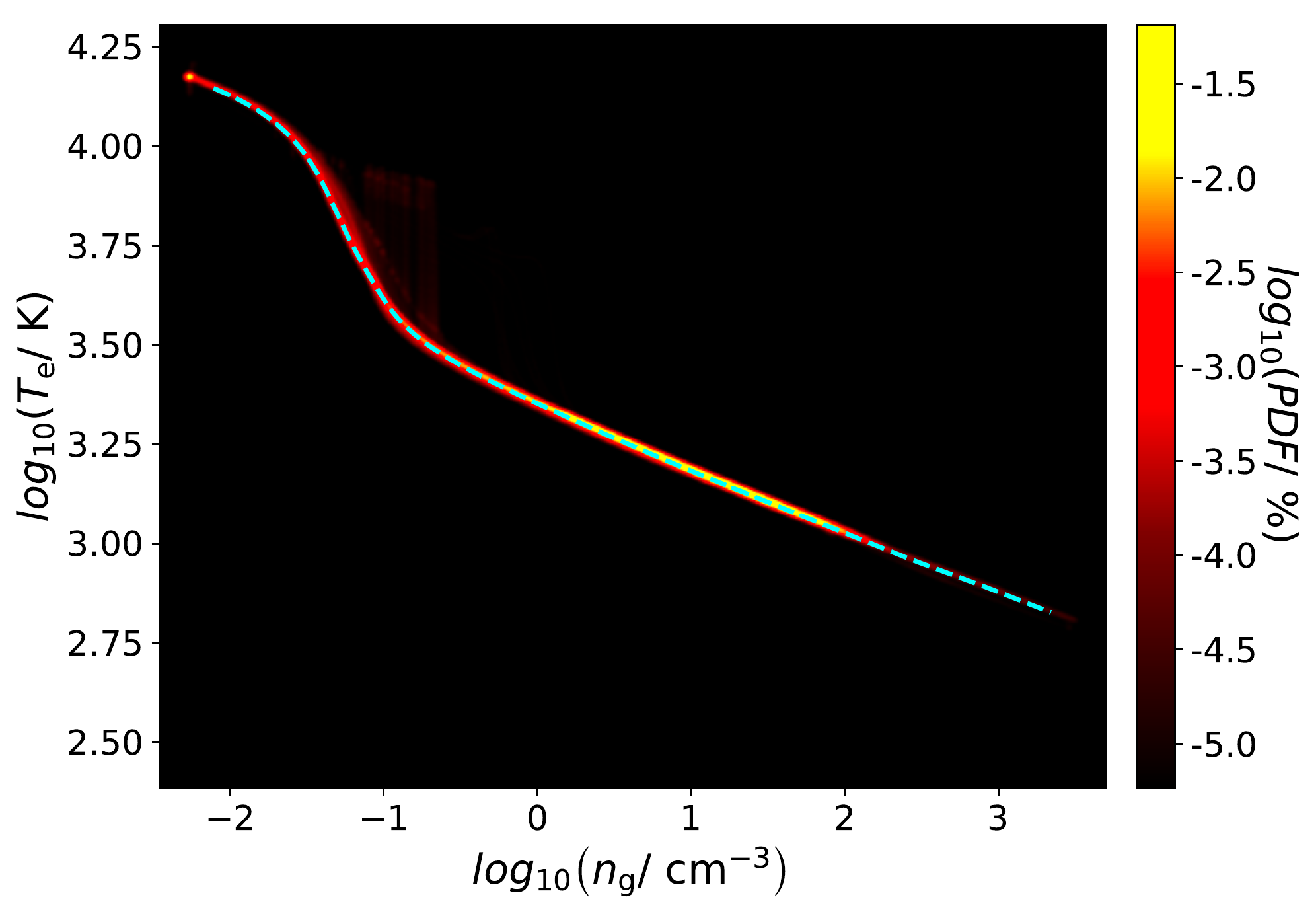}

                         \end{center}

                       \end{minipage} \end{center}  
 \caption{Probability distribution of thermal electron number densities $n_{\mathrm{th}}$ (left column) and electron temperatures $T_{\mathrm{e}}$ (right column) as a function of the gas number density $n_{\mathrm{g}}$. Dashed lines indicate the median values. The top row shows the full parameter set whereas the bottom row shows the results for all regions with a distance $d \geq 75\ \mathrm{pc}$ away from any cluster. We refer to the latter as the ISM condition.}
 \label{fig:ElDistribution}
\end{figure*}

\begin{figure*}
 \begin{center}
         \begin{minipage}[c]{1.0\linewidth}
                         \begin{center}
                                 \includegraphics[width=0.49\textwidth]{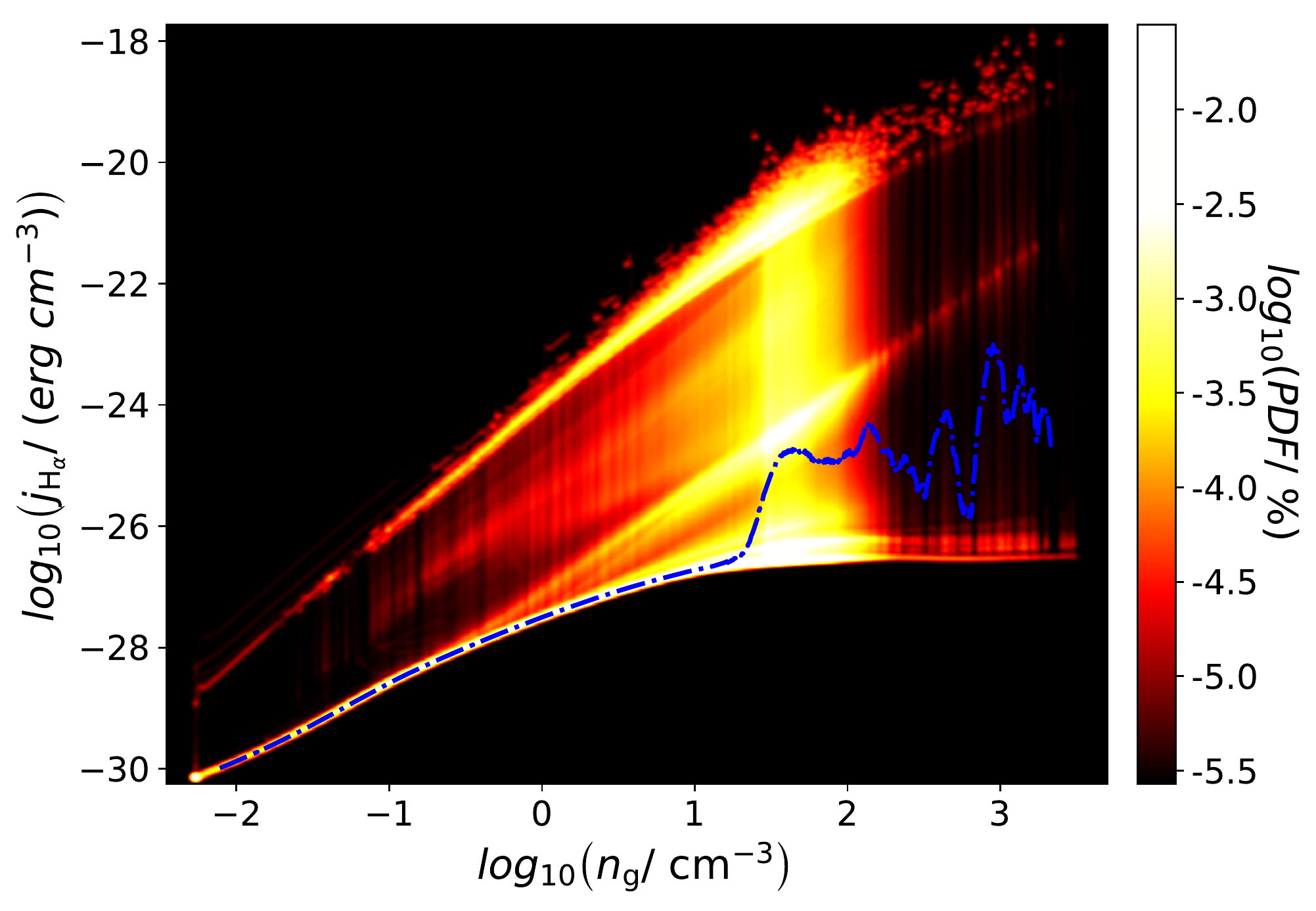}
                                 \includegraphics[width=0.49\textwidth]{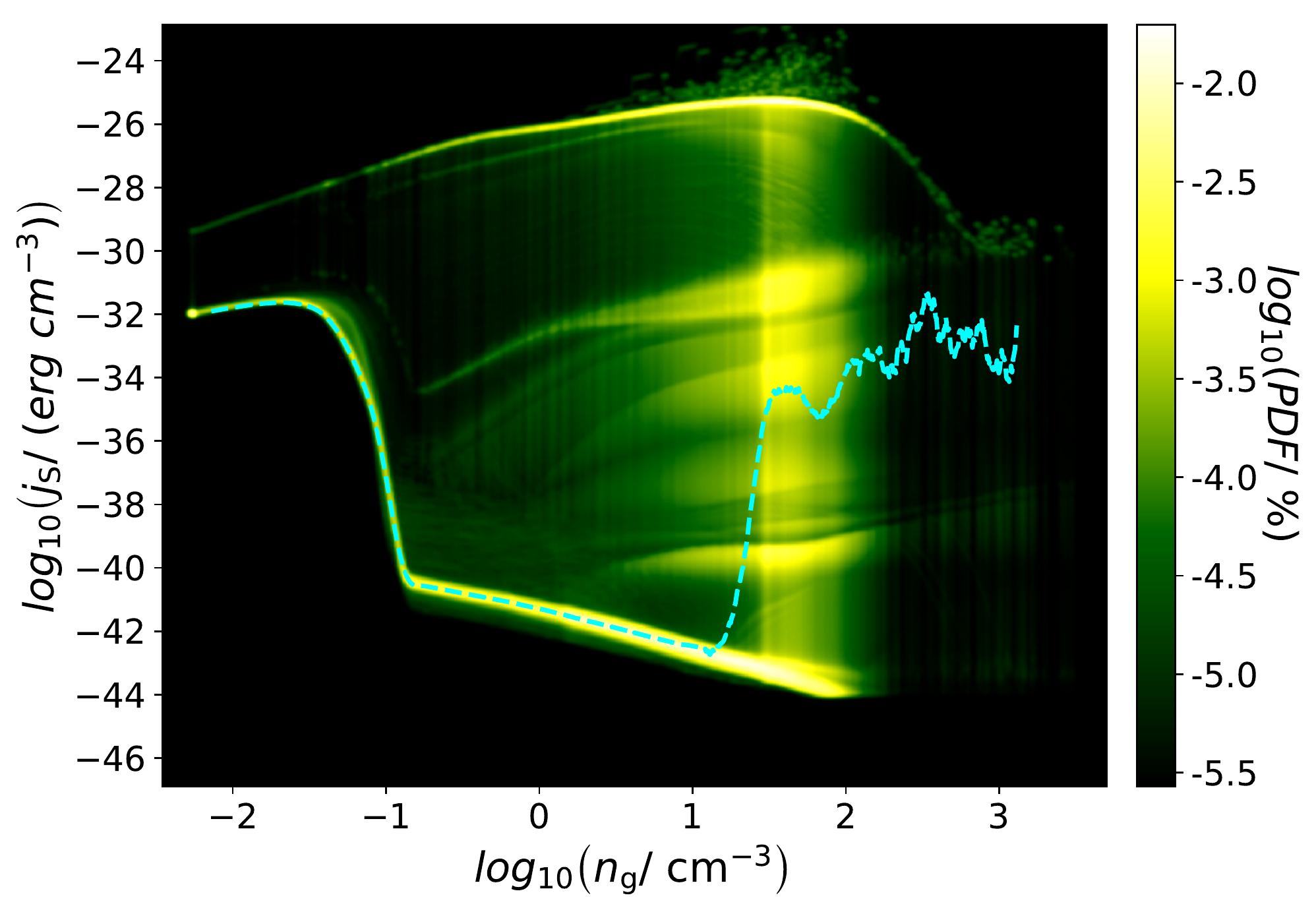}
                         \end{center}
                       \end{minipage} 
         \begin{minipage}[c]{1.0\linewidth}
                         \begin{center}
                                 \includegraphics[width=0.49\textwidth]{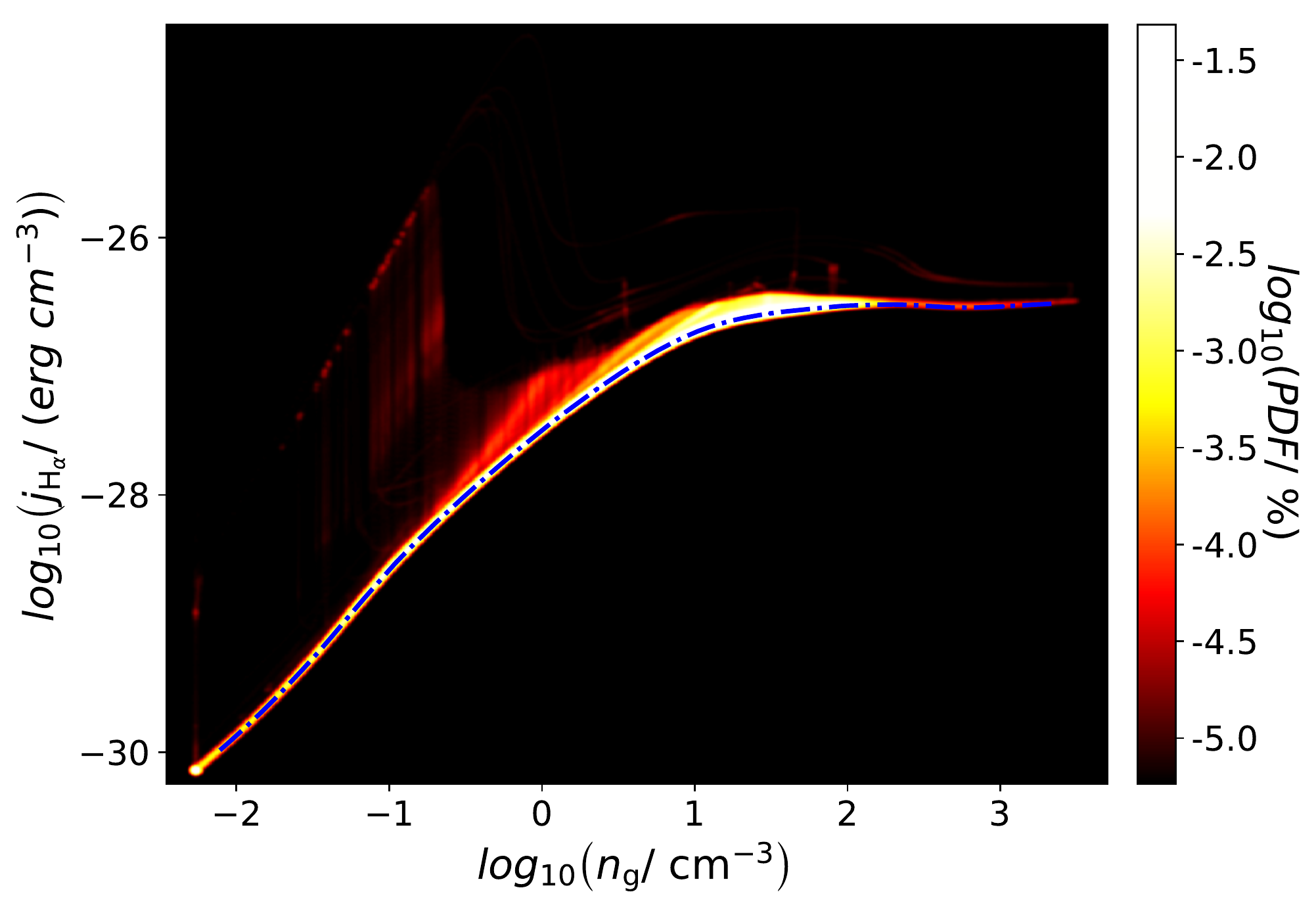}
                                 \includegraphics[width=0.49\textwidth]{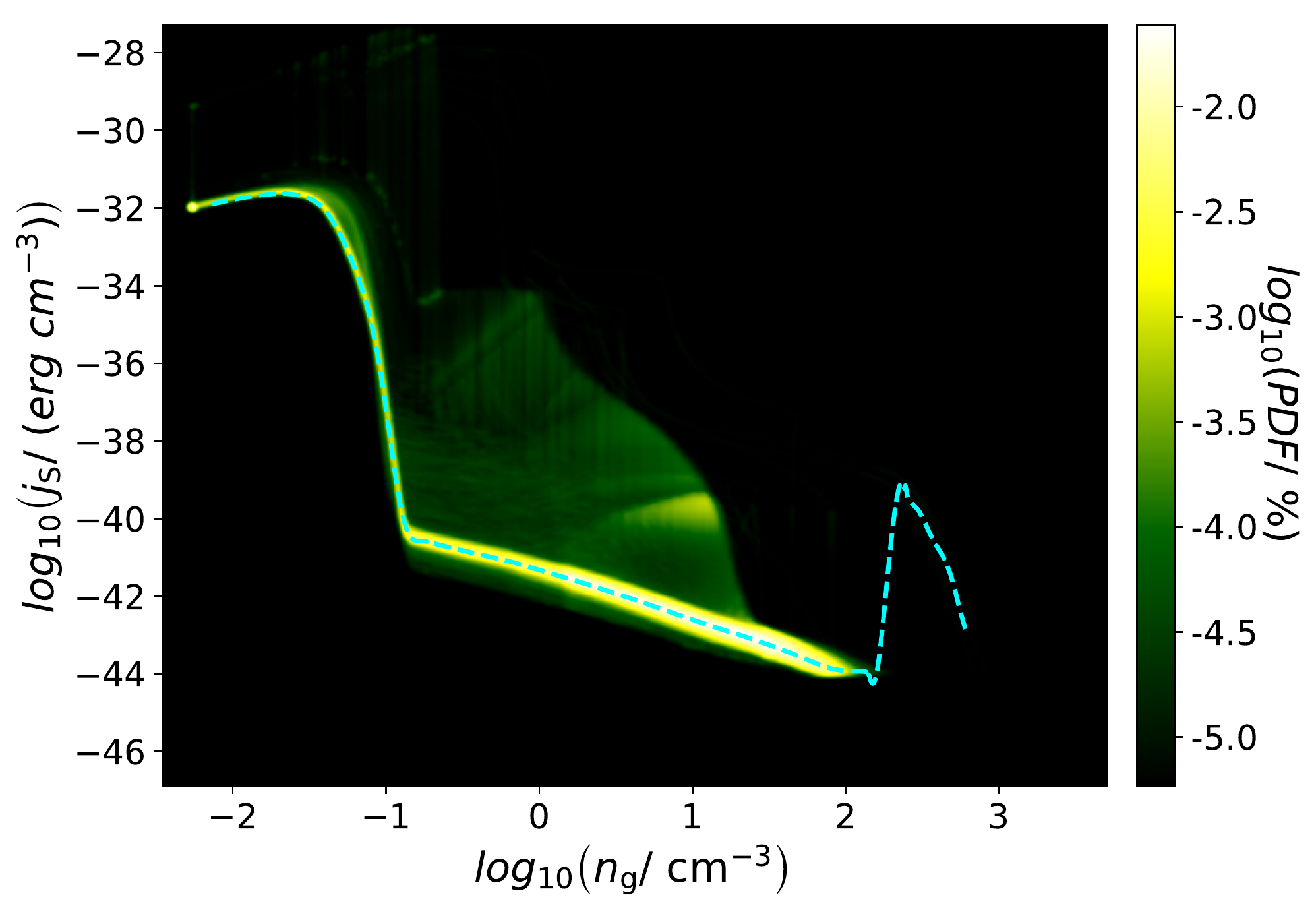}
                         \end{center}
                       \end{minipage} 
                     \end{center}  
 \caption{The same as Fig. \ref{fig:ElDistribution} for the emissivities of \Ha (left column) and \SIII (right column). The large scatter at constant density reflects the partially ionized nature of the diffuse gas between \hii regions. Of note, much of the scatter in the \SIII emissivity comes from partially ionized gas distributed above and below the disk. Ionizing radiation from massive clusters can travel much larger distances in this direction than in the disk midplane, and so one finds gas here with a much wider range of densities and ionization parameters than in the midplane.}
 \label{fig:EmDistribution}
\end{figure*}

\begin{figure*}
 \begin{center}
         \begin{minipage}[c]{1.0\linewidth}
                         \begin{center}
                                 \includegraphics[width=0.49\textwidth]{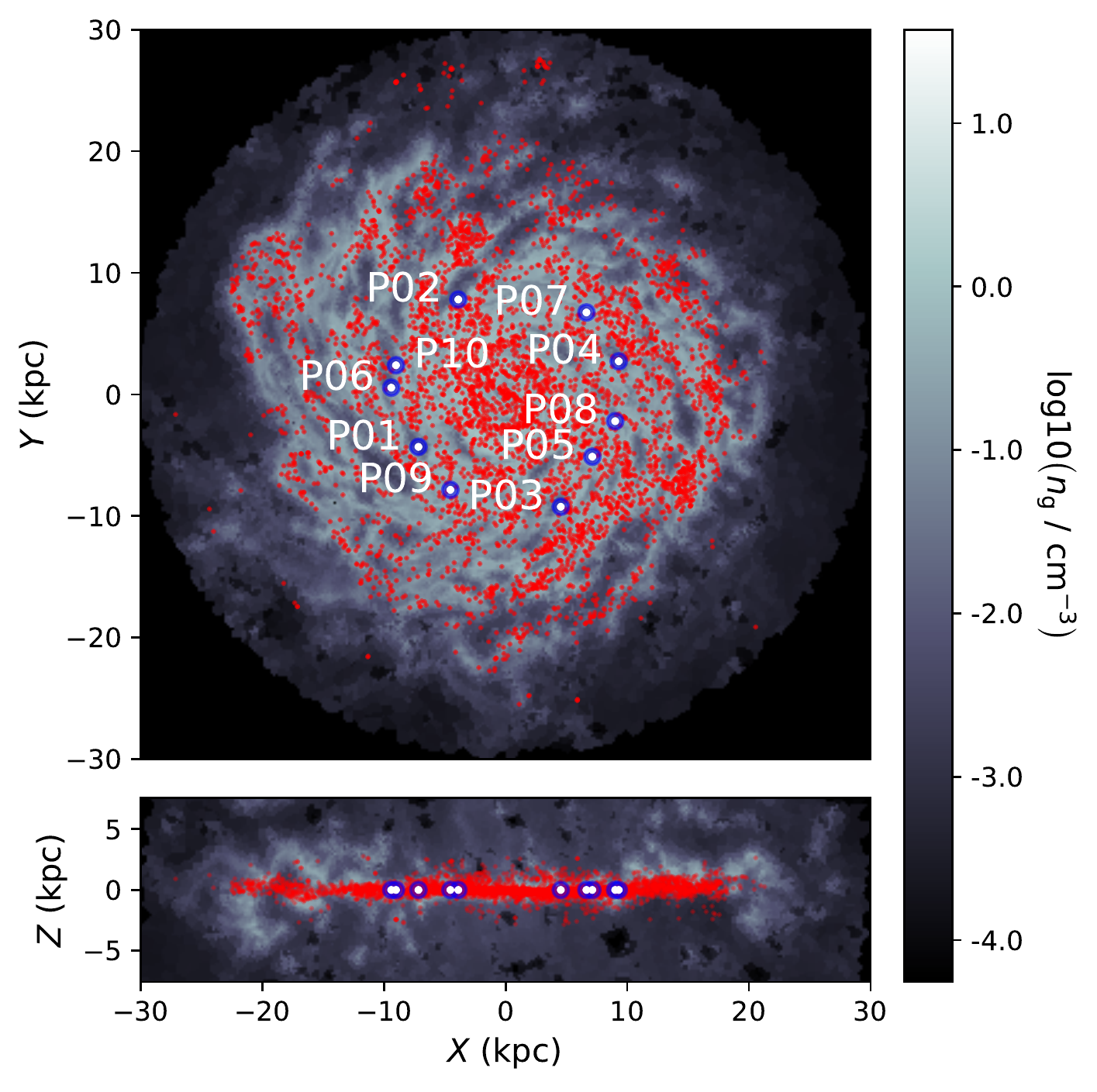}
                                 \includegraphics[width=0.49\textwidth]{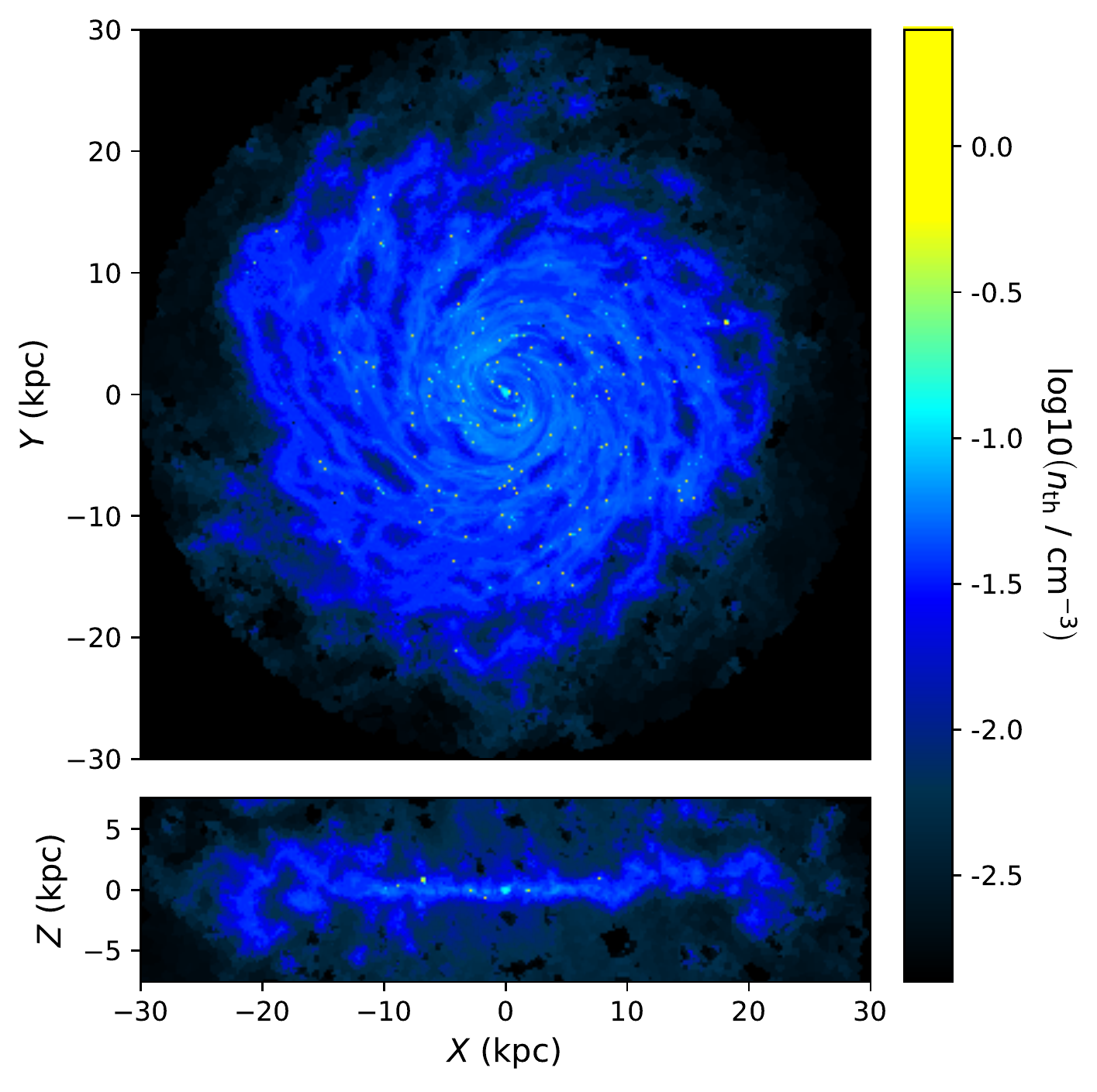}
                                 
                                  \caption{Left panel: Midplane gas distribution $n_{\mathrm{g}}$  of the Auriga Au-6 galaxy \citep[][]{Grand2017} at $z=0\ \mathrm{pc}$ in the $x$-$y$ plane (top) and $y = 0 \, \mathrm{pc} $ in the $x$-$z$ plane (bottom). Red dots represent all the sampled cluster positions projected on the plane while the blue dots with labels indicate the distinct observer positions (see Section~\ref{sect:Observations}). Right panel: The same as the left panel, but for the thermal electron number densities $n_{\mathrm{th}}$ as derived in Sect. \ref{sect:ClusterMassDistribution} and Sect. \ref{sect:Photoionization}.}
                                   \label{fig:MidDensGas}
                         \end{center}
                       \end{minipage} 
         \begin{minipage}[c]{1.0\linewidth}
                         \begin{center}
                                 \includegraphics[width=0.49\textwidth]{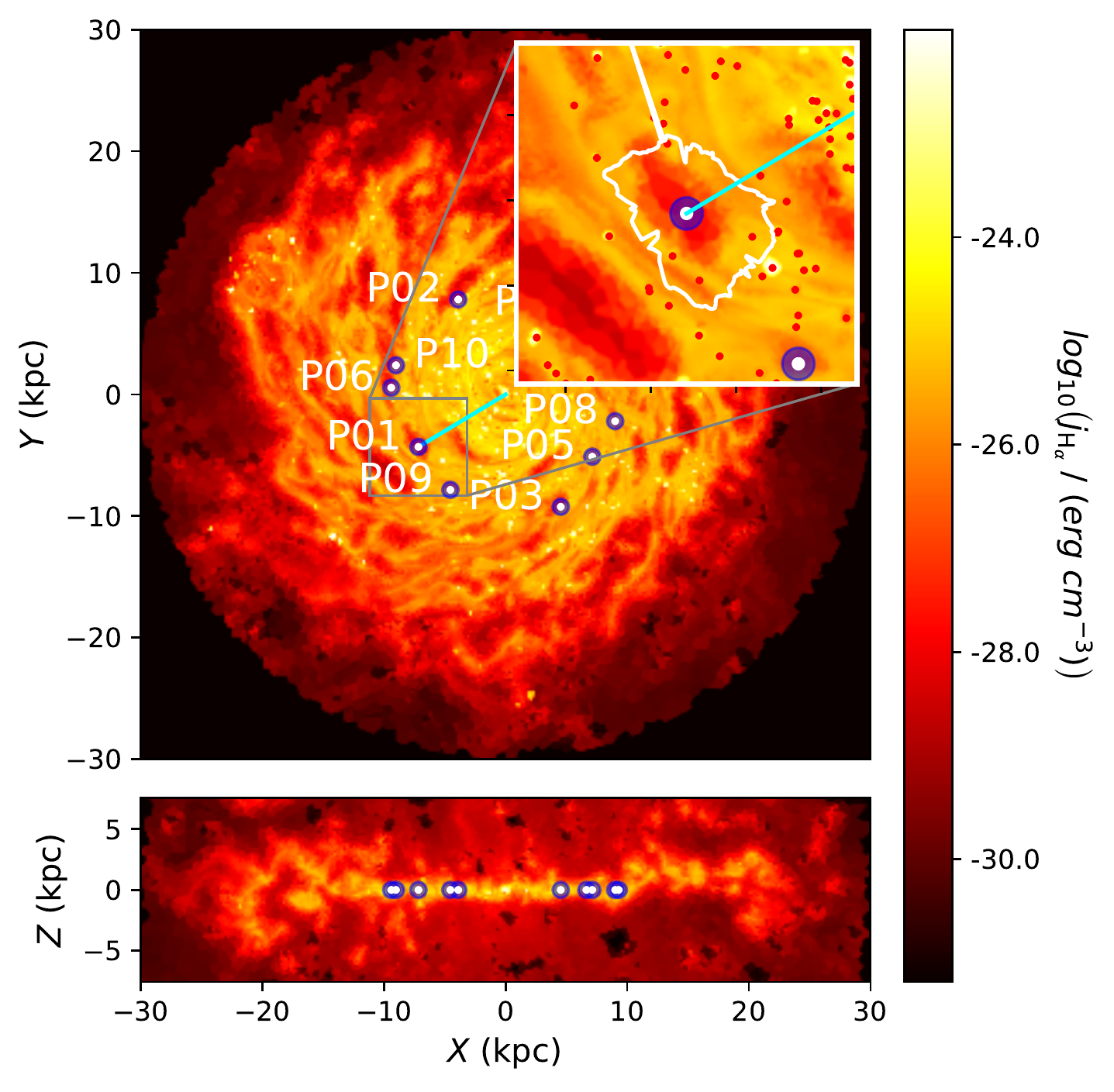}
                                 \includegraphics[width=0.49\textwidth]{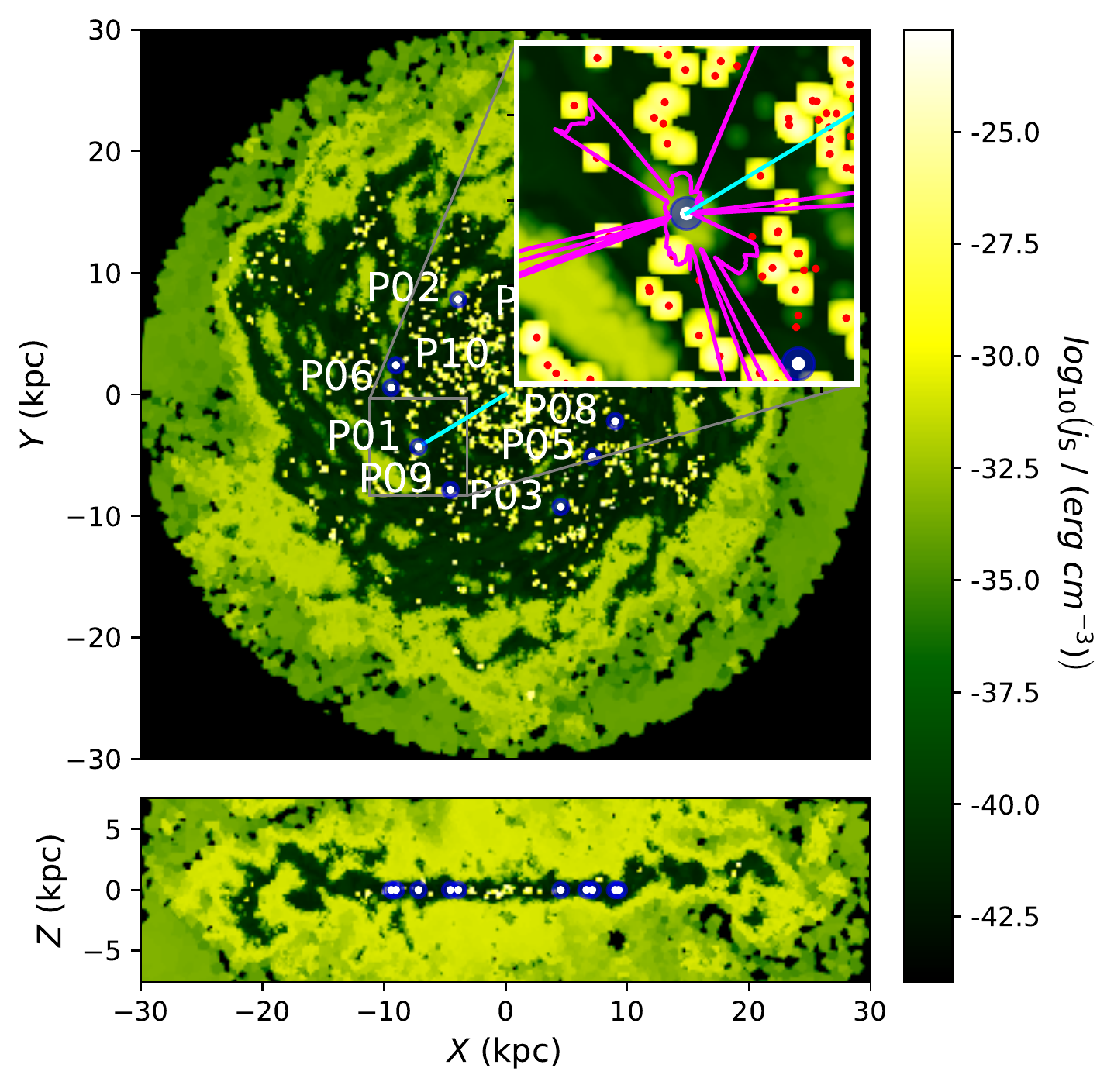}
                                 
                                  \caption{Left panel: The same as Fig.~\ref{fig:MidDensGas} for the \Ha emissivity. The cyan lines represent the direction towards the galactic center. Zoom-in panel: Red dots indicate the sampled cluster positions in the very midplane ($|z|<80\ \mathrm{pc}$). The marked area around the observer position P01 is the profile of the \Ha emissivity-weighted distance $\left\langle d_{\mathrm{H}_{\alpha}} \right\rangle$ (see Eq. \ref{eq:WeightedDistance}) along the galactic longitude (see Section~\ref{sect:ObservationsSource} and compare Fig. \ref{fig:HealDistanceHa} below). Right panel: The same as the left panel for the \SIII emissivity. For the pink marked region in the zoom-in panel see also Fig. \ref{fig:HealDistanceS} below.}
                                  
                                  \label{fig:MidEmHaSII}
                         \end{center}
                       \end{minipage} 
 \end{center} 
\end{figure*}

\begin{figure*}
 \begin{center}
         \begin{minipage}[c]{1.0\linewidth}
                         \begin{center}
                                 \includegraphics[width=0.49\textwidth]{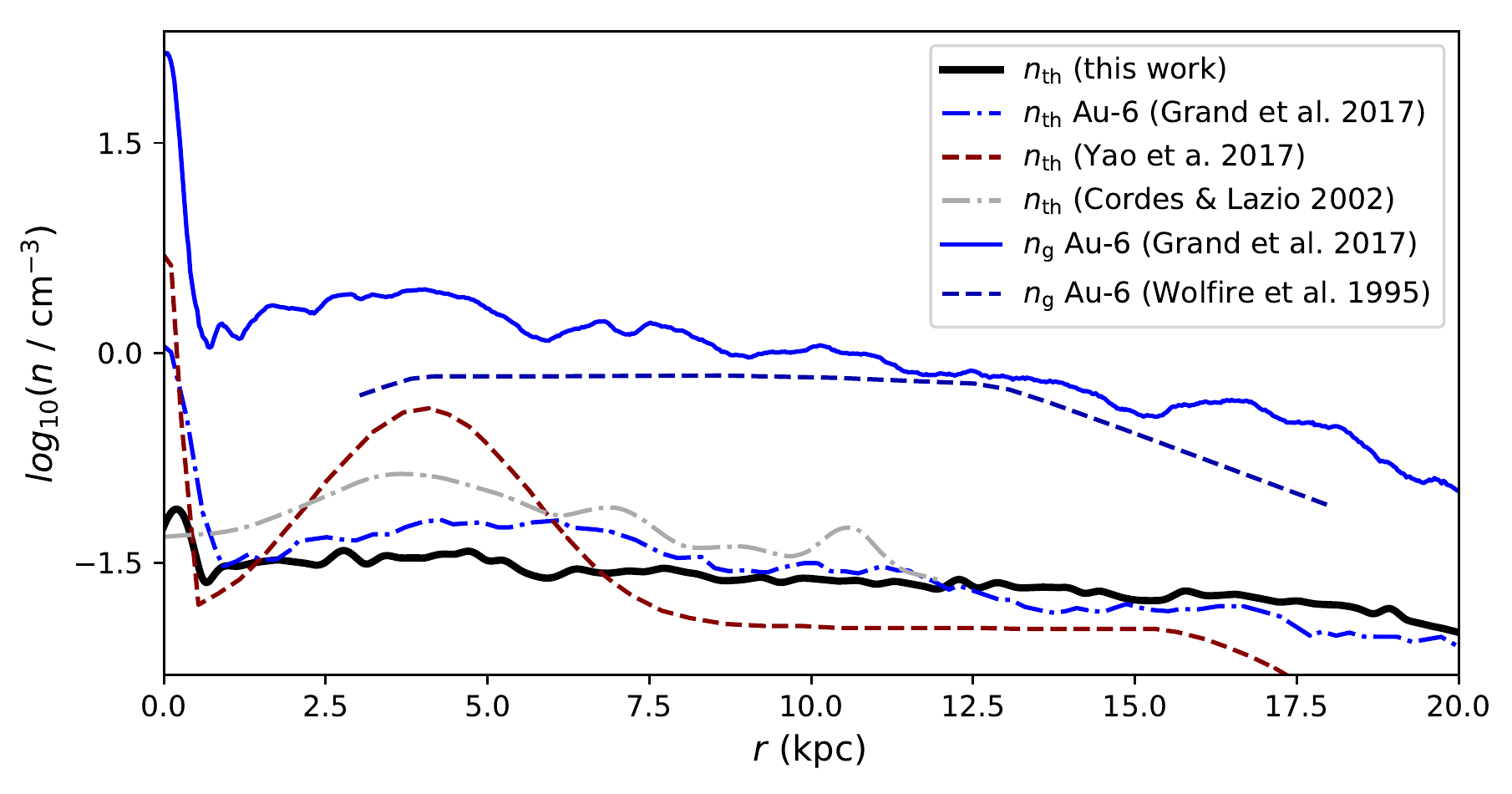}
                                 \includegraphics[width=0.49\textwidth]{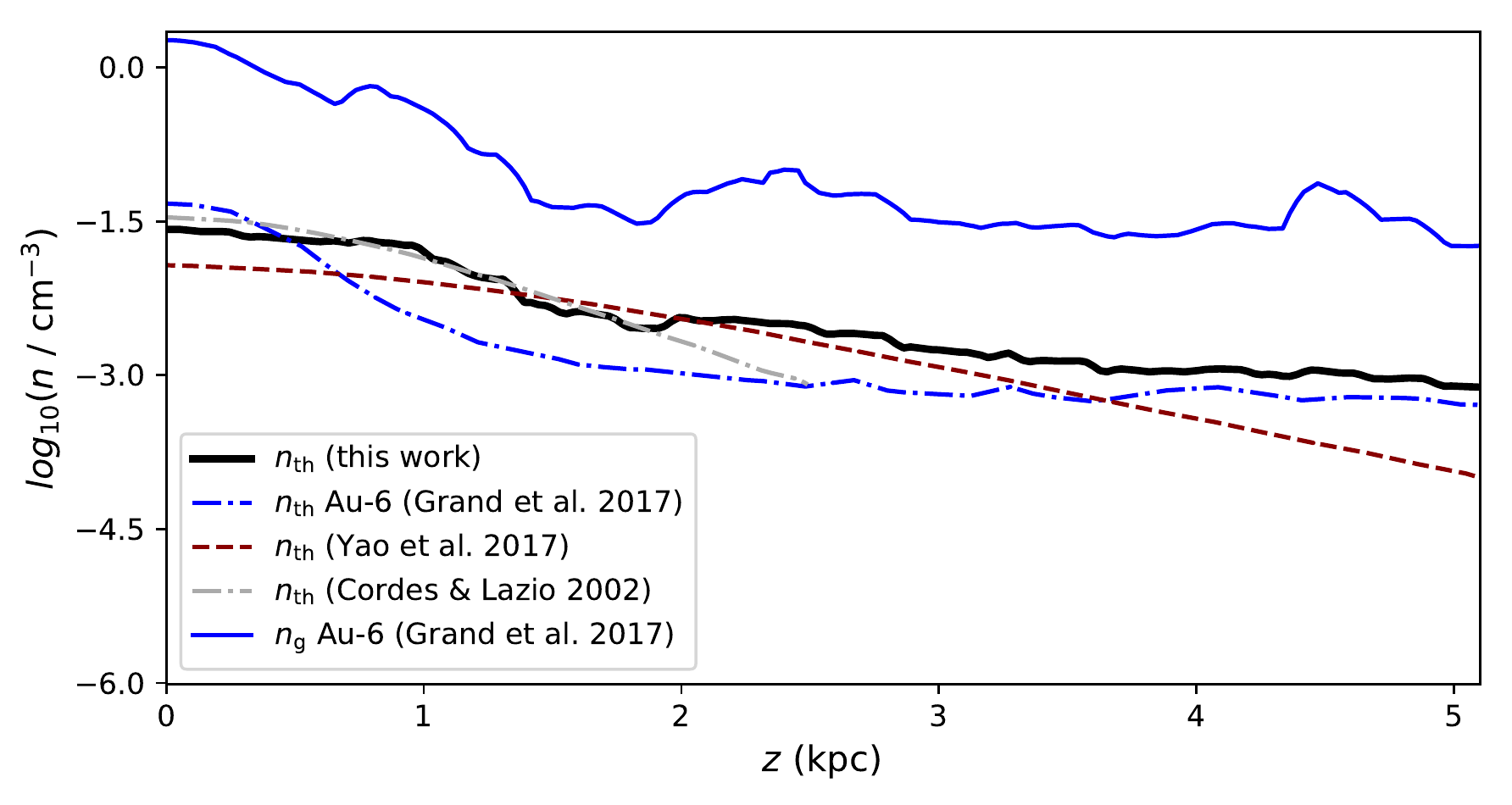}
                                 
                                  \caption{Left panel: Averaged distributions of gas density $n_{\mathrm{g}}$ (blue solid line) and thermal electron density  $n_{\mathrm{th}}$ (black solid line), corresponding to Figure \ref{fig:MidDensGas}, along the Galactic radius $r$ at $z=0\ \mathrm{pc}$. For comparison, we also show the radial profile presented in \protect\cite{Wolfire1995} for the gas density in the Milky Way (blue dashed line) and the models of \protect\cite{Yao2017} (dashed-dotted gray line) and \protect\cite{Cordes2002} (dashed red line) for the radial distribution of $n_{\mathrm{th}}$ in the Milky Way. Right panel: As in the left panel, but showing the averaged vertical distribution at a radius of $r=8\ \mathrm{kpc}$.}
                                   \label{fig:ProfDens}
                         \end{center}
                       \end{minipage} 
         \begin{minipage}[c]{1.0\linewidth}
                         \begin{center}
                                 \includegraphics[width=0.49\textwidth]{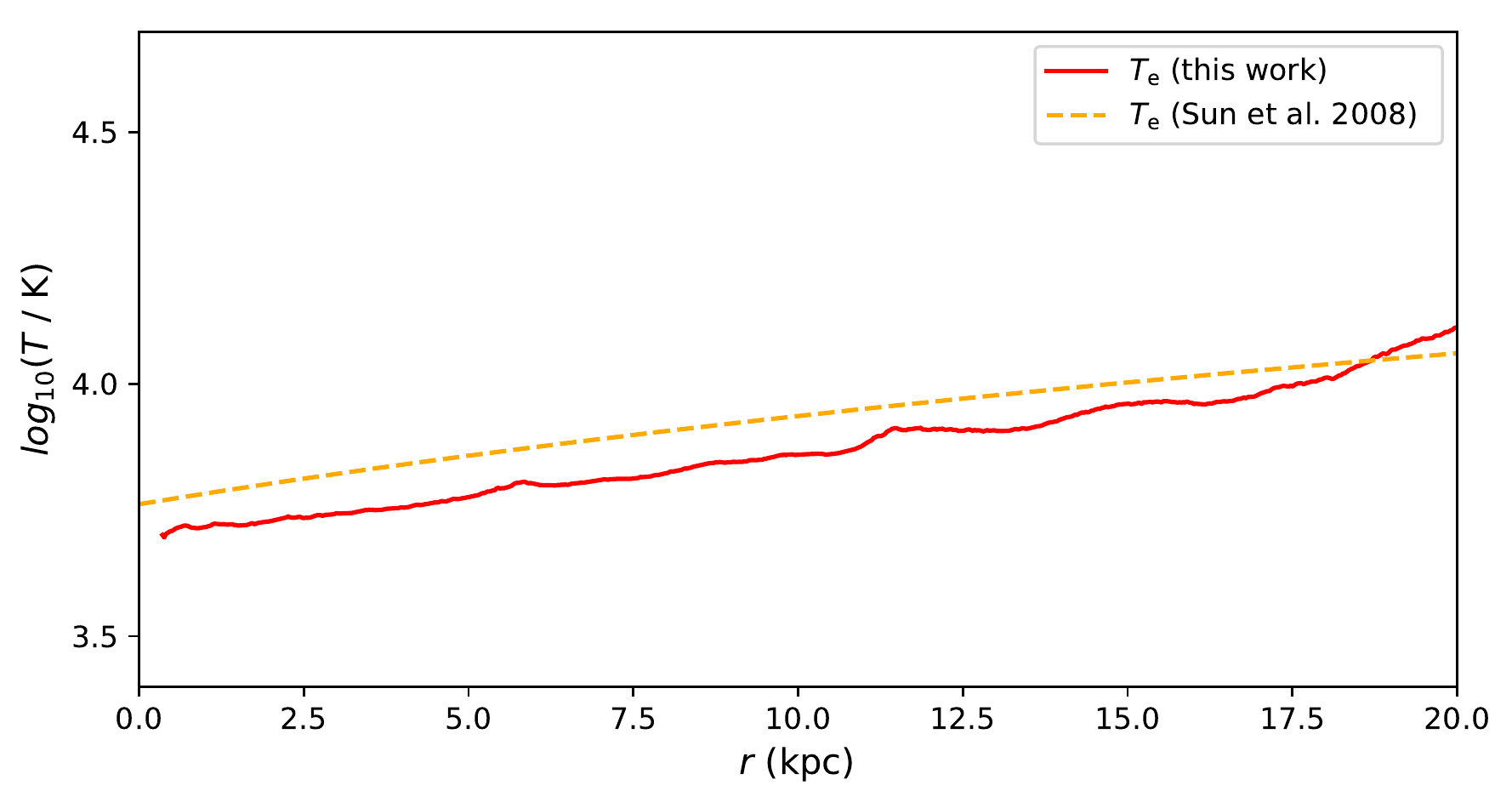}
                                 \includegraphics[width=0.49\textwidth]{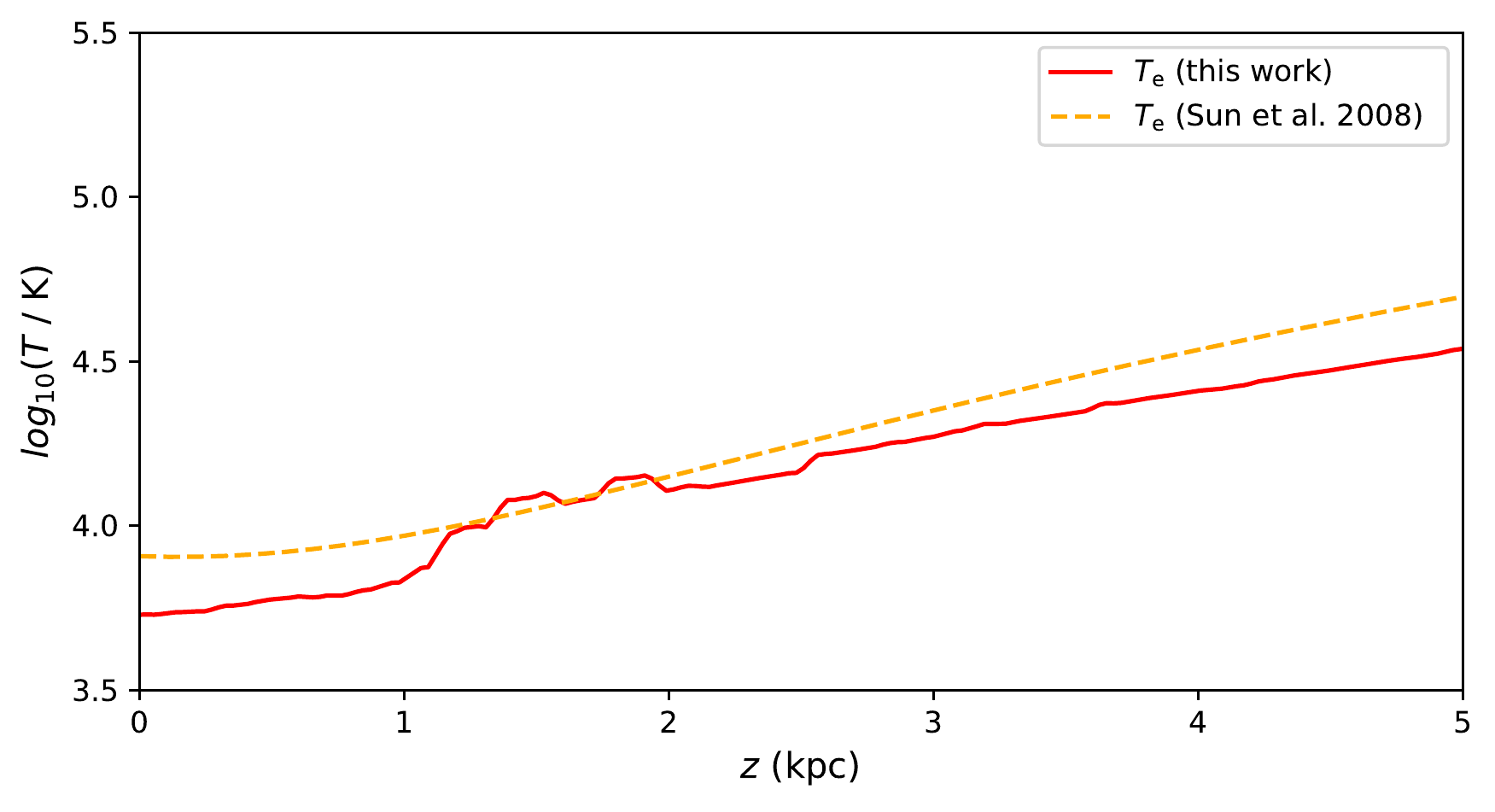}
                                 
                                  \caption{The same as Fig. \ref{fig:ProfDens} for the electron temperature $T_\mathrm{e}$ (red) and the Galactic temperature parametrization (dashed yellow) presented in \protect\cite{Sun2008}.}
                                  
                                  \label{fig:ProfThermal}
                         \end{center}
                       \end{minipage}
         \begin{minipage}[c]{1.0\linewidth}
                         \begin{center}
                                 \includegraphics[width=0.49\textwidth]{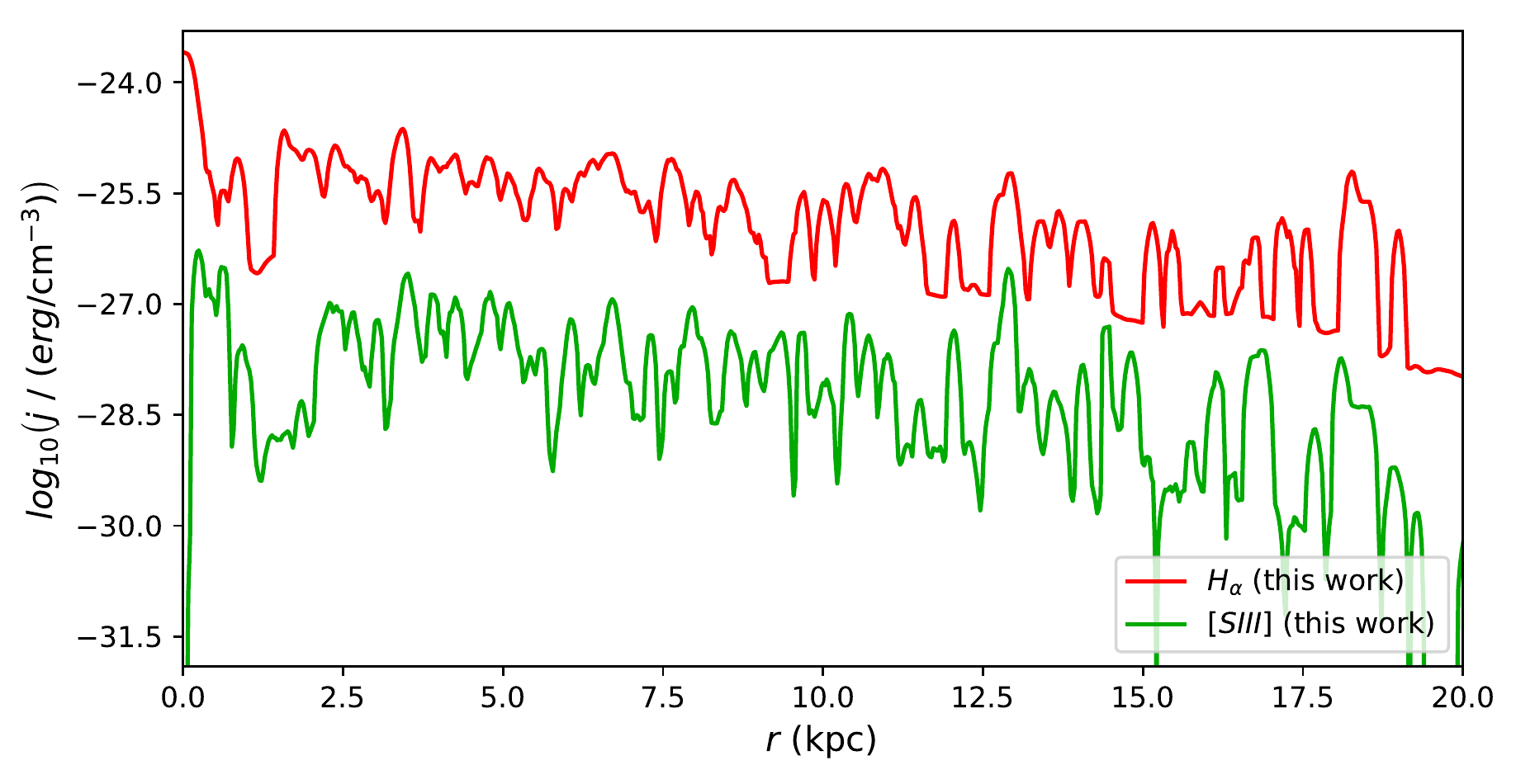}
                                 \includegraphics[width=0.49\textwidth]{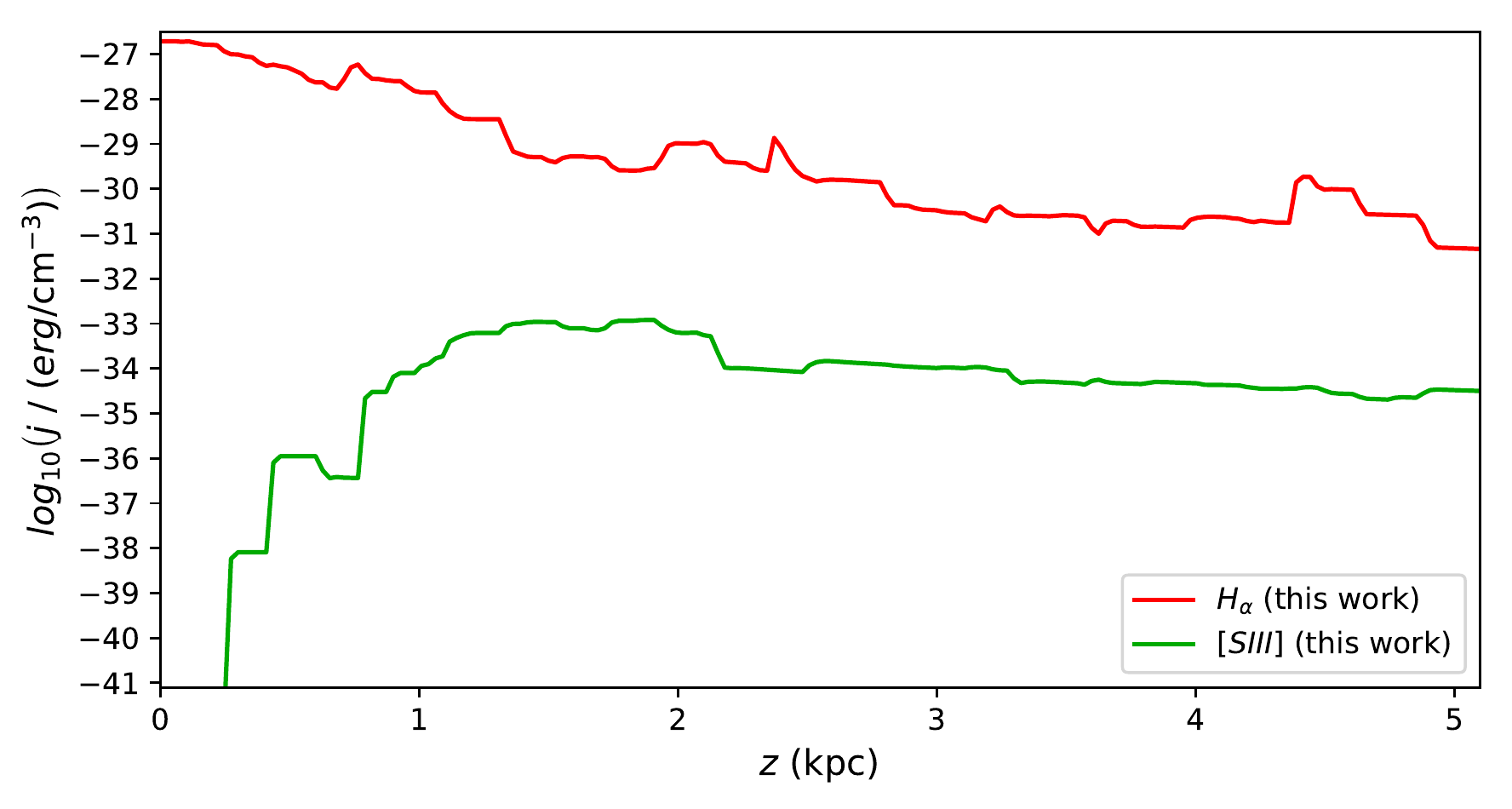}
                                 
                                  \caption{The same as Fig. \ref{fig:ProfDens} for the \Ha emissivity (red) and \SIII emissivity (green), respectively.}
                                  
                                  \label{fig:ProfEm}
                         \end{center}
                       \end{minipage} 
 \end{center} 
\end{figure*}

\subsection{Galaxy-wide spatial distribution of electrons, temperatures, and emissivities}
\label{sect:Spatial distributions}
In the top row of Fig. \ref{fig:ElDistribution} we present the PDFs of electron number density $n_{\mathrm{th}}$ and electron temperature $T_{\mathrm{e}}$ that we obtain by applying our post-processing scheme to the Au-6 galaxy at redshift $z=0$. We see immediately that there is a clear bimodality in the electron density distribution. The high electron density branch corresponds to gas that is almost completely ionized. This component is produced by ionizing photons escaping from the clusters and hence is found in close vicinity to them. The low electron density branch, on the other hand, arises primarily from gas that is located at large distances from young clusters and that is only slightly ionized. The ionization of this component is brought about primarily by cosmic rays. A similar bifurcation is recognizable for the distribution of $T_{\mathrm{e}}$. However, compared to the density PDF the high temperature branch is less pronounced.

Figure \ref{fig:EmDistribution} shows the PDFs for the synthesized emissivities $j_{\mathrm{H\alpha}}$ for the \Ha line and $j_{\mathrm{S}}$ for \SIII. For the total dataset (top row), there is a range of emissivity values at constant gas density resulting from a range of ionization fractions for \Ha. \SIII is more complex due to its higher ionization potential, and the fact that it is not a recombination line, but rather is collisionally excited. In contrast, diffuse interstellar gas between clusters (Figure \ref{fig:EmDistribution}, bottom row) shows a well defined correlation between emissivity and gas density. This is expected because this gas component is dominated by the diffuse ISRF, and thus there is a one-to-one correlation between density and ionization state.

In Figure~\ref{fig:MidDensGas}, we present cuts in the $z = 0$ and $y = 0$ planes through the gas density distribution in our chosen snapshot from the Auriga Au-6 simulation. The corresponding distributions of electron density and electron temperature are consistent with photoionization by stellar sources. The large-scale spiral morphology of the galaxy is apparent in all three plots, but close examination of Figures \ref{fig:MidDensGas} reveals a number of additional spots with high electron density, corresponding to the high ionization regions in the immediate vicinity of each star cluster. This ionization structure strongly influences the local emissivities of the different species. In Figure~\ref{fig:MidEmHaSII} we show the local emissivity of \Ha (left) and \SIII(right), respectively. While the spiral arms are visible in diffuse \Ha emission because they are partially ionized, the gas is too dense to allow for an appreciable abundance of $\rm S^{2+}$, and so they are not visible in \SIII emission. Within the inner 20~kpc of the galaxy, the \SIII emission is completely dominated by the bright spots of emission associated with the individual \hii regions. 

In Figures~\ref{fig:ProfDens} \& \ref{fig:ProfThermal} we illustrate how the gas density, electron density and electron temperature vary as a function of radius and as a function of the height above the mid-plane. The radial plots show azimuthally-averaged values computed in the mid-plane ($z = 0$~pc), while the vertical plots show azimuthally-averaged values computed at a radial distance $r = 8$~kpc (i.e.\ a position roughly corresponding to the location of the solar neighborhood in the Milky Way). 

The gas density in the mid-plane in Au-6 is roughly comparable with that in the Milky Way at $r \sim 12.5$~kpc, but is clearly higher at both larger and smaller $r$. This simply reflects the fact that Au-6 is somewhat more gas rich than the Milky Way. However, the size of the discrepancy is not large: we find at most a factor of a 40\% difference between the mid-plane gas density in the model and the value inferred for the Milky Way by \citet{Wolfire2003}, meaning the effects of porosity and an increased scale height spread out the extra mass.

For the thermal electron distribution, we find very good agreement with the results of \cite{Yao2017} at radii $7.5 < r < 15$~kpc, but we do not reproduce the increase in the electron density that they find at smaller radii. However, we note that at small radii, any comparison between Au-6 and the real Milky Way will be strongly affected by the lack of a bar in the simulated galaxy, and so it is perhaps not surprising that we do not get good agreement in this regime.  

The range of thermal electron temperatures matches well with the parametrization for our own Milky Way presented in \cite{Sun2008} up to $r = 20\ \mathrm{kpc}$ away from the Galactic center, but we note an underestimation of the temperature along $z$. However, the overall magnitude and trends demonstrate the predictive capability of our population synthesis model.

\section{Producing synthetic observations}
\label{sect:RT}
We create synthetic maps of line emission in order to further quantify the quality of our population synthesis model. For this we make use of the RT code {\sc polaris} \citep[][]{Reissl2016}, which is capable of dust polarization calculations \citep[][]{Reissl2017,Reissl2018,Seifried2019} as well as RT with atomic and molecular lines including Zeeman splitting \citep [][]{Brauer2017,Brauer2017B,Reissl2018}. {\sc polaris} solves the RT problem on the native Voronoi grid of the post-processed Au-6 data set considering both plane external detectors as well as observations inside the grid on a spherical detector, pixellated using Healpix \citep{Healpix}. 

Excluding polarization effects such as non-spherical dust grains or line Zeeman splitting, the RT equation for a velocity channel and along a certain path length $d\ell$, with emissivity $j_{\nu}$ and opacity $\kappa$ simply reads
\begin{equation}
\frac{{\rm d}I_{\nu}}{{\rm d}\ell}=j_{\nu}-\kappa_\nu I_{\nu}\, .
\end{equation}
In this paper, we produce synthetic maps of H$\alpha$, H$\beta$  and \SIII line emission. All of these lines are optically thin and so in this case line attenuation is dominated by dust extinction and 
$\kappa = \kappa_{\mathrm{dust}}$. Here, we apply the canonical ISM dust grain mixture \citep[][]{Mathis1977,Draine2001} with $37.5\ \%$ graphite and $62.5\ \%$ silicate grains following a power-law size distribution ${n(a) \propto a^{-3.5}}$ distributed over a grain size range of ${a\in \left[5\ \mathrm{nm};250\ \mathrm{nm}\right]}$.  
We calculate the dust density in each cell with
\begin{equation}
\rho_{\mathrm{dust}}= \delta_{\rm DGR} \rho_{\mathrm{gas}}\left(1.0 - 0.9 \times \frac{\rho_{\mathrm{H^{+}}}}{\rho_{\mathrm{H}}}  \right),
\end{equation}
where $\rho_{\rm H^{+}}$ is the mass density of ionized hydrogen and $\rho_{\rm H}$ is the total mass density of hydrogen in all forms (H$^{+}$, H or H$_{2}$).

Here, we account for the fact that the  amount of dust is lower in close proximity to ionizing sources \citep{Pellegrini2011, Pellegrini2009}.  One modification we need to make to the simulation concerns the dust-to-gas ratio $\delta_{\rm DGR}$. We apply $\delta_{\rm DGR}=0.003$ to reproduce the magnitude of line emission and structure observed in the Milky Way (see Sect.~\ref{sect:ObservationsAllSky}). This is a little lower than the usual ratios of $1/100\ -\ 1/300$ \citep[see e.g.][]{Whittet1992,Boulanger2000}. However,  the simulated Au-6 galaxy is $2.4\times$ more massive than the Milky Way. Thus, in order to get the correct dust attenuation, the reduction in dust abundance is needed to match the Milky Way opacity per unit length. Note that if we were not interested in comparing with Milky Way observations, or were using a model galaxy that was a closer match to the Milky Way in terms of gas mass, this modification of the dust-to-gas ratio would not be necessary. Further details regarding our treatment of the dust are given in \cite{Reissl2016,Reissl2018X}. 

To generate these maps with {\sc polaris} we use emissivities calculated as described in Section~\ref{sect:Photoionization}. These are the frequency-integrated values, and so to get the emissivity at any particular frequency, we need to multiply them by an appropriate line profile function. We assume that Doppler broadening determines the line profile within each cell and also account for the bulk velocity of the cell using the velocity field from the Au-6 simulation. For any given line, we therefore have
\begin{equation}
j_{\mathrm{\nu}} = j_{\mathrm{X}} \frac{c}{\sqrt{\pi}a_{\rm tot}\nu_{\mathrm{X, 0}}}\exp\left( -\frac{c^2(\nu_{\rm X}(v)-\nu_{\mathrm{}})^2}{a_{\mathrm{tot}}^2\nu_{\mathrm{X, 0}}^2}\right),
\end{equation}
where $j_{\rm X}$ is the frequency-integrated emissivity of the line, with the $\mathrm{X}$ representing either \Ha,  \Hb, or \SIII emission, $\nu_{\mathrm{X, 0}}$ is the line-center frequency, $\nu_{\rm X}(v)$ is the line-center frequency shifted by the  velocity $v$ of the cell relative to the observer, and $a_{\rm tot}$ is the line broadening parameter. 

For $a_{\rm tot}$, we consider a mix of thermal and micro-turbulent broadening and assume that the two contributions are in equipartition,
\begin{equation}
a_{\rm tot}^2 = a_{\rm turb}^{2} + a_{\rm th}^{2} = 2 a_{\rm th}^{2}.
\end{equation}
We therefore have
\begin{equation}
a_{\mathrm{tot}}^2=\frac{4k_{B}T_{\mathrm{gas}}}{m_{\mathrm{X}}}, \end{equation}
where $T_{\rm gas}$ is the gas temperature (which we assume to be equal to $T_{\rm e}$) and $m_{\rm X}$ is the mass of the atom. Note that the assumption of equipartition between turbulent and thermal motions would be a bad approximation if we were treating molecular emission (e.g.\ CO) from cold gas, but it is a much better approximation for the ionized gas tracers we are interested in here, since their emission comes primarily from much hotter regions in which we expect the turbulence to be transonic or only mildly supersonic.

One simplification that we are making here is the assumption that the emissivity within each grid cell is constant. In practice, this is a reasonable assumption. Regions with higher emissivity gradients typically also have higher densities and hence are well resolved in the simulation. On the other hand, in the large cells above and below the plane, the spatial resolution is poor but the emissivity gradient is small, so we still make little error by assuming a constant emissivity. We note, however, that the {\sc warpfield} models of the individual star forming regions still retain their full sub-grid resolution. 

Finally, we note that {\sc polaris} itself also allows one to compute atomic and molecular level populations and line emissivities \citep[see e.g.][]{Brauer2017}, although we do not make use of this capability in our present study.

\section{Analysis of synthetic observations} 
\label{sect:Observations}
Using the method outlined in the previous sections,  
we  compute synthetic maps of \Ha and \SIII emission as seen by observers located at different locations within the simulation volume and we examine the properties of these ``all-sky'' maps. Furthermore, we discuss what would be seen by distant observers looking along a line of sight perpendicular to the disk or at any other arbitrary angle.

\subsection{Galactic all-sky emission maps}
\label{sect:ObservationsAllSky}
For the all-sky maps we select ten distinct positions within the Auriga Au-6 simulation in environments with parameters close to our own solar neighborhood environment. We identify regions with density comparable to the local bubble our Sun is located in \citep[see e.g.][]{Fuchs2009,Liu2017,Alves2018}. These lie within the Galactic plane and at a distance of about $8\ \mathrm{kpc} \leq r \leq 10\ \mathrm{kpc}$ from the center. 
The exact positions are indicated by blue circles in Fig.~\ref{fig:MidDensGas}. Our selection ensures that the resulting all-sky maps can be meaningfully compared to real all-sky observations on Earth, and that they are not dominated by signals from nearby dense clouds or young massive clusters that are not present in the real data.

There are three resolutions to consider when comparing synthetic emission maps to real observations. First is the resolution of the observed \Ha map, which is a combination of data from multiple surveys, with a range in angular resolution from a few arcmin to 1 degree. Next is the resolution of the Healpix pixelation scheme used in {\sc polaris}. 
The all-sky maps are calculated with rays distributed according the Healpix pixelation scheme with $N_{\mathrm{side}}=256$ subdivisions per side of the 12 base pixels, resulting in a total of $N_{\rm pix} = 12 \times 256^2 = 786432$ pixels over the entire sky. This corresponds to an angular resolution of about $13.7\ \mathrm{arcmin}$, or a physical scale $\sim 40$~pc at a distance of 10~kpc. The last resolution is the resolution of the Voronoi grid from the Auriga simulation. This varies as a function of position and is also seen in projection at different distances, so that the same physical size of grid cell corresponds to very different angular sizes depending upon whether it is close to or far from the observer. In practice, our Healpix resolution is sufficient to ensure that even the smallest cells in the Au-6 Voronoi mesh are sampled with one or more rays.

\begin{figure*}
   \begin{center}
      \begin{minipage}[c]{1.0\linewidth}
           \begin{center}
           \includegraphics[width=0.75\textwidth]{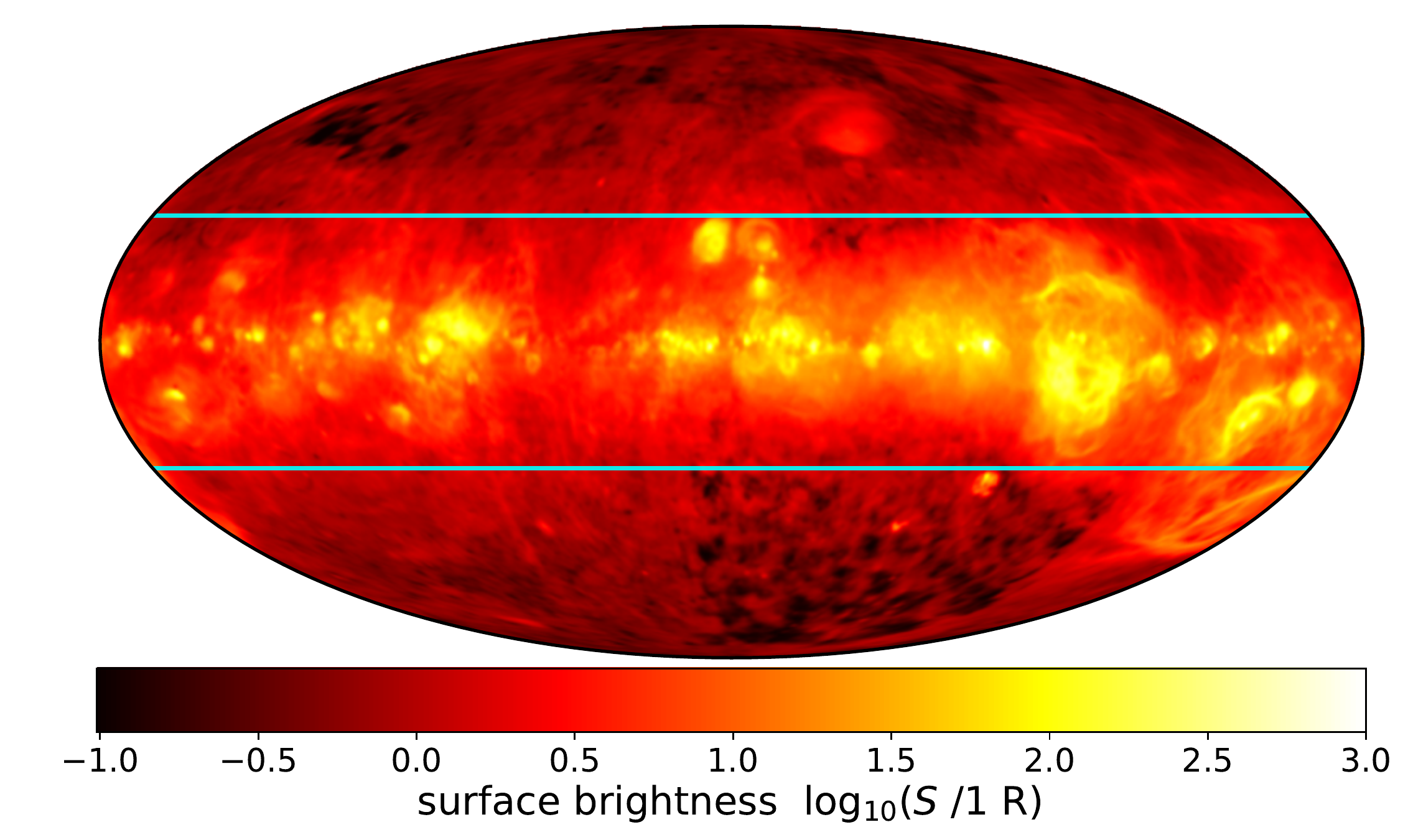}
           \end{center} 
           \caption{Observed all-sky \Ha map as presented in \protect\cite{Finkbeiner2003}. Blue horizontal lines indicate the disk region within  $\left|b \right|<30^\circ$ that we consider when analyzing the structure of the emission maps.}
         \label{fig:IMap1}
      \end{minipage}
      \begin{minipage}[c]{1.0\linewidth}
         \begin{center}
           \includegraphics[width=0.75\textwidth]{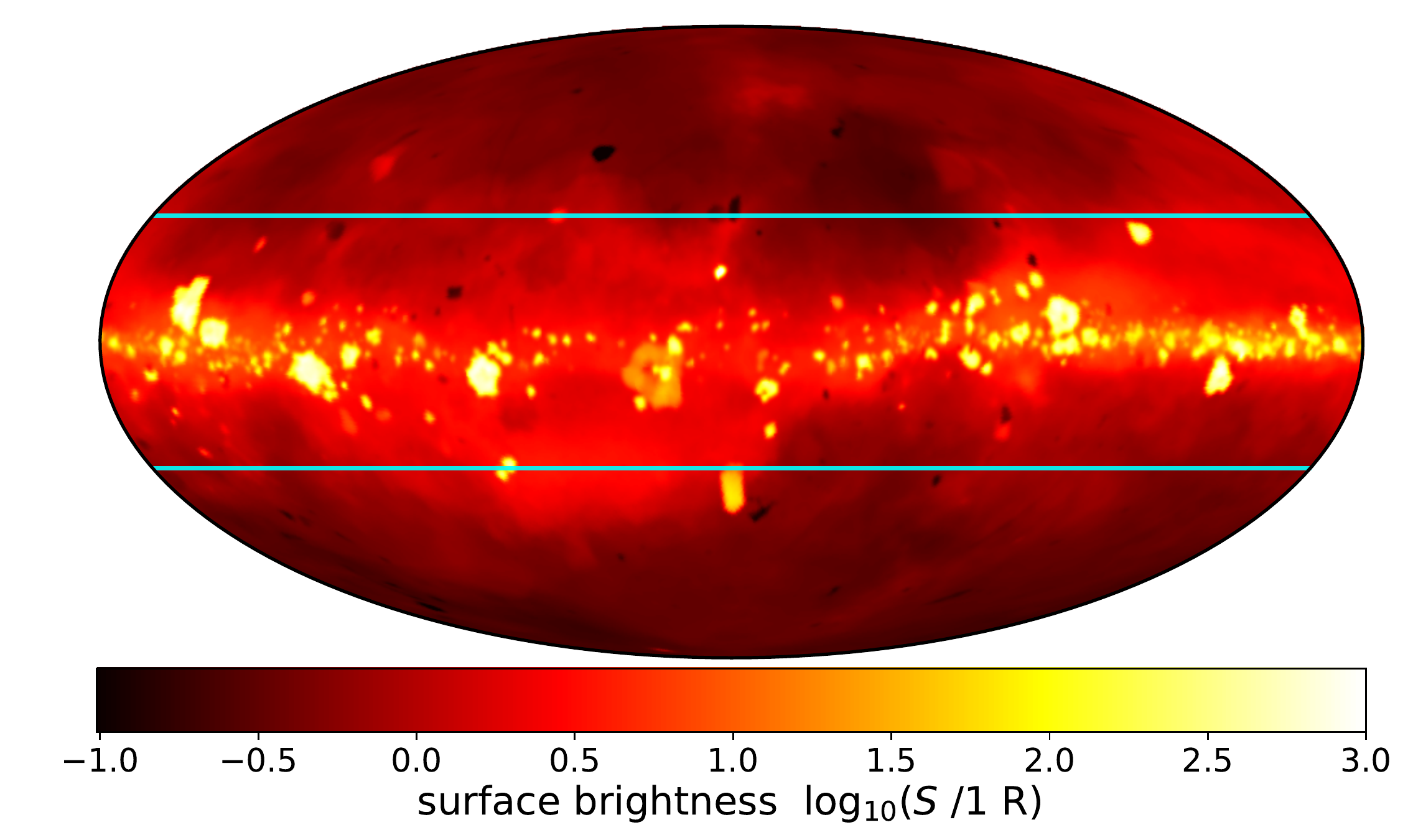}
           \end{center} 
           \caption{As Fig.~\ref{fig:IMap1}, but showing the synthetic all-sky \Ha emission map as seen by an   observer located at position P01 in the model disk. The map is integrated over a velocity range of $\pm 200\ \mathrm{km/s}$.}
         \label{fig:IMap2}
      \end{minipage}
   \end{center}   
\end{figure*}

Figure \ref{fig:IMap1} shows a reprocessed all-sky \Ha map based on data from the VTSS and SHASSA surveys and centered towards the center of Milky Way, as presented by \cite{Finkbeiner2003}. The map has a resolution of about $3.4\ \mathrm{arcmin}$ ($N_{\mathrm{side}}=1024$) and shows characteristic multi-scale patches of glowing hot ionized hydrogen gas surrounding star-forming regions. Figure \ref{fig:IMap2} presents the corresponding synthetic \Ha map of the post-processed Au-6 galaxy for the  observer position P01. The overall structure of \Ha patches as well as the magnitude of the emission match rather well. However, the angular nature of some of the bright and dark patches in the synthetic maps still reflects the underlying Voronoi grid geometry rather than any actual physical effect. This demonstrates that grid resolution is one of the limiting factors in this kind of synthetic image calculation. We note that in the anti-center directions a large fraction of the disk is observed at lower resolution than our predictions, and this could account for the higher contrast seen in some of our star-forming regions. 

Of course, our post-processing method is not limited to lines such as \Ha that have already been widely observed in the Milky Way. As an example, in Fig. \ref{fig:IMap3} we show our prediction for what an all-sky map of \SIII emission would look like. The large scale structure is similar to the \Ha maps. However, the \SIII emission is less affected by the diffuse line emission and dust extinction, allowing it to probe deeper into the Galactic disk (see also Section~\ref{sect:ObservationsSource}). We emphasize that this result comes without any fine-tuning of our post-processing pipeline. Instead, it is a direct result of Galactic population synthesis modeling from first principles.

A quantitative analysis of the structure of the synthetic \Ha all-sky emission maps is provided in Fig. \ref{fig:SpectrumEM}. Here, we apply a multipole analysis as outlined in Appendix \ref{app:Multipole}. The multipole moment is plotted down to a resolution of 1 degree. Due to the variable angular resolution in the observed map, it is difficult to meaningfully compare our synthetic map with the observations on scales smaller than this. 
Probing this issue farther will be possible with forthcoming \Ha surveys at half-arcmin resolution. 

For the analysis we mask regions above and below the plane with a galactic latitude of $\left|b \right|>30^\circ$. The inner region is defined to be the Galactic disk region. We have three motivations for focusing on this region:
\begin{enumerate}
	\item The Milky Way observations farther from the disk begin to become dominated by a combination of noise and survey artifacts related to tiling. 
	\item The Milky Way observation outside the disk are contaminated by extragalactic sources such as the Large and Small Magellanic Clouds that are obviously absent from our synthetic maps.
	\item In the Au-6 simulation, lines of sight toward the poles are dominated by low density gas, which is represented by a relatively small number of physically large Voronoi cells. We therefore see an increase in the number of grid artifacts as we look towards the poles in the synthetic all-sky maps. 
\end{enumerate}

The resulting multipole spectra in Fig. \ref{fig:SpectrumEM} reveal a characteristic zigzag pattern for the large scale structure. A similar pattern can be observed in observations and synthetic images of Milky Way synchrotron emission \citep[][]{Haslam1981A,ReisslA}, with the magnitude slowly declining towards small scales. In contrast to Galactic synchrotron emission, the overall trend for \Ha emission is the opposite with a minimum at a multipole moment of $l=0$ and an increase towards smaller scales. This trend and the zigzag pattern is common to all considered observer positions, although some positions have more small-scale power compared to the Milky Way. We attribute this difference in structure to the local conditions surrounding each observer position. The number of clusters in the proximity of each observer ranges from one to several dozens. This influences the multipole fitting significantly. Another contributory factor is the dust extinction. The depth to which one can see in the mid-plane is sensitive to the local dust distribution, and so from some of our example observer positions one can see clusters at much greater distances than from other observer positions. As more distant clusters have small angular scales, this translates into a considerable variation in the amount of small-scale power seen from each position.
It explains the excess in magnitude from $l\approx 20$ onward compared to the spectrum of the Milky Way that we find for many (but not all) observer positions.

This result also demonstrates the necessity of accounting for dust extinction when making this kind of synthetic image. 
Ignoring the dust extinction in the RT calculations would lead to an increase in the small-scale structure of the \Ha maps. This is because without extinction, the \Ha emission penetrates the entire disk, meaning that all of the clusters present within the disk would contribute to the image. 

\subsubsection{Source of the Galactic emission}
\label{sect:ObservationsSource}

\begin{figure*}
  \begin{center}
     \begin{minipage}[c]{1.0\linewidth}
        \begin{center}
           \includegraphics[width=0.75\textwidth]{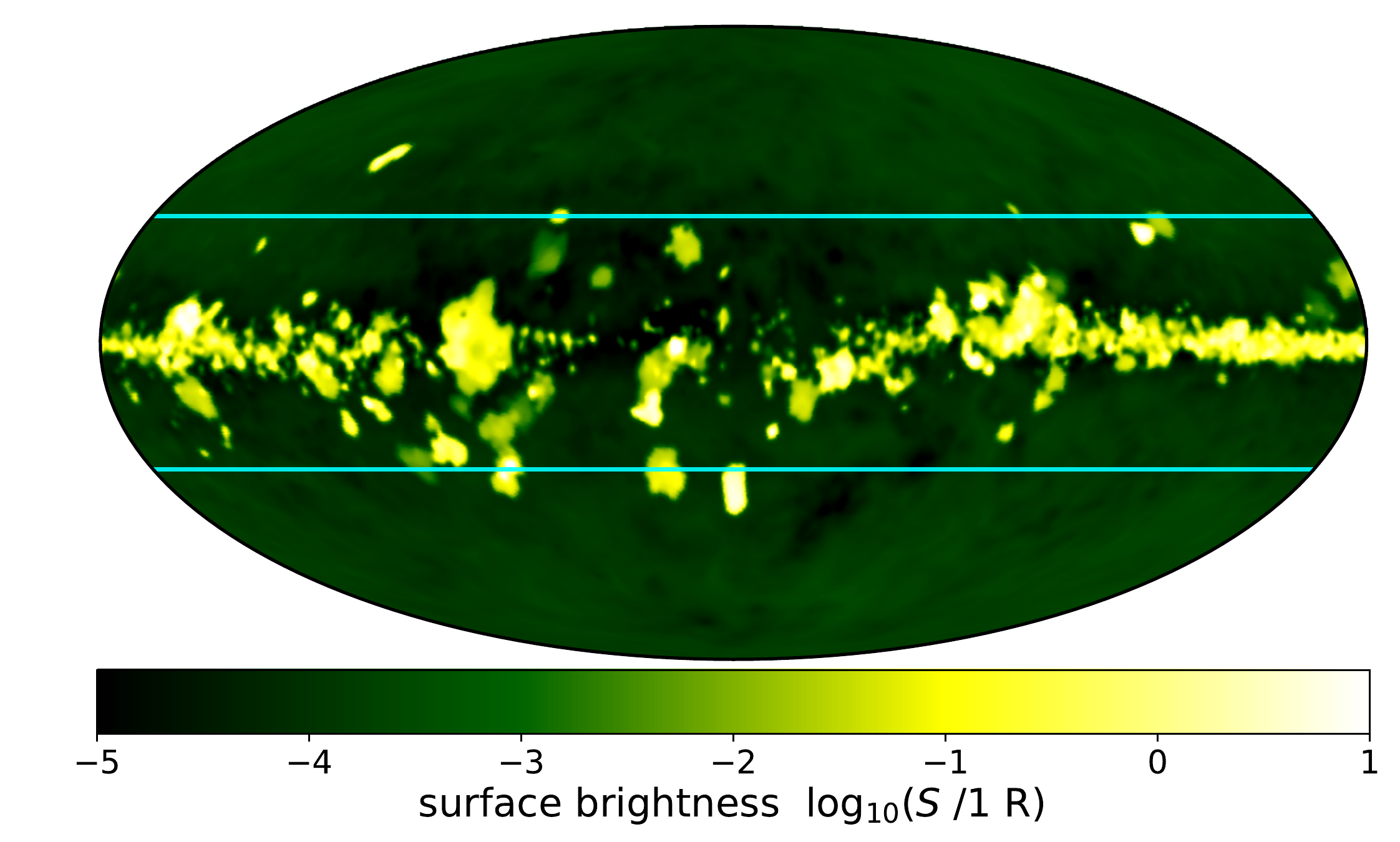}
        \end{center} 
        \caption{Same as Fig. \ref{fig:IMap1}, but showing the synthetic all-sky \SIII emission map as seen by an   observer located at position P01 in the model disk.}
        \label{fig:IMap3}
      \end{minipage}
   \end{center}   
\end{figure*}

\begin{figure}
   \begin{center}
      \begin{minipage}[c]{1.0\linewidth}
         \begin{center}
             \includegraphics[width=1.0\textwidth]{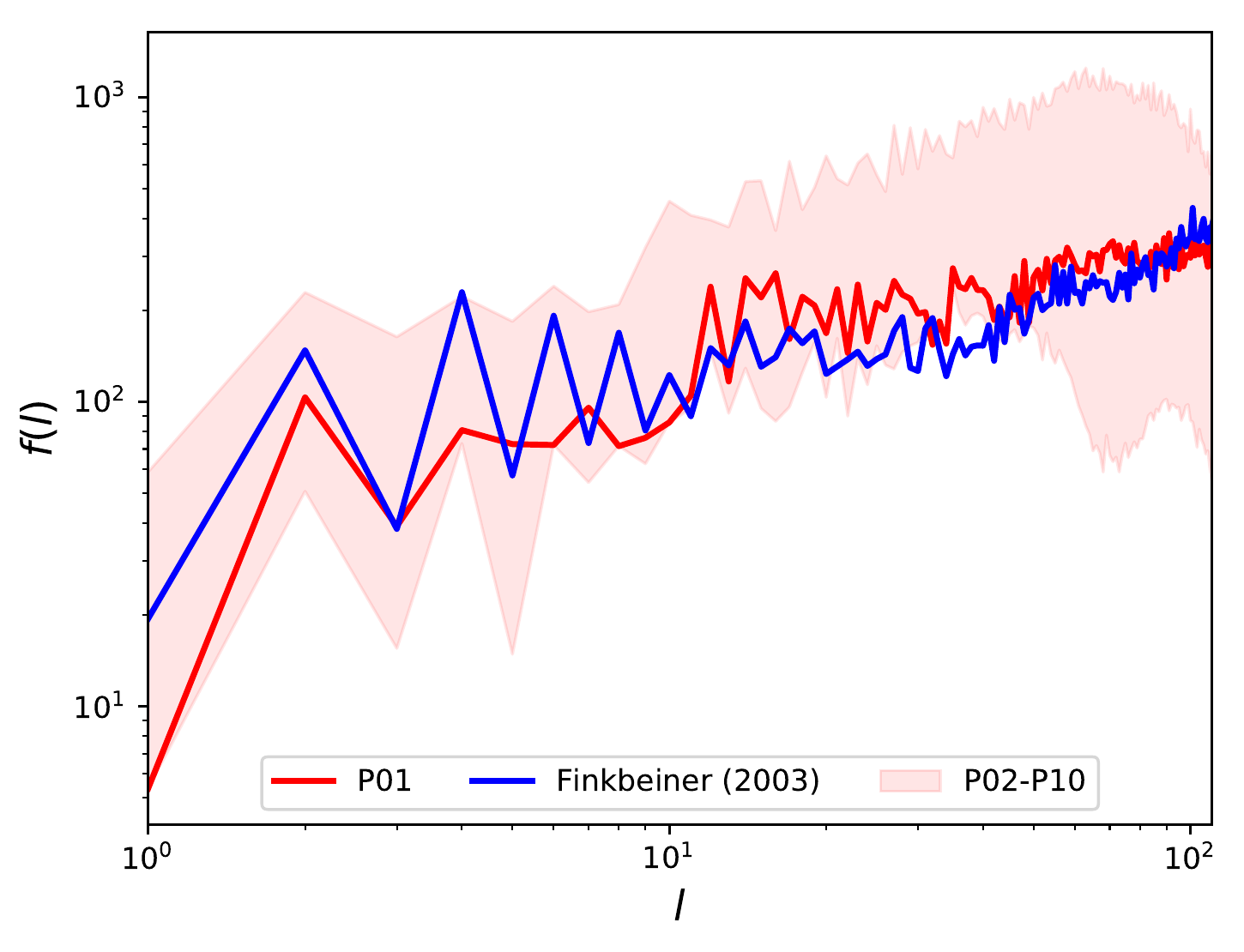}
         \end{center} 
         \caption{Multipole spectrum as a function of multipole moment $l$ of the synthetic \Ha map for the different observer positions P01 to P10. For comparison, we also show the spectrum derived from the map presented in \protect\cite{Finkbeiner2003}. The analysis is done for the designated Galactic disk with a latitude of $|b|<30^{\circ}$.}
         \label{fig:SpectrumEM}
      \end{minipage}
   \end{center}
\end{figure}

In Fig. \ref{fig:IMap3} we show a synthetic all-sky map for Galactic \SIII emission. The large-scale structure is different than in the \Ha maps, as it is less extended due to the lower relative emissivity of diffuse gas. Individual star-forming regions have high contrast because there is less diffuse emission and lower dust extinction, allowing the \SIII line to probe deeper into the Galactic disk. As a measurement of the origin of the \Ha and \SIII emission within the grid, we define the emissivity-weighted distance along the line of sight as
\begin{equation}
\left\langle  d \right\rangle = \frac{\int_{\ell'}^0   d(\ell)\times j_{\nu}\exp\left(-\tau_{\nu}(\ell) \right)  \ d\ell   }{\int_{\ell'}^0   j_{\nu}\exp\left(-\tau_{\nu}(\ell) \right)\ d\ell  }\, 
\label{eq:WeightedDistance}
\end{equation}
with a distance $d(\ell)$ between the observer and the particular position $\ell$ and an optical depth of $\tau_{\nu}(\ell)$ from the observer position up to $\ell$.

\begin{figure}
   \begin{center}
         \begin{minipage}[c]{1.0\linewidth}
         \begin{center}
            \includegraphics[width=1.0\textwidth]{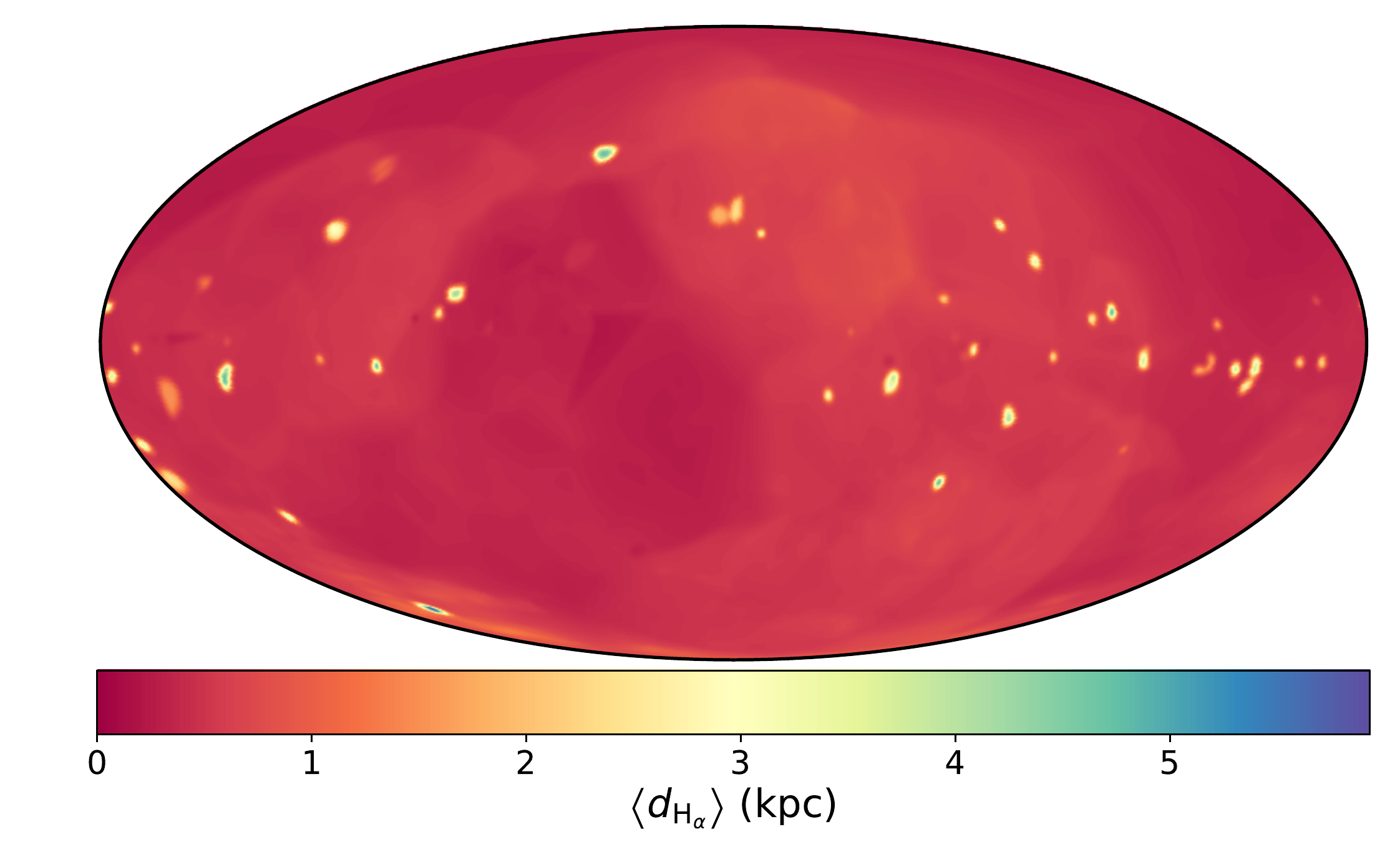}
         \end{center} 
               \caption{All-sky map of the \Ha emissivity-weighted $\left\langle d_{\mathrm{H}_{\alpha}} \right\rangle$ as seen from position P01.}
      \label{fig:HealDistanceHa}
      \end{minipage}
      
       \begin{minipage}[c]{1.0\linewidth}
         \begin{center}
            \includegraphics[width=1.0\textwidth]{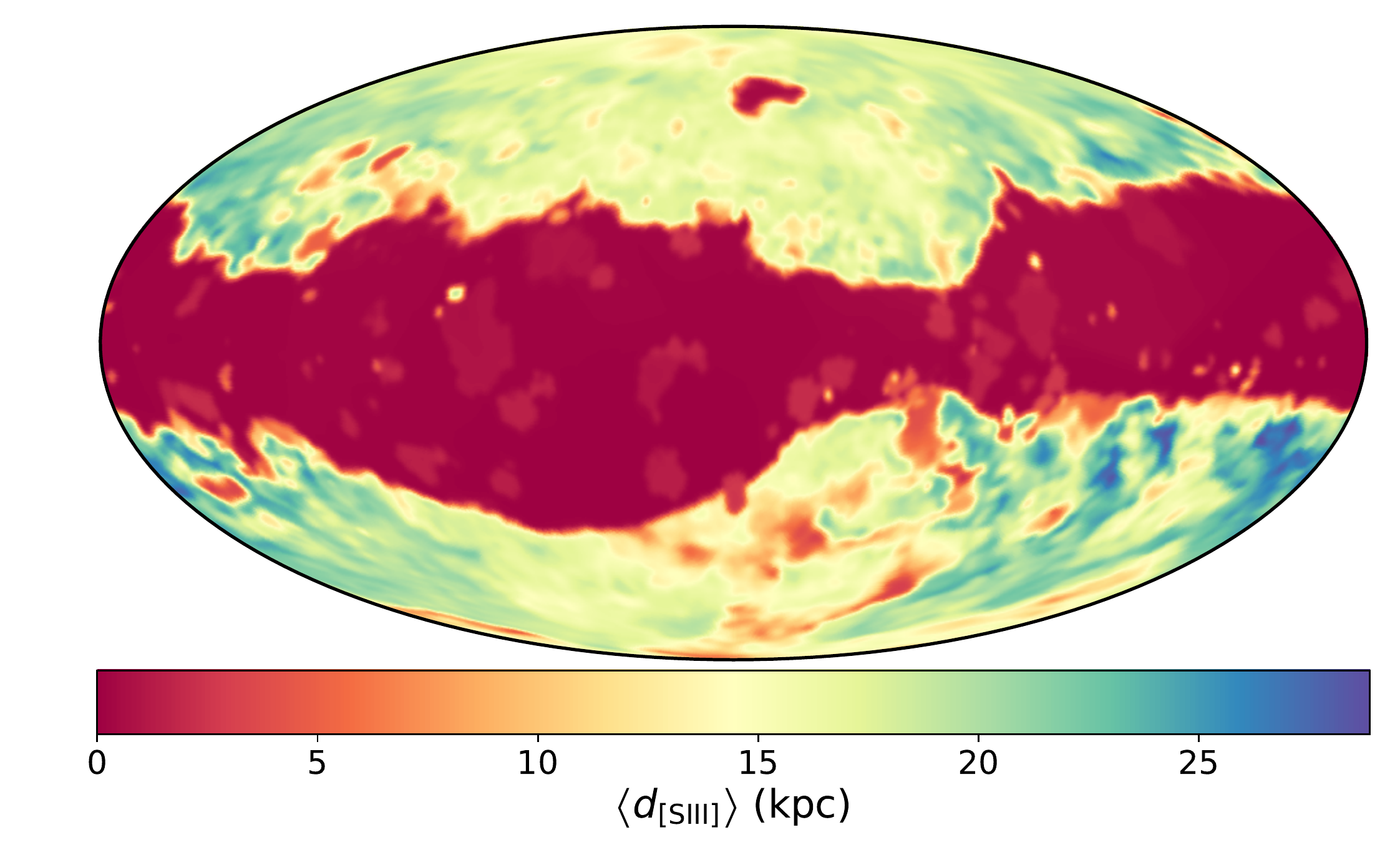}
         \end{center} 
               \caption{The same as Fig.  \ref{fig:HealDistanceHa} for $\left\langle d_{\mathrm{[S_{III}]}}\right\rangle$. }
      \label{fig:HealDistanceS}
      \end{minipage}
      
      \begin{minipage}[c]{1.0\linewidth}
         \begin{center}
            \includegraphics[width=1.0\textwidth]{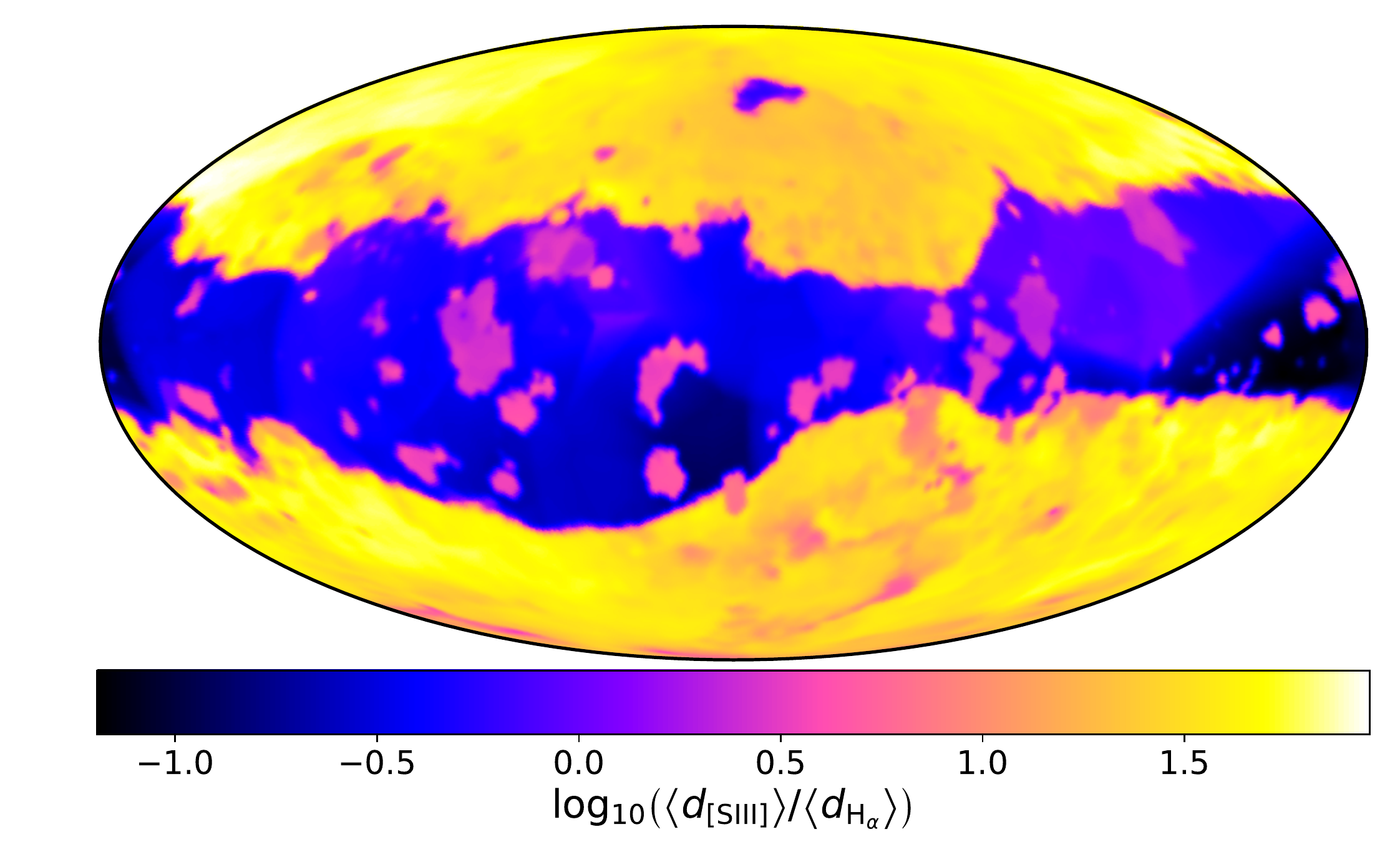}
         \end{center} 
               \caption{All-sky map of the ratio $\left\langle d_{\mathrm{[S_{III}]}}\right\rangle$ to $\left\langle d_{\mathrm{H}_{\alpha}} \right\rangle$.}
      \label{fig:HealDistanceRatio}
      \end{minipage}

   \end{center}   
\end{figure}

In Figures~\ref{fig:HealDistanceHa} and \ref{fig:HealDistanceS}, we show an all-sky map of the emissivity-weighted distances for the \Ha and \SIII emission, which helps to highlight the lines of sight for which these tracers probe very different distances. In Figure~\ref{fig:HealDistanceRatio}, we show the ratio of emissivity-weighted distances. 

The \Ha emission seen by an observer located (as the Solar System is) within a low density bubble is dominated by diffuse ionized gas with relatively low extinction, and typical distances of about 1$ \, \mathrm{kpc}$. Along lines of sight toward bright, relatively nearby star-forming regions, their emission outshines the local diffuse emission, resulting in a higher emission-weighted distance of about $2\ \mathrm{kpc}$.\footnote{Note that as the diffuse gas still contributes along these sight-lines, this is an underestimate of the actual distance to these star-forming regions.}

The morphology of the \SIII effective distance map is markedly different. In the galactic plane, many sightlines are once again dominated by relatively nearby diffuse emission, while some are dominated by brighter, more distant star-forming regions. However, the characteristic distances of both components are larger, and so there is more structure in the diffuse map, and the compact \hii regions are typically farther away.
The origin of this is a matter of statistics and extinction. Statistically, the probability of finding more massive clusters, which are required for bright \SIII, is lower than that of finding lower mass clusters. This translates to a larger typical distance between massive regions, which leads to a larger ratio distances in Figures~\ref{fig:HealDistanceRatio}. We also see that there is a dramatic increase in the emission-weighted distance of the \SIII emission as we look out of the plane. This comes about because there is so little local emission in these directions that we start to become dominated by the faint signal from the hot gas in the halo, although we see from Figure~\ref{fig:IMap3} that in practice this signal is likely too weak to be detectable.

While purely theoretical now, planned missions to map the majority of the star forming disk in all optical emission lines (including \SIII) are under construction \citep[e.g. SDSS-V; see][]{kollmeier2017}. We find \SIII will make it possible to trace obscured high mass star formation up to five times farther in the disk than \Ha, partly due to extinction, and partly due to less confusion with diffuse gas, which is much fainter in \SIII than \Ha, due to reduced diffuse \SIII emission (see Figure~\ref{fig:MidEmHaSII}-right). Catalogs of star-forming regions, or selection functions based on \SIII will be less sensitive to galactic structure, and more sensitive to star formation and population characteristics. 

\begin{figure}
   \begin{center}
      \begin{minipage}[c]{1.0\linewidth}
         \begin{center}
             \includegraphics[width=1.0\textwidth]{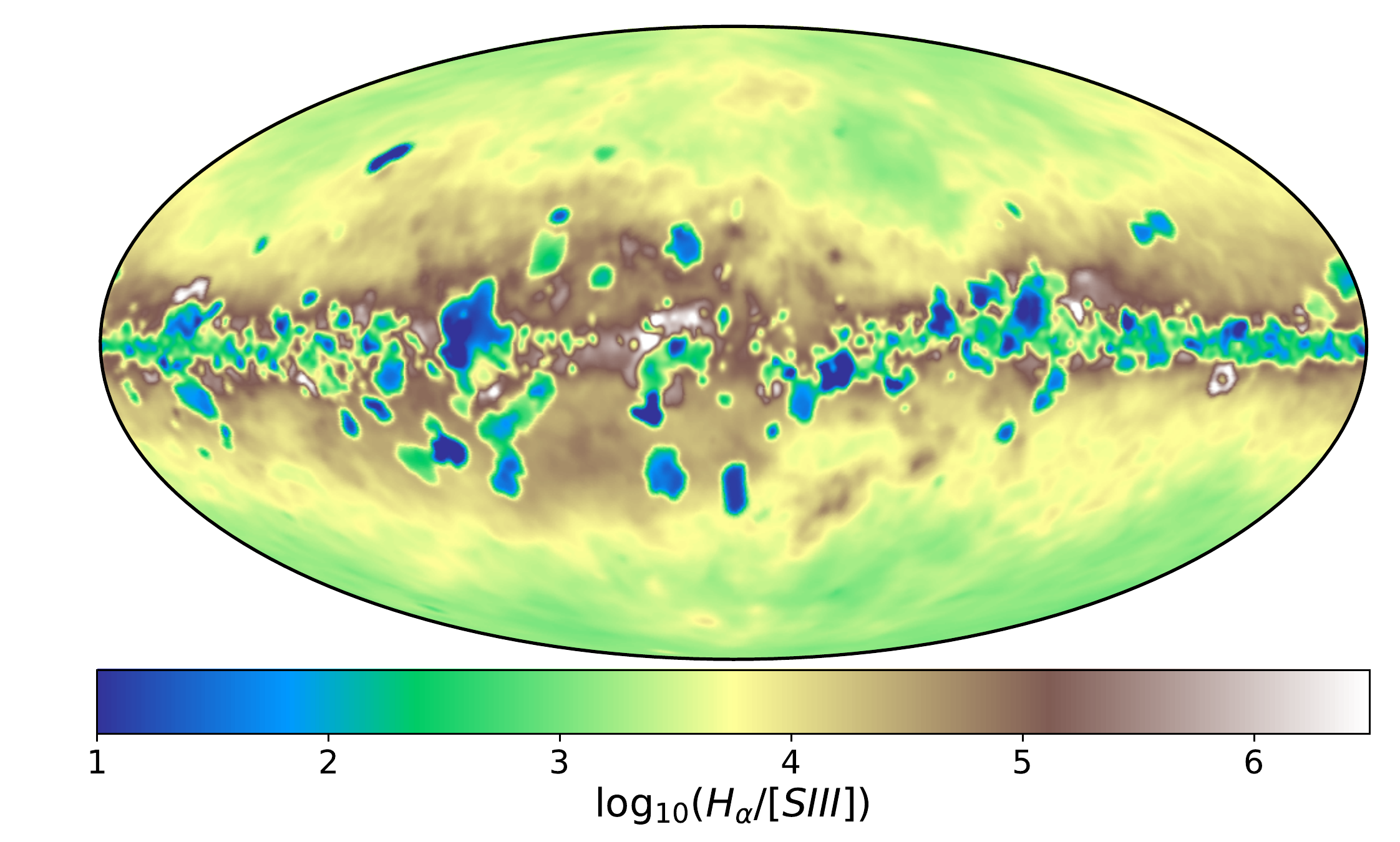}
         \end{center} 
         \caption{All-sky map of the \Ha /\SIII ratio as seen from position P01.}
         \label{fig:Ratio}
      \end{minipage}
   \end{center}
\end{figure}

\subsubsection{\Ha/[S {\sc III}] Ionization Parameter Mapping}

Spatially resolved emission ratio maps employing high-to-low ionization potential tracers \citep{Pellegrini2012}, as well as dust tracers sensitive to photon flux \citep{Oey2017, Binder2018}, can be used to map local variations in the ionization parameter $U = n_{\rm ion} / n_{\rm H}$, i.e.\ the ratio of the ionizing photon number density to the atomic hydrogen number density.
Such ratio maps depend on the intensity and spectral shape of the ionizing radiation, and ISM structure in and around star-forming regions. These dependencies make such maps useful for seperating distinct star-forming regions from each other \citep{Pellegrini2012}, as well as separating these regions from larger-scale galactic structure. On small scales, local variations are dominated by local gas ionization structure which depends on the optical depth to Lyman continuum radiation of individual star-forming regions. This makes it possible to quantify the relative number of optically thick (radiation bounded) to optically thin  (density bounded) \hii regions, as well as to measure the covering fraction of blister type \hii regions. $\rm H^+$ ionization fronts have a width equal to the mean free path of ionizing photons.  Due to the high cross section for ionizing radiation, this is typically short ($\ll {\rm pc}$) \citep{agn2006}. As ionizing photons are depleted due to radiative transfer, the local ionization degree of the gas drops. The relative abundance of more highly ionized ions decreases first, owing to their faster recombination rates, resulting in large changes in emission line ratios. 
Emission line ratio maps can therefore be used to highlight the locations where the ionization fraction of the gas is rapidly decreasing, allowing them to reveal the existence of an ionization front in dense gas \citep{Pellegrini2012}.

Emission line ratio maps are often interpreted in terms of large-scale ISM evolution. However, this intepretation assumes that 
the emission traces gas in the ISM which has dynamically responded to feedback (e.g.\ by being driven into large shells). However, without the aid of simulations there is no direct way to know if the termination of the flow of radiation in a given direction occurs in the natal gas cloud in which the massive stars have formed, or if this cloud has already been fully ionized, leading to the ionizing radiation being absorbed farther away by 
surrounding material not directly associated with the star formation and unaffected by mechanical feedback (winds, supernovae etc.) from the young stars.

Figure~\ref{fig:Ratio} shows the all sky \Ha/\SIII ratio. Despite the spherical symmetry of the input {\sc warpfield} models, the ratio maps reveal low ionization gas surrounding more highly ionized regions, with irregular morphologies. 
The significance of this is twofold. First, it means that even when the evolution of individual star-forming regions is idealized as spherically symmetric (which will tend to underestimate the escape fraction of ionizing photons compared to what one would find in a more irregularly structured cloud), the emergent radiation is strong enough to ionize the surrounding ISM, thus creating a diffuse ionized gas (DIG) component that contributes to the observed total flux. The surrounding gas is dense enough to absorb the radiation over a short distance, making the DIG appear as a continuation of the now ionized natal cloud. This introduces a difficulty in testing physics by comparing models of isolated individual clouds and their evolution to observations, even when the observations are able to spatially resolve individual star-forming regions. These comparisons require making an observational determination of what gas is involved in star formation, and distinguishing between gas that has been accelerated and swept into shells by feedback versus gas which is only illuminated by the escaping radiation. For these reasons, a proper comparison of observations to models must not only include a self-consistent population of star-forming regions, but also the extinction and emission from the environment in which they form.

\subsection{Applications to non-Milky Way galaxies}
\label{sect:ObservationsExtra}

Viewed from an external position, our model galaxy should be equivalent to a typical $L_*$ spiral galaxy.
Figure~\ref{fig:ExtMapHaHatrue15} and  Figures~\ref{fig:ExtMapHaHatrue0deg}-\ref{fig:ExtMapHaHatrue90deg} in the Appendix show the galaxy as seen from different orientations, ranging from face-on (0$^\circ$) to edge-on (90$^\circ$). These maps were generated with a projected grid of 1024 pixels covering a total area of $60 \times 60$~kpc and hence are sensitive to structures on scales down to $\sim 60$~pc. For comparison, recent IFU observations of NGC~628 provide spectroscopic information on many \hii region diagnostic lines at a spatial resolution of $\sim 40$~pc \citep{Kreckel2018,RN2018}, and the ongoing PHANGS-MUSE\footnote{\url{https://sites.google.com/view/phangs/projects}} and forthcoming SIGNALS\footnote{\url{http://www.signal-survey.org}} surveys will do the same for a much larger sample of nearby spiral galaxies at similar resolutions. 
We expect our models to play an important role in helping to guide the interpretation of these forthcoming surveys.

\subsubsection{Extinction}

When interpreting extragalactic observations it is necessary to understand simultaneously the role of dust, diffuse emission and crowding. \Ha observations play an important role in understanding star formation \citep{Kennicutt1998}, and the luminosity function of star-forming regions helps to constrain the initial mass function of clusters, and SFRs. In the absence of dust attenuation, \Ha emission is roughly proportional to the SFR. In practice, \Ha is attenuated by dust extinction, which must be corrected for in order to recover an unbiased measure of the SFR. This correction is typically carried out by observing \Hb in addition to \Ha. As the \Ha and \Hb lines have difference frequencies, they suffer from different amounts of dust extinction. Therefore, if the intrinsic ratio of \Ha to \Hb is known, the observed ratio can be used to infer the amount of dust absorption.

A full parameter space exploration of the effect of both dust and crowding is beyond the scope of this paper, and so we can only give a broad overview here. In order to quantify the effects of differential extinction for our model galaxy we calculate the radial profile of \Ha and \Hb emission measured by an external observer for inclination angles ranging from 0$^\circ$ to 90$^{\circ}$ at 15$^{\circ}$ intervals. We also compute the ``true'' \Ha and \Hb fluxes, neglecting all dust absorption. We note that we do not neglect the effects of dust absorption when calculating the ionization state of the gas or the emissivity of the underlying model. Instead, we merely set the dust opacity to zero when ray-tracing the resulting \Ha and \Hb emission. 

\begin{figure*}
	\begin{center}
		\begin{minipage}[c]{1.0\linewidth}
			\begin{center}
				\includegraphics[width=1.0\textwidth]{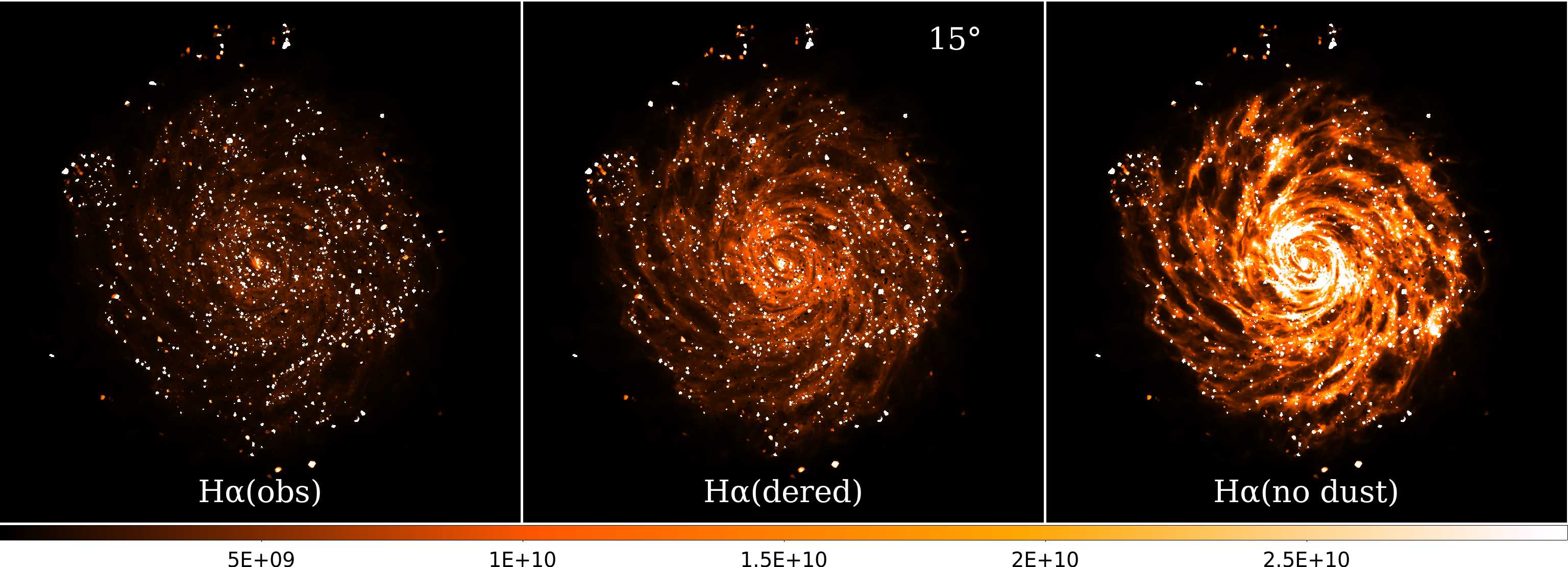}
				\includegraphics[width=1.0\textwidth]{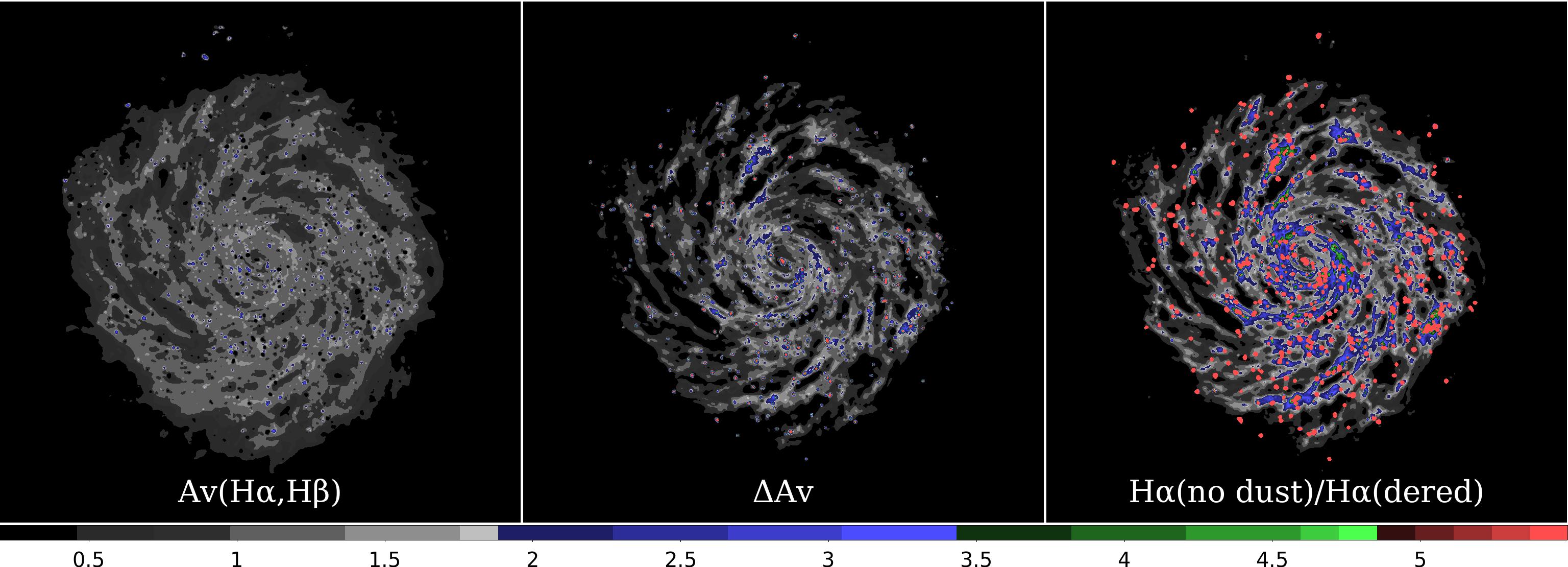}
			\end{center} 
		\end{minipage}
	\end{center}
	\caption{Log of the \Ha emission maps, with a range of $10^7 - 5\times10^{10}~{\rm Jy}$, viewed at  an inclination of $i=15^\circ$ (top row) from left to right: observed, de-reddened using the Balmer decrement, and with no dust absorption. In the bottom row, we show the derived extinction in ${\rm A_V}$ magnitudes (left), $\Delta \rm{A_{V}}$ (center), and the ratio of the dust free \Ha emission to that recovered with standard Balmer decrement de-reddeing.}
	\label{fig:ExtMapHaHatrue15}
\end{figure*}

As an observer would also do, we compute the dereddened \Ha flux from the Balmer decrement. We assume that the extinction can be modelled as a foreground screen, and that an $R_{\rm V}=3.1$ reddening law applies. We also assume that the intrinsic \Ha/\Hb ratio is 2.86.\footnote{In reality, the ratio is a weak function of the temperature and density of the ionized gas, but we neglect this complication here.} The extinction is calculated (and dereddening applied) at the native resolution of our ray-traced images, namely 60~pc. The different \Ha emission and extinction maps are shown in Figure~\ref{fig:ExtMapHaHatrue15}. We define the deviation ${\rm \Delta} A_{\rm V}$ as the difference between the value derived from the ratio of dust-free and observed \Ha fluxes and $A_{\rm V}$ derived from the \Ha/\Hb ratio,  
\begin{equation}
\Delta A_{\rm V}  = 2.5/f_{\rm H\alpha} \times {\rm log_{10}}~(F({\rm H\alpha})_{\rm no~dust}/F({\rm H\alpha})_{\rm obs}) - A_{\rm V}\;,
\end{equation}
where $F({\rm H\alpha})_{\rm no~dust}$ and $F({\rm H\alpha})_{\rm obs}$ are the \Ha flux in the dust-free map and the full calculation, respectively, and $f_{\rm H\alpha} = 0.818$ is the ratio of the extinction at the wavelength of \Ha to the extinction in the V band, assuming a standard $R_{\rm V}=3.1$ reddening law.

We see from the Figure that the standard de-reddening correction does a relatively good job in regions where the gas density is relatively low (e.g.\ at large galactocentric radius). However, it tends to systematically underestimate the true amount of obscuration in regions of high gas density, with this effect becoming particularly pronounced as one nears the center of the galaxy. 

\begin{figure}
	\begin{center}
		\begin{minipage}[c]{1.0\linewidth}
			\begin{center}
				\includegraphics[width=1.0\textwidth]{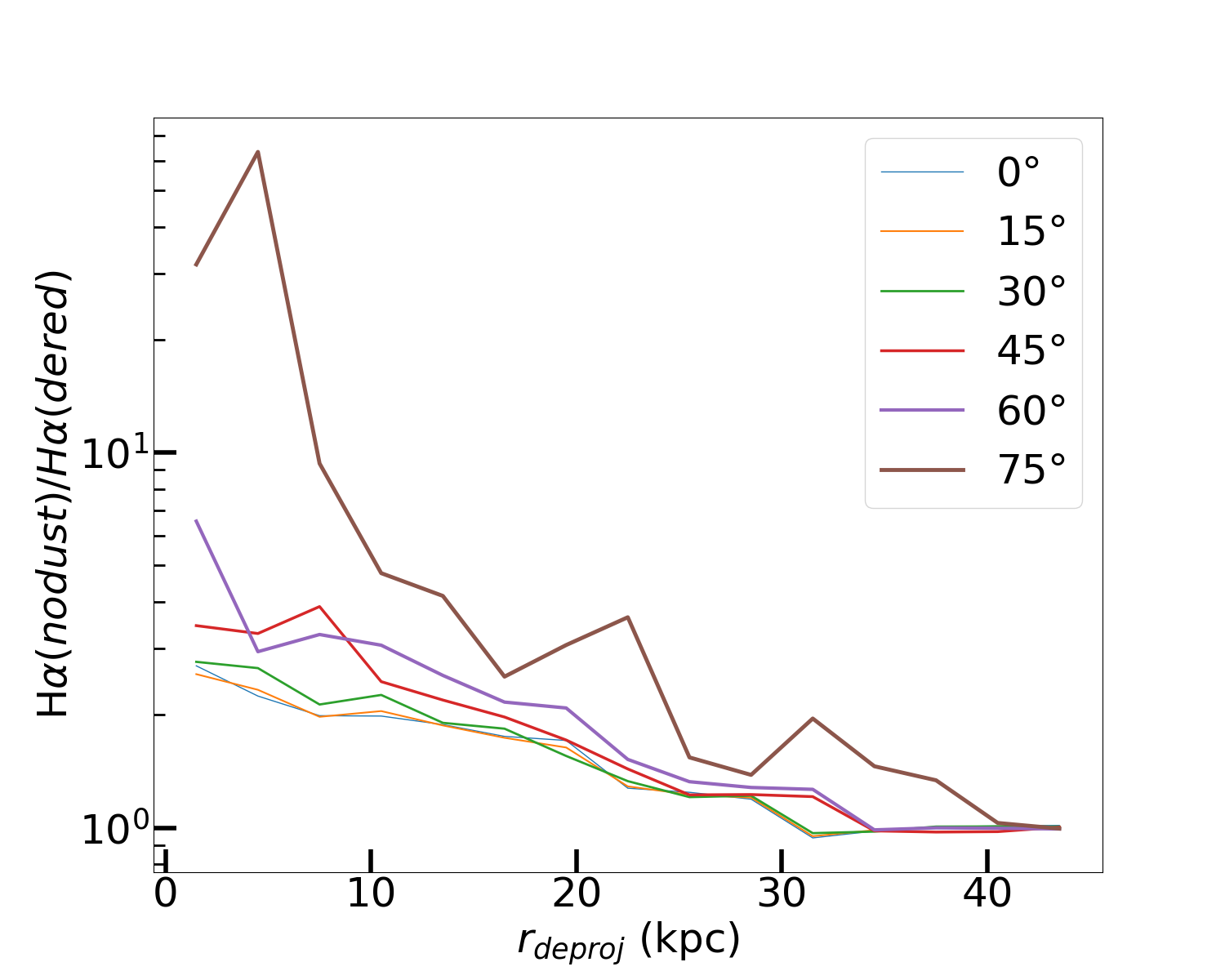}
			\end{center} 
			\caption{The ratio of de-reddened, deprojected intrinsic no-dust \Ha flux measured in annular apertures, to that recovered using the \Ha/\Hb ratio, for inclination angles ranging from 0 to 75$^{\circ}$. Note: we ratio the fluxes after integrating over the aperture, as opposed to calculating the average ratio in the aperture.}
			\label{fig:Ha_true_dered_profile}
		\end{minipage}
	\end{center}
\end{figure}

In Figure~\ref{fig:Ha_true_dered_profile}, we show how the ratio of the intrinsic to the de-reddened flux varies as a function of deprojected galactocentric radius (i.e.\ the radius as measured after correcting for inclination) for a range of different galactic inclinations. The values depicted are those obtained after averaging over an annulus of thickness ${\rm d}r = 3~{\rm kpc}$, which has the effect of smoothing out small-scale variations, e.g., associated with spiral arms. When de-projecting the ``observed'' maps, we assume that all objects are in an infinitely thin plane. We also extend the outer radii to 45~kpc so that \hii regions at significant heights above the galactic midplane do not fall outside of the deprojected image.

Several different physical effects influence the form of these profiles. Our {\sc warpfield} models have internal extinctions which directly affect the amount of emergent \Ha emission and the intrinsic surface brightness of the \hii region. As galactic inclination increases, the amount of extinction along the line-of-sight to these clouds decreases their flux. Simultaneously, along the same line-of-sight diffuse gas emission begins to compete until the pixel is dominated by DIG emission, and individual \hii regions become hidden. 

So long as the emergent light of the \hii regions is brighter than the DIG, we can use the standard de-reddening technique to correct for the effects of extinction and recover a good estimate of the intrinsic \Ha emission. When the inclination angle is low, the surface brightness of the \hii regions remain high in general and the Balmer correction is effective. Nevertheless, even in this case, some clusters are so embedded that they are effectively hidden by the DIG. This is a rare occurrence in the outer reaches of the galaxy, but becomes more common as we move towards the center, resulting in a steady increase in the ratio of intrinsic to de-reddened flux with decreasing galactocentric radius. 

However, as the inclination of the galaxy increases, a point is reached where emission from the foreground DIG dominates the emission in both the \Ha and \Hb lines, with the latter being more affected. Once this occurs, the measured Balmer decrement simply traces conditions in a low optical depth layer of the DIG, and hence no longer allows us to accurately correct the flux from the \hii regions. As a result, the de-reddened \Ha flux can in this case dramatically underestimate the intrinsic flux, by a factor of ten or more (see e.g.\ the behaviour of the $i = 75^\circ$ galaxy at small deprojected radii in Figure~\ref{fig:Ha_true_dered_profile}). As one would expect, this is a much bigger problem at small radii, where the diffuse emission is bright and the forground extinction is considerable, than at large radii, where the diffuse emission is fainter, and gas density is lower. 

\section{Results}
\label{sect:results}

We have analyzed our Milky Way analog in three ways. First, in terms of local conditions, such as the gas temperature and ionization state, where we see significant variations resulting from the variation in the fraction of the local radiation field coming from young hot stars versus the older, evolved stellar population. 
Second, as observed in two ionized gas tracers from locations representative of the position of the Solar System in the real Milky Way. Finally, as 
it would appear to an observer seeing the Milky Way from the outside. In this section, we summarize and discuss a few of the key results of this analysis. 

\subsection{Synthetic All-Sky Milky Way Emission Maps}

To demonstrate the power of our new approach, we have presented synthetic \Ha and \SIII emission maps produced using a simulated Milky Way-like galaxy. We have compared the synthetic \Ha maps to observations of \Ha in the real Milky Way. The same comparison cannot be done for \SIII, since no large-scale maps of this line yet exist for the Milky Way, and so in this case our results are predictions for what will be observed by future large surveys such as the Local Volume Mapper project.\footnote{\url{https://www.sdss.org/future/lvm/}}

We find relatively good agreement between our synthetic \Ha maps and observations of the Milky Way. Since the observed \Ha flux depends both on the internal structure and extinction of the individual \hii regions and also the larger-scale distribution of gas density and ionization within the galaxy, the fact that we find good agreement with the observations suggests that our model is doing a reasonable job of capturing both of these features of the real galaxy.

To produce these results, we adopt a mean cloud density\footnote{Note that this value is the mean density of the entire cloud, including any CO-dark molecular gas or atomic gas associated with it. The mean density of any portions of the cloud traced by bright CO emission could plausibly be somewhat higher.} of $n = 100 \: {\rm cm^{-3}}$ and a star formation efficiency on cloud scales of $\epsilon \approx 1\%$. With these parameters, we find that a significant number of the clouds undergo re-collapse and form multiple stellar populations with age spreads of a few Myr.
We have not explored in detail the sensitivity of our results to variation of these parameters, but we can nevertheless make some qualitative statements. For example, if we had assumed a much higher average cloud density, such as $n = 1000 \: {\rm cm^{-3}}$, then many more clouds would have undergone re-collapse. Moreover, cloud expansion would typically have stalled at a much earlier stage, resulting in far higher internal extinctions and little resulting \Ha emission. In this case, our average \hii regions would have been significantly fainter, meaning that we would no longer match the observations on both large and small scales.
In this way, we see that input parameters such as the mean cloud density that are difficult to directly constrain from observations can be indirectly constrained by finding the range of values for which our synthetic maps match the real ones.

Our maps of \SIII emission show that it typically penetrates to much greater distances in the galactic disk than \Ha and is also less confused by foreground emission. Both of these features can be easily understood in terms of the basic physics of the \SIII line. Ionizing sulphur to S$^{++}$ requires significantly higher energy photons than ionizing hydrogen to H$^{+}$, and so \SIII emission primarily traces regions where the flux of these energetic photons is large, i.e.\ regions close to massive clusters, with little being produced in ionized gas lying far away from the clusters. Once emitted, the \SIII photons propagate further than \Ha photons simply because the difference in their wavelengths  makes them much less susceptible to attenuation by dust. 

The fact that \SIII penetrates the disk of a galaxy much more than \Ha, and that it is less confused by foreground emission means that at low angular or spatial resolution, simple single object photo-ionization models will almost certainly fail to reproduce its relative intensity compared to other emission lines, since the observations will be probing emission produced by the superposition of many distant sources with small angular size. This is much less of an issue for \Ha because the sources probed by that line are typically much closer, and hence have larger angular sizes and suffer less from confusion.  

A variable, but significant escape fraction of ionizing radiation from individual star-forming regions can also create locally bright diffuse ionized gas with morphologies indistinguishable from traditional \hii regions. The emission of the illuminated DIG can compete with that from our inserted {\sc warpfield} models, leading to potentially ambiguous interpretations of the expansion of star-forming regions, absent kinematic data. To account for this, it is important for models of the coupling of feedback with individual molecular clouds to also account for the escape of ionizing radiation and its impact on the larger-scale surrounding environment, as we aim to do in the method presented here.

\subsection{Extragalactic Systems}
We have produced \Ha and \SIII emission maps to illustrate the connection between small scale physics and kpc-scale observations. Projection of the model galaxy from the view point of an exterior observer makes it possible to compute the emission from diffuse ionized gas (DIG) and from individual \hii regions simultaneously in a self-consistent fashion. We find the intrinsic-to-recovered \Ha emission to systematically vary with galactic radii, depending on local extinction, DIG emissivity, and critically, on the fraction of objects in the deeply embedded phase, which in turn depends on cloud evolution and the local star formation rate. Winds and radiation have very different impact on the evolution of clouds with different masses. For the cloud population in an entire galaxy, this becomes a function of the cloud mass distribution and the global star formation rate, as outlined in Sect.~\ref{sect:ClusterMassDistribution}. For some tracers, this may result in the emission from part or all of the galaxy becoming highly stochastic in the case where the emission of the tracer is dominated by the youngest and most massive clusters. This will have 
significant implications for the interpretation of observed emission line ratios and the derivation of physical parameters such as the overall SFR and the gas metallicity \citep[see e.g.][]{Richardson2019, Kewley2001, Dopita2016}. The characterization of this uncertainty is highly challenging and requires the use of a time-dependent population synthesis model as introduced in this study. We plan to explore this issue in more detail in a future paper.

\subsection{Impact of deeply embedded star clusters} 
Our model predicts that a Milky Way type galaxy should contain a significant population of faint and deeply-embedded star-forming regions ($\rm A_V \ge 5$) at any given time. These potentialy spend up to tens of Myr forming stars in a cycle of collapse and expansion. This result depends not only on the detailed modeling of all relevant stellar feedback processes, as implemented in {\sc warpfield}, but also on how nature samples density, mass, and star formation efficiency within the galaxy. Because this population of young star clusters is very difficult to measure it makes the interpretation of the observed line fluxes more difficult and introduces uncertainty to many inferred physical parameters such as SFR or metallicity. When observing external galaxies we could demonstrate that  we are able to recover the total intrinsic dust free \Ha flux to better than 50\% error by using the \Ha/\Hb ratio for viewing angles of less than $45^{\circ}$. This is in line with expected values for dereddened galaxies \citep{Calzetti2001}. 
However, for more edge-on galaxies, the uncertainty can be as large as a factor of ten. We note that direct application of such results to observations should be done with caution. It would be desirable to derive a correction factor that only depends on the relative viewing angle to the galactic disk. However, this will be difficult as the result may also depend on natal cloud density, star formation efficiency and metallicity, as well as galactic morphology. This requires further investigation.

\section{Summary}
\label{sect:Summary}
Existing methods for modelling emission lines from individual star-forming regions \citep[e.g.][]{Kewley2001} or entire galaxies \citep[e.g.][]{Ceverino2019} most often rely on relatively unconstrained parameters to describe the regions. These simplify the distribution of complex structure, variations  of the ionization parameters, densities, and object ages that make up real galaxies. However, the underlying emission properties of individual clusters are set by the relative thickness of the shell surrounding the \hii region, which in turn is set by the balance of winds and radiation in a complex non-linear manner (Pellegrini et al., in prep.). Consequently,  accurate models of the emission from entire galaxies cannot simply assume values or precomputed distributions for all of these parameters, but must instead derive them from accurate models of the full feedback-induced dynamical evolution of each \hii region. Without such a physical basis to ground the models there is an infinite number of possibly degenerate permutations. 

In this paper, we have demonstrated that it is both possible and necessary to combine physically self-consistent models of individual star-forming regions with the results of cosmological simulations of star formation and to use the result to make predictions of the corresponding line emission on scales ranging from tens of parsecs to the size of the entire galaxy. Our small-scale models include the effects of winds, radiation and supernova feedback, as well as the influence of gravity. The evolution and emission spectrum of each source is deterministic, depending on cloud parameters, not our detailed treatment of feedback. Our use of clouds with finite masses and physical scales (as opposed to dimensionless models of SF regions) allows us to calculate the emergent radiation into the galaxy.
This combination of physical scales and physics is beyond the reach of any large-scale simulation to date, with the important caveat that a 1D geometry is assumed. 

Models that parameterize \hii region emission in terms of the ionization parameter $U$, metal abundance $Z$ and number density $n_{\rm H}$ \citep[e.g.][]{Kewley2001} have the drawback that they may include results that are non-physical, in the sense that they would not be reached during the dynamical evolution of any real star-forming region. Because we explicitly follow the dynamical evolution of the \hii regions under the influence of all of the relevant feedback mechanisms, we avoid this problem. One consequence of this is that the ionization parameter is a prediction of our model rather than a tunable parameter. Instead, the key tunable parameters in our model are 
the mean natal cloud density, cloud mass, metallicity, and star formation efficiency (or, alternatively, the cluster mass). Since the models are relatively fast to run, it is reasonable to contemplate varying these parameters and finding which values best fit observations of real galaxies, but this is a topic that lies beyond the scope of this introductory paper. 

We find that the origin of the observable flux in the tracers that we examine in this introductory study (\Ha, \SIII) arises from a complex distribution of cluster masses and ages in different galactic environments. Realistic models therefore need to take the full time evolution of the contributing star-forming region into account in concert with the appropriate environment, such as in the {\sc warpfield-pop} approach presented here. Attempts based on simple self-similarity solutions fail to account for the enormous complexity of star formation and the impact of stellar feedback on the resulting emission from the multi-phase interstellar medium in galaxies across cosmic time. 

\appendix
\section{Multipole fitting}
\label{app:Multipole}
We quantify the structure of our all-sky emission maps by computing their angular power spectrum. Here, we briefly outline how we go about this.
The pattern projected on the sky can be written as a series of spherical harmonics:
\begin{equation}
\label{eq:sph}
S(\vartheta,\varphi) \simeq \sum_{\rm l=0}^{\rm N}\sum_{\rm m=-l}^{\rm l}{a_{\rm l,m}Y_{\rm l,m}(\vartheta,\varphi)}\, .
\end{equation}
Here, $S$ stands for any signal we presented in this work so far, $Y_{\rm l,m}(\vartheta,\varphi)$ is the spherical harmonic and $a_{\rm l,m}$ is the fit coefficient. Mathematically, Equation~\ref{eq:sph} is exact only when we allow the sum over $l$ to go to infinity, but in any actual fitting procedure, the computation has to stop at a distinct value $N$.  All-sky signals can then be quantified in terms of the fit coefficients $a_{\rm l,m}$ for each multipole $l$.
The resulting spectrum is usually quantified by the single parameter function
\begin{equation}
f(l)= \frac{l(l+1)C_{\rm l}}{2\pi},
\end{equation}
where $C_{\rm l}=\rm{Var}\left( \left| a_{\rm l,m} \right| \right)$ is the variance of the magnitude of the complex fit parameter $a_{\rm l,m}$ over all possible values of $m$. We perform this kind of analysis with the implementation provided by the python package {\sc healpy}\footnote{http://healpix.jpl.nasa.gov}.

\section{Extragalactic Emission Maps}

\begin{figure}
	\begin{center}
		\begin{minipage}[c]{1.0\linewidth}
			\begin{center}
				\includegraphics[width=1.0\textwidth]{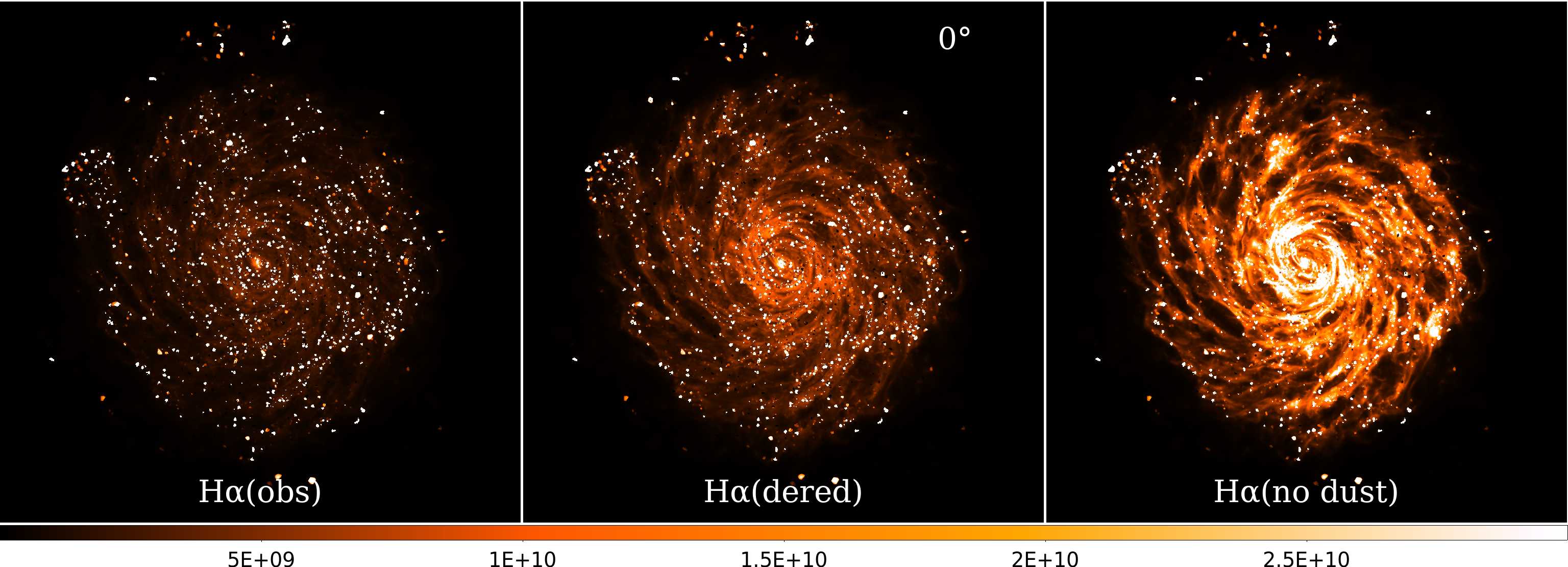}
				\includegraphics[width=1.0\textwidth]{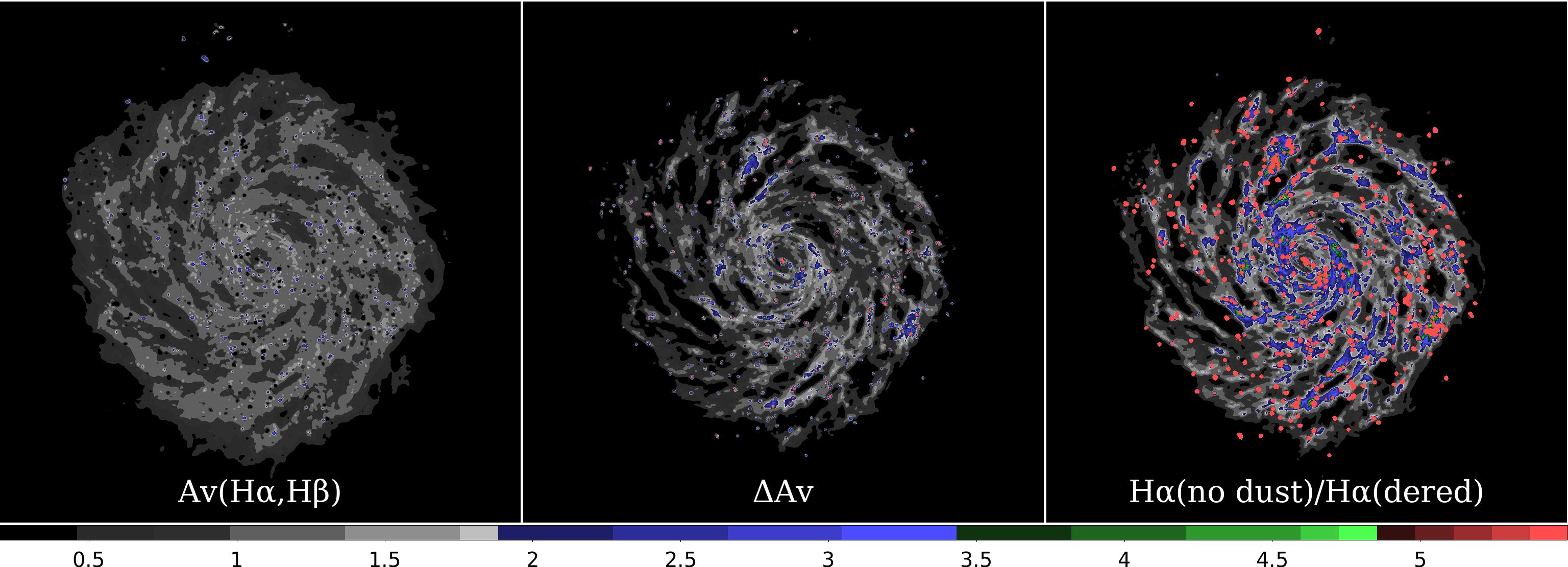}
			\end{center} 
		\end{minipage}
	\end{center}
	\caption{Same as Figure \ref{fig:ExtMapHaHatrue15} but at 0$^{\circ}$.}
	\label{fig:ExtMapHaHatrue0deg}
\end{figure}

\begin{figure}
	\begin{center}
		\begin{minipage}[c]{1.0\linewidth}
			\begin{center}
				\includegraphics[width=1.0\textwidth]{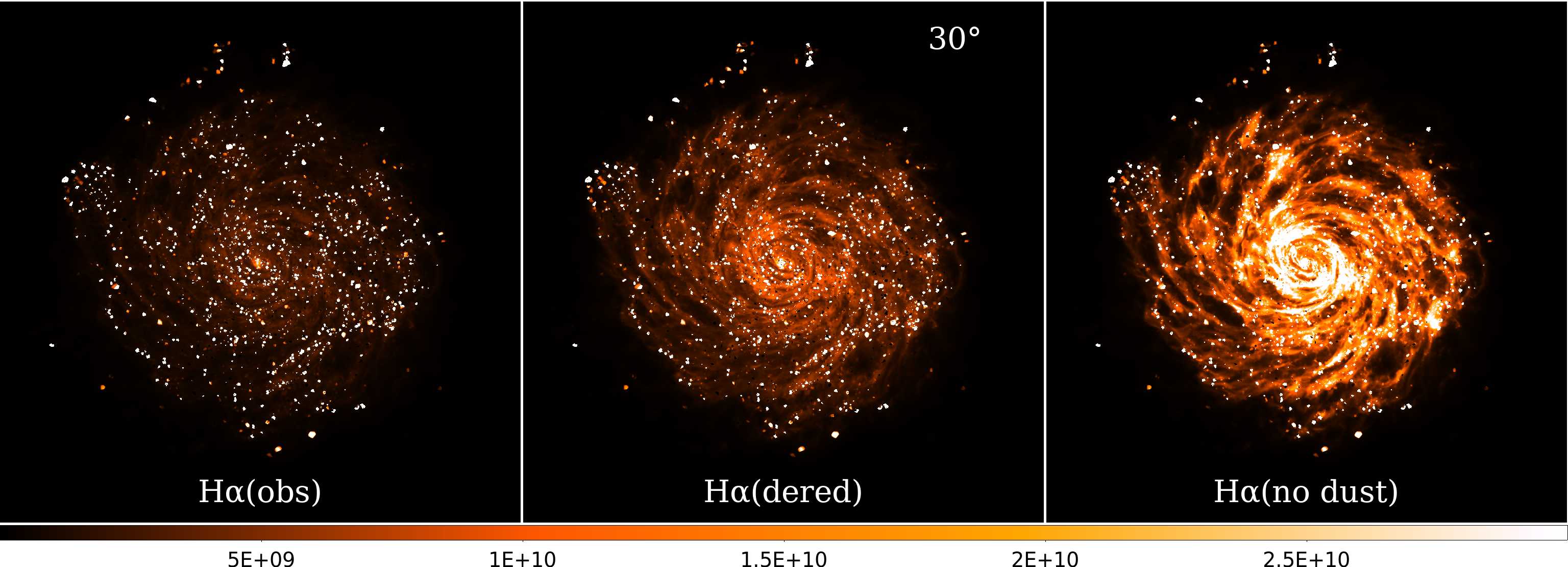}
				\includegraphics[width=1.0\textwidth]{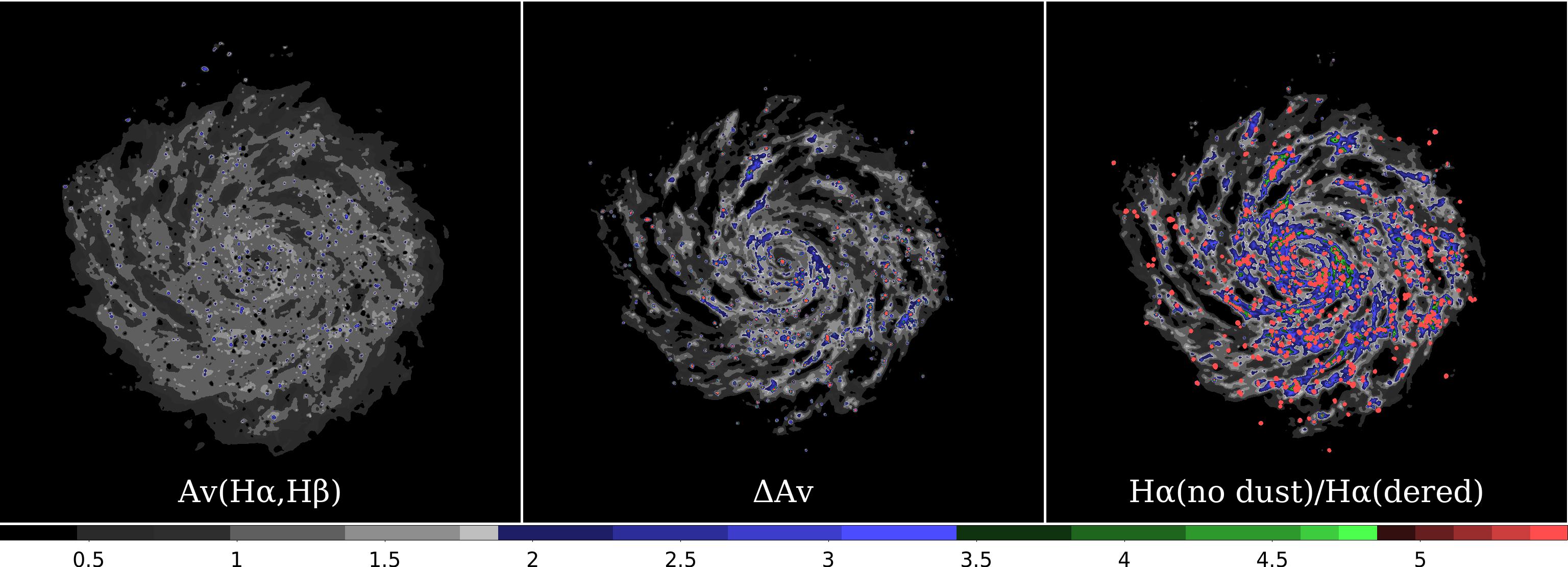}
			\end{center} 
		\end{minipage}
	\end{center}
	\caption{Same as Figure \ref{fig:ExtMapHaHatrue15} but at 30$^{\circ}$.}
	\label{fig:ExtMapHaHatrue30deg}
\end{figure}

\begin{figure}
\begin{center}
	\begin{minipage}[c]{1.0\linewidth}
		\begin{center}
				\includegraphics[width=1.0\textwidth]{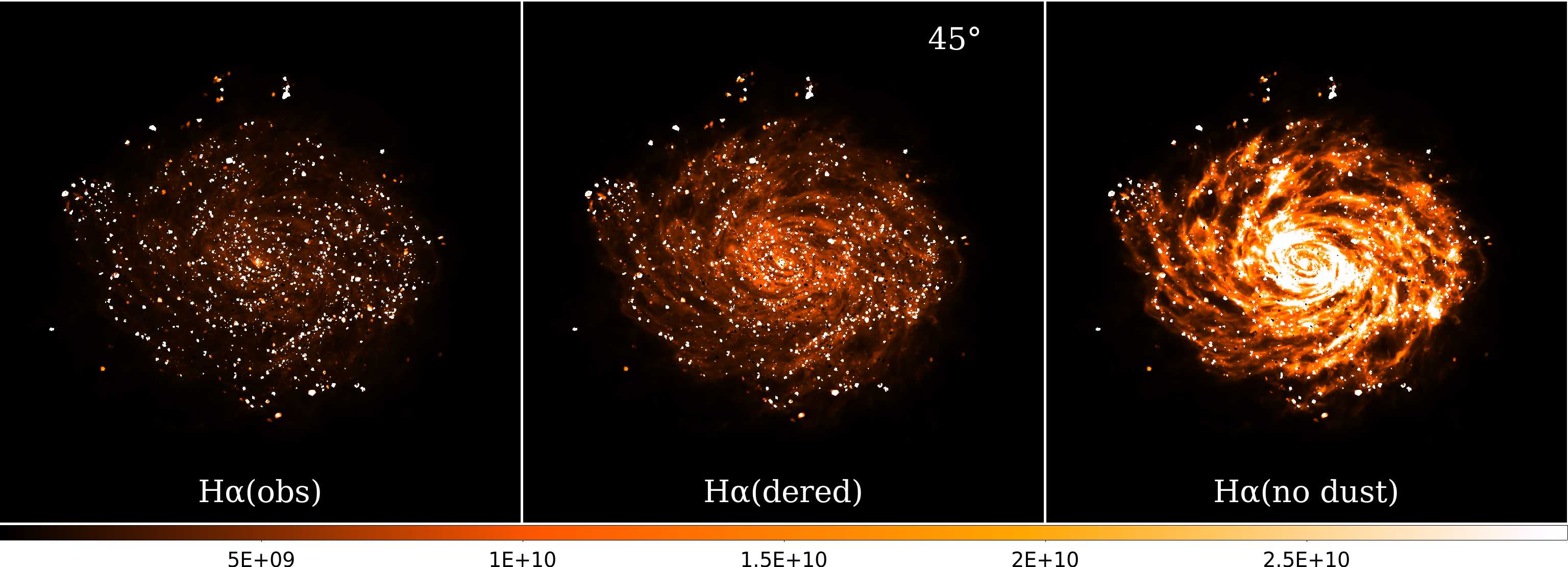}
				\includegraphics[width=1.0\textwidth]{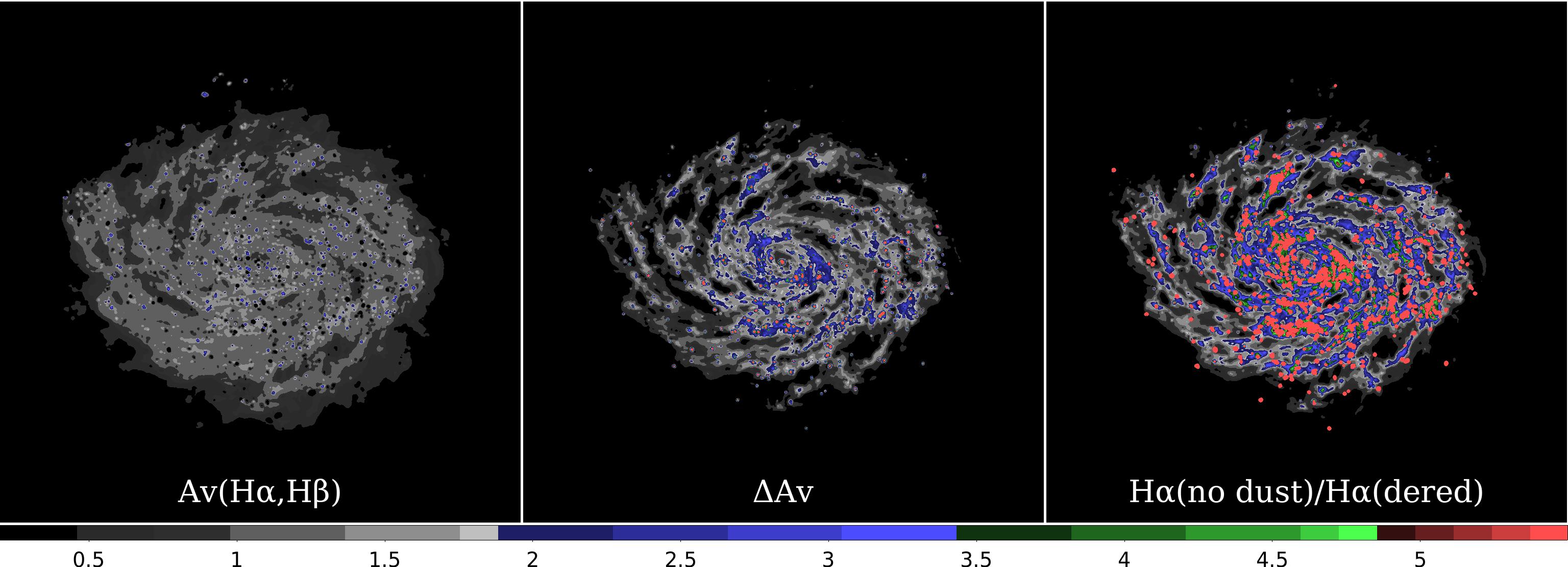}
		\end{center} 
	\end{minipage}
\end{center}
	\caption{Same as Figure \ref{fig:ExtMapHaHatrue15} but at 45$^{\circ}$.}
\label{fig:ExtMapHaHatrue45deg}
\end{figure}

\begin{figure}
\begin{center}
	\begin{minipage}[c]{1.0\linewidth}
		\begin{center}
				\includegraphics[width=1.0\textwidth]{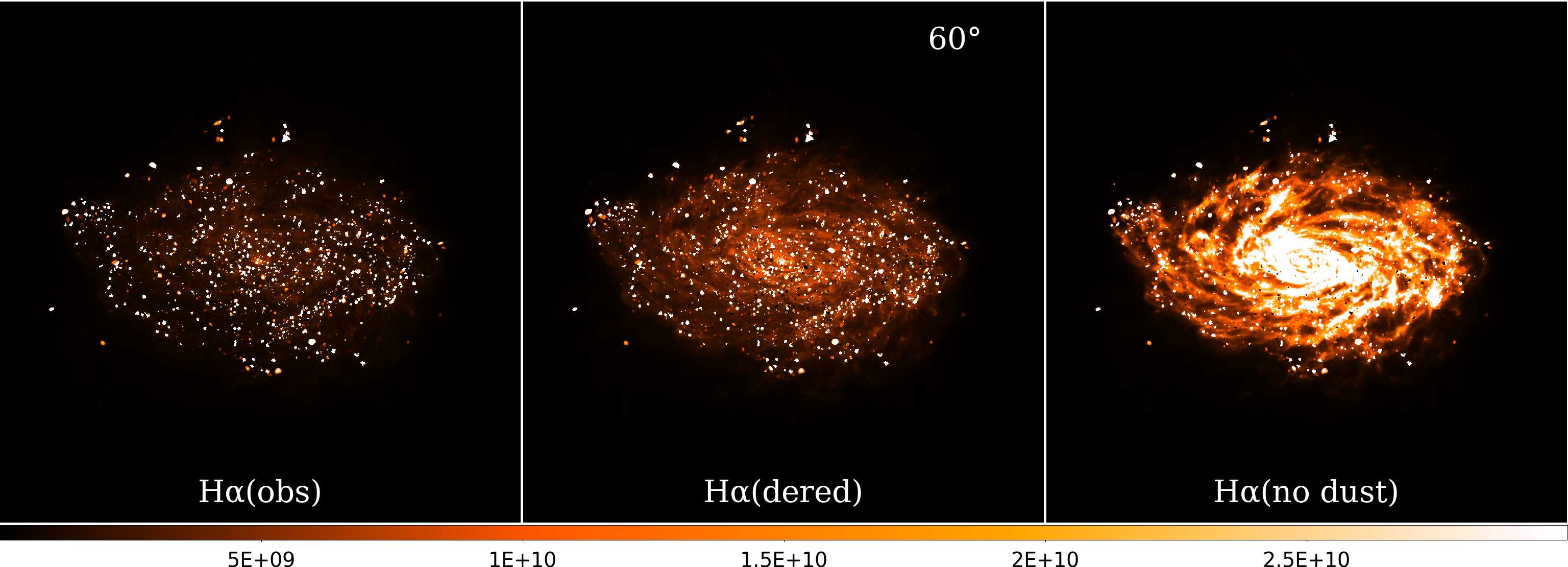}
				\includegraphics[width=1.0\textwidth]{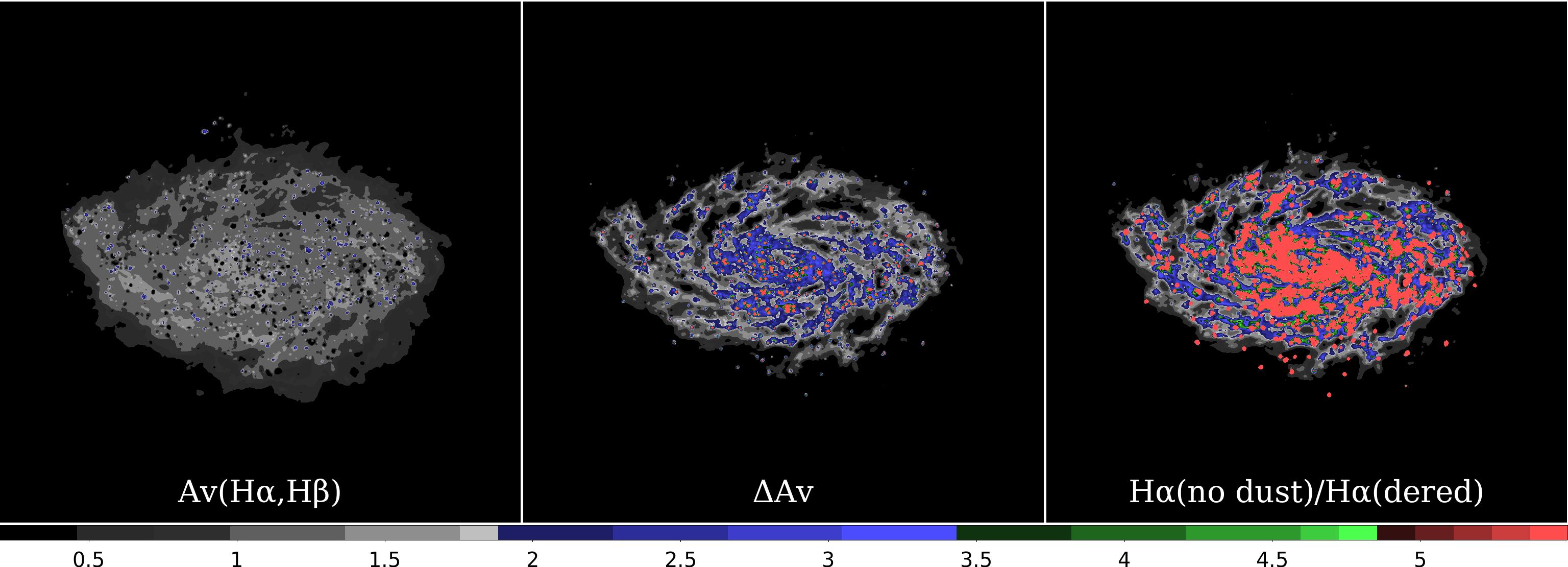}
		\end{center} 
	\end{minipage}
\end{center}
	\caption{Same as Figure \ref{fig:ExtMapHaHatrue15} but at 60$^{\circ}$.}
\label{fig:ExtMapHaHatrue60deg}
\end{figure}

\begin{figure}
\begin{center}
	\begin{minipage}[c]{1.0\linewidth}
		\begin{center}
				\includegraphics[width=1.0\textwidth]{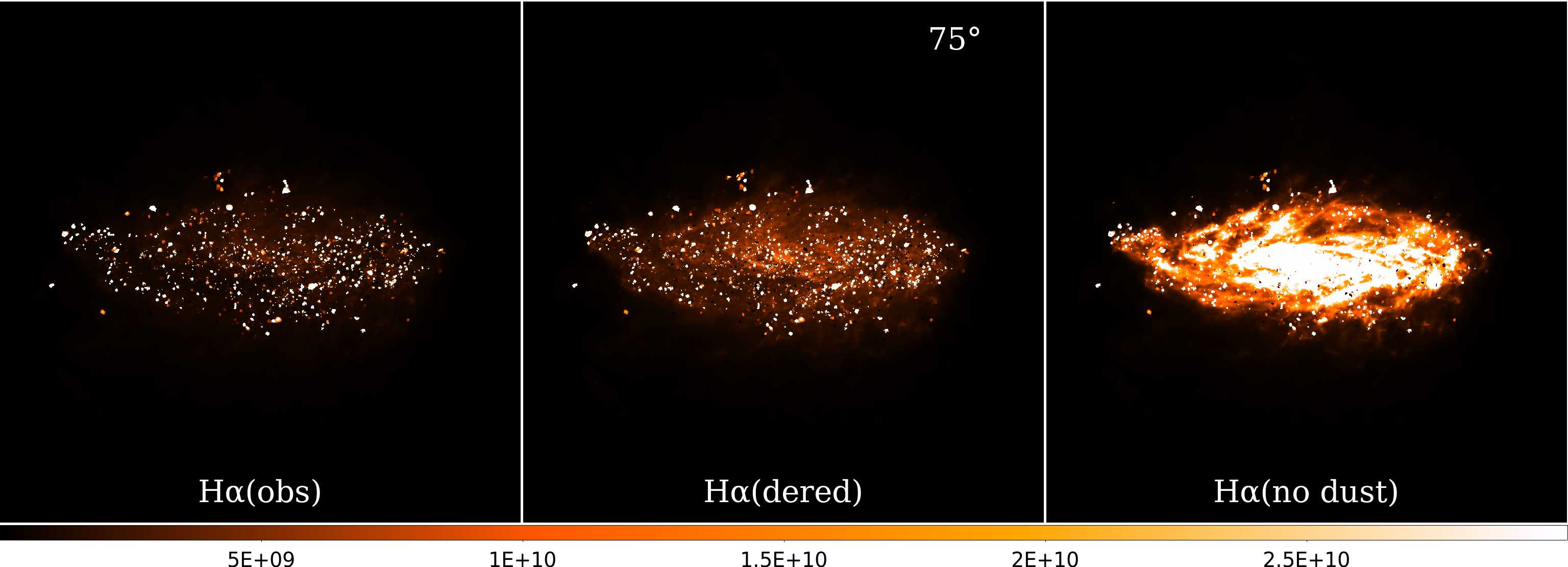}
				\includegraphics[width=1.0\textwidth]{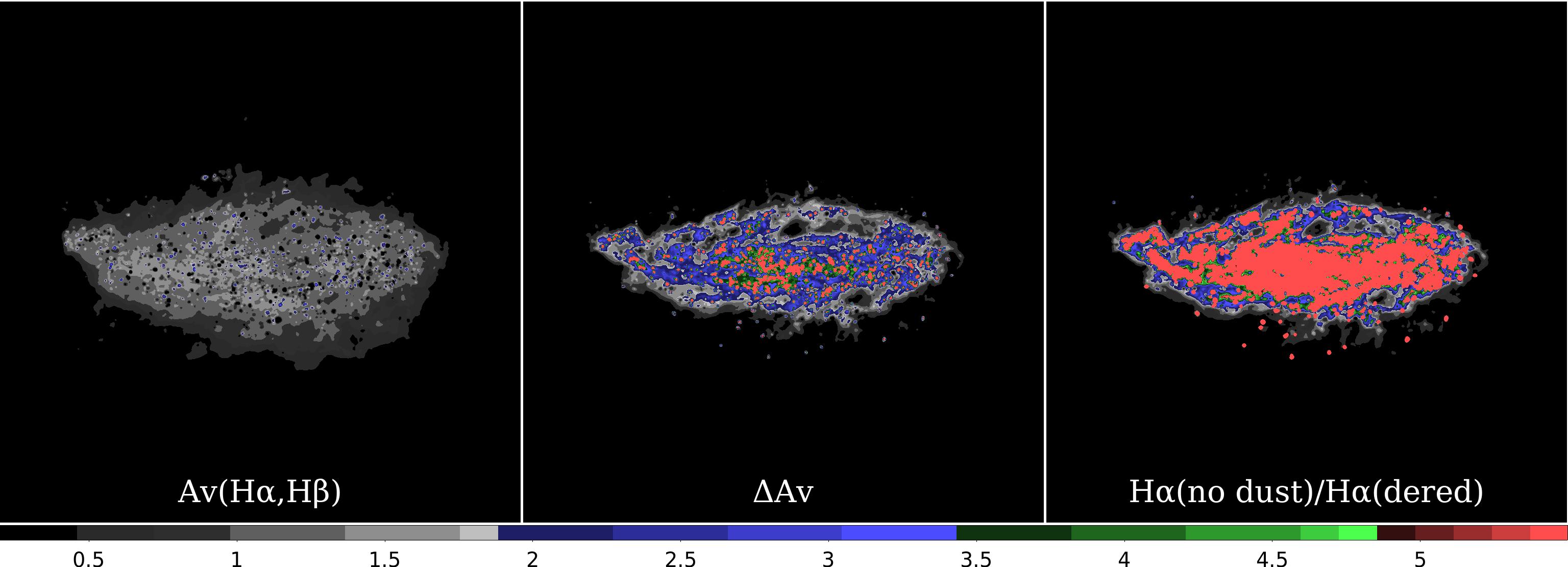}
		\end{center} 
	\end{minipage}
\end{center}
	\caption{Same as Figure \ref{fig:ExtMapHaHatrue15} but at 75$^{\circ}$.}
\label{fig:ExtMapHaHatrue75deg}
\end{figure}

\begin{figure}
\begin{center}
	\begin{minipage}[c]{1.0\linewidth}
		\begin{center}
				\includegraphics[width=1.0\textwidth]{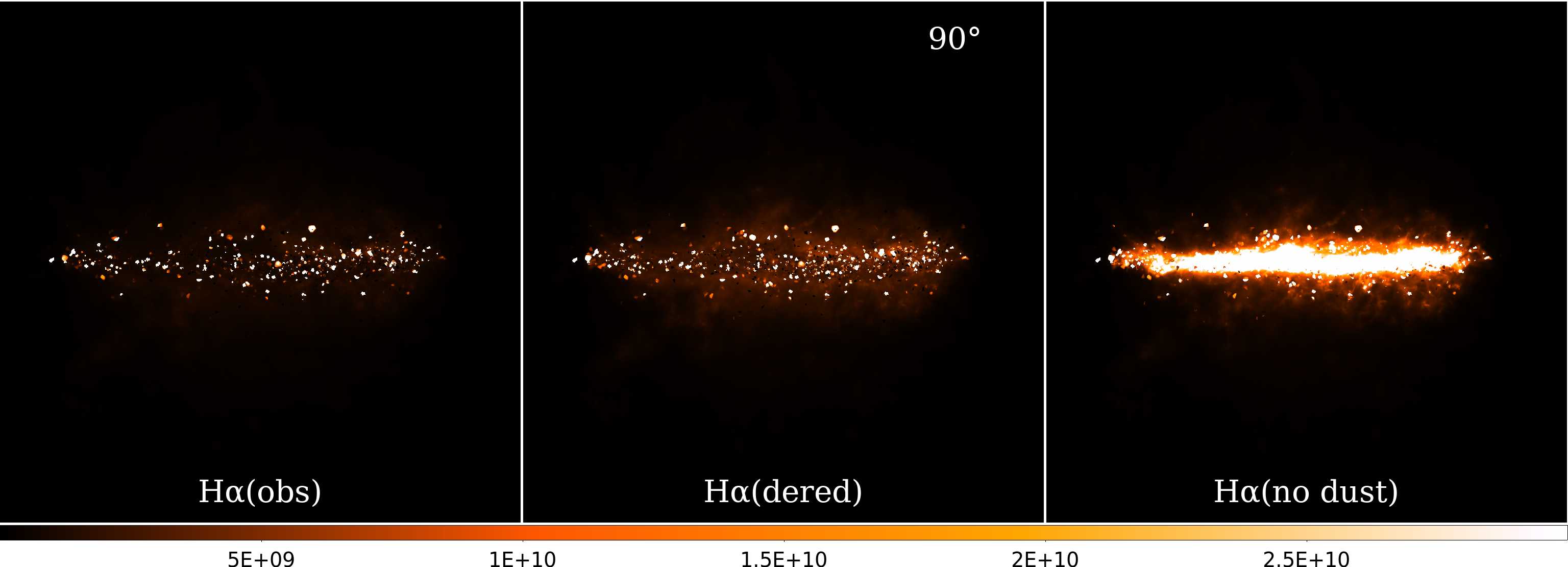}
				\includegraphics[width=1.0\textwidth]{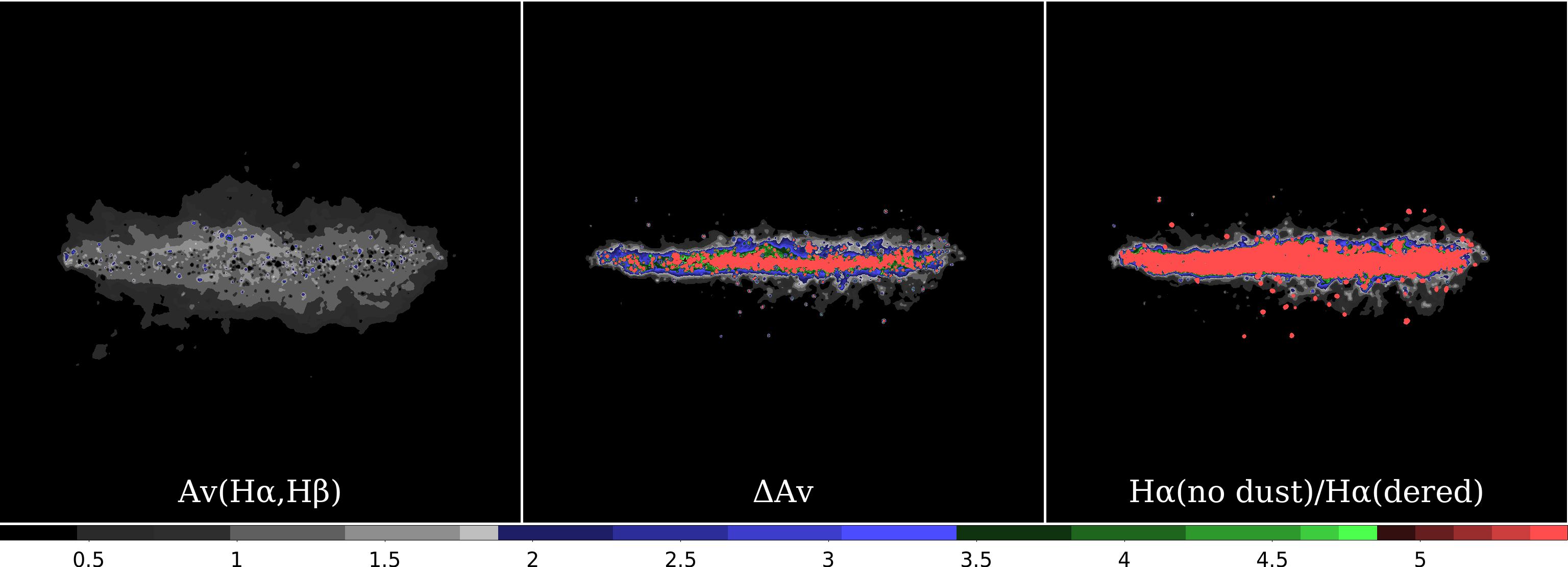}
		\end{center} 
	\end{minipage}
\end{center}
	\caption{Same as Figure \ref{fig:ExtMapHaHatrue15} but at 90$^{\circ}$.}
\label{fig:ExtMapHaHatrue90deg}
\end{figure}

For completeness we include the extragalactic projections at different inclinations angles used to calculate the profiles in Figure~\ref{fig:Ha_true_dered_profile} as  Figures~\ref{fig:ExtMapHaHatrue0deg}-\ref{fig:ExtMapHaHatrue90deg}. Apart from the differing inclinations, the details of these figures are the same as for Figure~\ref{fig:ExtMapHaHatrue15}, including both the physical and the intensity scale.

\newpage
 
\section*{Acknowledgements}
\footnotesize{Special thanks goes to the Auriga collaboration for generously sharing their data. We also thank Mattia Sormani for useful discussions regarding population sampling. S.R., E.W.P., R.S.K, D.R.\ and S.C.O.G.\ acknowledge  support  from  the  Deutsche  Forschungsgemeinschaft via the Collaborative Research Center (SFB 881) ``The Milky Way System'' (subprojects B1, B2, and B8) and the Priority Program SPP 1573 ``Physics of the Interstellar  Medium''  (grant  numbers  KL  1358/18.1,  KL  1358/19.2).}

\bibliographystyle{mnras}
\bibliography{bibtex} 

\bsp	
\label{lastpage}
\end{document}